\documentclass[review]{elsarticle}
\usepackage{graphicx} 
\usepackage{amssymb} 
\usepackage{amsmath} 
\usepackage{amsfonts} 
\usepackage{slashed} 
\usepackage[dvipsnames]{xcolor} 
\usepackage{lineno,hyperref}
\usepackage{epstopdf}
\modulolinenumbers[5]

\journal{Journal of \LaTeX\ Templates}

\begin{document}

\begin{frontmatter}

\title{The impact of the LHCb $B^{\pm}$ production cross-section on parton distribution functions and determining the mass of heavy quarks}

\author{A. Aleedaneshvar}
\author{A. Khorramian}
\address{Faculty of Physics, Semnan University, P.O. Box 35131-19111, Semnan, Iran}


\begin{abstract}
In this work, the impact of recent measurements of heavy-flavour production in $pp$ collisions
on parton distribution functions (PDFs) and their uncertainties is studied. In this regard, the absolute and normalised cross sections of beauty hadron production measured by
LHCb at center-of-mass energy of 7 TeV and 13 TeV are separately included in the
next-to-leading order (NLO) global QCD analysis together with the measurements of the inclusive
and heavy-flavour production cross sections at HERA. It is illustrated that the heavy-flavour data of the LHCb experiment impose additional constraints
on the PDFs, especially on the gluon distribution at low partonic fractions $x$ of the proton momentum. One of the most important results of the present analysis is the significant reduction of the gluon uncertainties in the region less than $x=10^{-4}$ that can play a crucial rule in many areas of high energy physics investigations. 
\end{abstract}


\end{frontmatter}

%\linenumbers
\section{Introduction} 
\label{sec:one}
An accurate understanding of the nucleon structure and also hadronic properties is
one of the important tasks of modern particle physics, in particular, in 
the theory of quantum chromodynamics (QCD). According to the factorization theorem~\cite{Collins:1989gx,Brock:1993sz} of QCD,
for a wide range of hard processes in high-energy lepton-nucleon scatterings and nucleon-nucleon collisions,
the cross sections can be divided in two parts: the universal sets of parton distribution functions (PDFs) which
are nonperturbative objects, and short distance partonic processes which can be calculated perturbatively. 
In this regard, an accurate knowledge of PDFs is essential for predictions at hadron colliders, especially at the large hadron collider (LHC)~\cite{Perez:2012um,Forte:2013wc,Rojo:2015acz,Butterworth:2015oua}. 

From a conceptual point of view, PDFs represent probability densities to find a parton of 
longitudinal fraction $ x $ of the nucleon momentum at a factorization scale $ \mu_f $. 
It is well known now that the $ x $-dependence of PDFs cannot be derived from first principles of QCD
so that they should be extracted from a wide range of experimental data through a global QCD analysis~\cite{Abramowicz:2015mha,Jimenez-Delgado:2014twa,Dulat:2015mca,Harland-Lang:2014zoa,Ball:2014uwa,Schmidt:2015zda,Accardi:2016qay,Wang:2016sfq,Alekhin:2017kpj,Ball:2017nwa}. The scale evolution of PDFs is driven by Dokshitzer-Gribov-Lipatov-Altrarelli-Parisi (DGLAP) 
equations~\cite{Altarelli:1977zs,Gribov:1972ri,Dokshitzer:1977sg}.
In order to make precise theoretical predictions for Standard Model (SM) processes at the LHC
and also searching for new physics, the reduction of the uncertainty of PDFs is a very important issue.

Overall, the deep inelastic scattering (DIS) measurements, like the inclusive~\cite{Abramowicz:2015mha}
and heavy-flavour production cross sections~\cite{Abramowicz:2014zub,Abramowicz:1900rp} measured at HERA, are our main sources to obtain essential information on the proton PDFs since they cover a broad range 
in $ x $ and $ \mu_f $. Nevertheless, in order to gain a better flavour separation and 
decrease PDF uncertainties, especially for the sea quarks and gluon, the inclusion of the LHC measurements
from various processes such as the jet production~\cite{Aad:2010ad,Aad:2011fc,Aad:2013lpa,Chatrchyan:2012bja}, 
inclusive electroweak boson production~\cite{Aad:2011dm,Aad:2011fp,Chatrchyan:2011jz,Chatrchyan:2012xt,Chatrchyan:2013mza,Aaij:2012vn,Aaij:2012mda,Aaboud:2016btc} 
and top quark pair production~\cite{ATLAS:2012aa,Chatrchyan:2012bra,Chatrchyan:2013faa} in the global
analyses of PDFs is inevitably necessary.
Despite all of these measurements, constraints on the gluon PDF are limited to $ x\gtrsim 10^{-4}$
in the perturbative region, so that for smaller values of $ x $, the gluon distribution is poorly known
and there are large uncertainties resulting from the lack of direct experimental information.
Then, additional measurements are necessary to improve our knowledge in these kinematic ranges.
Actually, accurate determination of the low-$ x $ gluon distribution is important
for studies of parton dynamics, non-linear and saturation effects. In addition, neutrino astronomy~\cite{CooperSarkar:2011pa,Garzelli:2015psa,Gauld:2015kvh}
and cosmic ray physics~‍\cite{dEnterria:2011twh} are the other fields that precise gluon distribution at small $ x $ is really
required for them. Of course, making accurate theoretical predictions of physical observables will be
measured in the future higher-energy colliders~\cite{AbelleiraFernandez:2012cc,Mangano:2016jyj}, is of key importance by itself. 

It is well established now~\cite{Cacciari:2015fta,Zenaiev:2015rfa,Gauld:2015yia,Gauld:2016kpd} that the production of charm and bottom
quarks at forward rapidity in proton-proton collisions at the LHC is very sensitive to the gluon 
PDF at low $x$ and can put tighter constraints on it in this kinematic region.
Actually, since the heavy-flavour production at LHC is dominated by the 
gluon-gluon fusion process, the LHCb measurements of charm~\cite{Aaij:2013mga,Aaij:2015bpa,Aaij:2016jht} 
and beauty~\cite{Aaij:2013noa} production in forward rapidity can gain unique information on the gluon distribution
at $10^{-6} \lesssim x \lesssim 10^{-4}$. 
Theoretical prediction of the LHCb heavy-flavour production measurements can
be calculated by convoluting the partonic cross sections for
heavy quark pair production with the input PDFs and the relevant heavy quark fragmentation fractions which are describing the probability of a quark to fragment
into a particular hadron. From a phenomenological point of view, the impact of heavy-flavour measurements of the old LHCb data for charm production at 5, 7 and 13 TeV~\cite{Aaij:2013mga,Aaij:2015bpa,Aaij:2016jht}, and also beauty production at 7 TeV~\cite{Aaij:2013noa} has previously been investigated on the gluon PDF with different approaches. These investigations have performed both in a global QCD analysis~\cite{Zenaiev:2015rfa} together with inclusive and heavy-flavour production cross
sections at HERA, and also on the NNPDF3.0 gluon PDF~\cite{Gauld:2015yia} 
by using the Bayesian reweighting method~\cite{Ball:2010gb,Ball:2011gg}. We describe these studies more accurately in Sec.~\ref{sec:two}. 

Fortunately, the new measurements of the $B^{\pm}$ production cross-section in $ pp $ collisions at $\sqrt s =$ 7 and 13 TeV~\cite{Aaij:2017qml} have recently been presented by LHCb collaboration. These data have been measured as a function of the transverse momentum, $p_T$, and rapidity, $y$, in the region $ 0< p_\textrm{T}< 40 $ GeV and $ 2.0<y<4.5 $ and then can provide unique information on the structure of the proton. Therefore, in this paper, we are going to study the impact of new heavy-flavour measurements of the LHCb~\cite{Aaij:2017qml} 
at $ \sqrt s= $7 TeV and 13 TeV on PDFs, specially gluon distribution, and their uncertainties by performing some global analyses. In addition, the effects of these data on the mass of charm and beauty quarks are also investigated.

The paper is organised as follows. In Sec.~\ref{sec:two}, we briefly
discuss the LHCb measurement of heavy-flavour production 
and present an overview of phenomenological efforts performed in this subject. 
In Sec.~\ref{sec:three}, we introduce our QCD analysis framework 
including the experimental data sets, PDF parametrizations, kinematic cuts, and etc. In Sec.~\ref{sec:four}, we study the impact of recent LHCb measurements of heavy-flavour production cross sections on PDFs and the mass of heavy quarks, 
by including the absolute and normalised LHCb data at $ \sqrt s= $7 TeV and 13 TeV~\cite{Aaij:2017qml} into a global analysis. 
Finally, we summarize our results and conclusions in Sec.~\ref{sec:six}.
%%%%%%%%%%%%%%%%%%%%%%%%%%%%%%%%%%%%%%%%%%%%%%%%%%%%%%%%%%%%%%%%%%%%%%%%%%%%%%%%
\section{Charm and beauty production at the LHCb}
\label{sec:two}
As anticipated in the introduction, the measurements of charm and beauty production in multi-TeV $pp$ 
collisions at the LHC provide a powerful tool for testing QCD. Such measurements can also provide further constraints on PDFs especially
for the sea quarks and gluon if they are performed at forward rapidities~\cite{Zenaiev:2015rfa}.
Fortunately, there have always been a good agreement between the experimental data measured at LHC
and the related theoretical predictions within the estimated systematics. This section provides an overview of the framework for pQCD computations for heavy-flavour production in hadronic collisions
and also the measurements of heavy-flavour production at the LHCb experiment~\cite{Aaij:2013mga,Aaij:2015bpa,Aaij:2016jht,Aaij:2013noa,Aaij:2017qml}.

In any QCD analysis, the experimental measurements are compared to 
corresponding theoretical predictions, so having accurate theoretical calculations is very important to
extract physical quantities. Computationally, the calculation of heavy quark pair production in hadronic collisions at next-to-leading order (NLO)
approximation of QCD has been available for about three decades. These calculations 
include the total inclusive cross-sections~\cite{Nason:1987xz},
as well as the differential distributions~\cite{Nason:1989zy,Beenakker:1990maa,Beenakker:1988bq,Mangano:1991jk,MNRcode}.
The cross sections of inclusive heavy quark pair production have also been calculated at the next-to-next-to-leading order (NNLO)~\cite{Czakon:2012pz,Czakon:2013goa,Baernreuther:2012ws}. 
For the case of differential distributions, one can refer to the NNLO calculations 
of top quark production~\cite{Czakon:2014xsa,Czakon:2015pga} which are also applicable
to charm and bottom production. It is worth pointing out in this context that
the fixed-order calculations can be improved with the resummation of soft gluons at NLL~\cite{Bonciani:1998vc,Kidonakis:1997gm} and
NNLL~\cite{Czakon:2009zw,Ahrens:2010zv} accuracy.
Unfortunately, the perturbative calculations of heavy-flavour production at high energies at NLO include 
substantial theoretical uncertainties. These uncertainties are larger for the case of charm production.
Actually, theoretical uncertainties in heavy-flavour production cross sections can be generated from 
various sources including the renormalisation and factorisation scales dependence,
the value of the heavy-quark masses, and the uncertainties of the parton distribution functions. There are different approaches for making pQCD predictions of heavy quark pair production
including FONLL~\cite{Cacciari:1998it,Cacciari:2001td}, POWHEG~\cite{Nason:2004rx,Frixione:2007vw,Alioli:2010xd}
and M\textsc{ad}G\textsc{raph}5\_aMC@NLO~\cite{Alwall:2014hca}. A brief description of
these approaches can be found in Ref.~\cite{Gauld:2015yia}. It is just worth noting that
FONLL is a semi-analytical approach, while in Monte Carlo
programs POWHEG and M\textsc{ad}G\textsc{raph}5\_aMC@NLO, the fully exclusive description of
the final state has been provided with a feature for passing the results to the P\textsc{ythia}8 parton shower.

The LHCb experiment has wide physics program covering many important aspects of heavy-flavour, Electroweak and QCD physics.
By virtue of the LHCb detector~\cite{Alves:2008zz}, that is, a single-arm forward spectrometer covering the pseudorapidity range $2<\eta <5$,
it is possible to study the particle productions containing the heavy charm or bottom quarks.
In this way, one can explore parton densities in the nucleon especially of the gluon 
in a region which is not accessible with HERA data. The LHCb measurements of heavy-flavour production
have been performed both for charm~\cite{Aaij:2013mga,Aaij:2015bpa,Aaij:2016jht} 
and beauty~\cite{Aaij:2013noa,Aaij:2017qml} production. The charm production data include the measurements
at center-of-mass energy of $ \sqrt s= $ 5 TeV~\cite{Aaij:2016jht}, 7 TeV~\cite{Aaij:2013mga} and 13 TeV~\cite{Aaij:2015bpa}.
Note that, in terms of time, the measurements at 7 TeV had been released earlier
and are based on data corresponding to an integrated luminosity of 15 nb$^{-1}$.
In that analysis, the full reconstruction of decays of the charmed hadrons $D^0$, $D^{+}$, $D_s^+$, $D^{*+}$ and $\Lambda_c^{+}$ has been considered,
 and the cross sections have been measured 
as a function of the transverse momentum, $p_T$, and rapidity, $y$, of the reconstructed hadrons
in the region $ 0< p_\textrm{T}< 8 $ GeV and $ 2.0<y<4.5 $. However, the data sample at 5
and 13 TeV corresponds to an integrated luminosity of 8.60 and 4.98 pb$^{-1}$, respectively.
Moreover, the LHCb measurements at 5 and 13 TeV do not include the reconstruction of decays of $\Lambda_c^{+}$
and cover a somewhat wider range in $ p_\textrm{T} $. Note that in these analyses, the corresponding cross section ratios
$ R_{13/5} $ and $ R_{13/7} $ have also been presented. The ratios of cross sections between
different center-of-mass energies greatly have the advantage to reduce relative
experimental systematic uncertainties compared to the differential cross-sections and also the cancellations of several theoretical uncertainties.
The LHCb data on the production cross-sections of $B$ mesons in $pp$ collisions~\cite{Aaij:2013noa} 
have been measured at $ \sqrt s= $7 TeV and correspond to an integrated 
luminosity of 0.36 fb$^{-1}$. In this case, the $B^+$, $B^0$ and $B_s^0$ mesons 
(as well as their charge-conjugate states) are reconstructed in exclusive decays 
mainly containing a $J/\psi$ in final states. 

From the view point of phenomenology, there are two approaches to study the impact of a specific new experimental data sets on PDFs. A direct way is to perform a new 
global analysis of PDFs considering a reliable range of the experimental data, with and 
without those data we are looking for their effects, and comparing the results with each other.
However, in a different manner, one can use the parton distribution reweighting procedure 
to estimate the impact of desired new data on a special predetermined PDFs set.
To be more precise, this method allows quantifying the impact of new data in a set of
PDFs without needing us to perform the full global QCD
analysis again. It has been used so far in some studies concerning the impact on PDF fits from data
such as the top quark pair production~\cite{Czakon:2013tha} and polarised
$W^{\pm}$ and jet production~\cite{Nocera:2014gqa}.

Regarding the potential of LHCb data on determining the gluon distribution more accurately, various analyses have been done. In Ref.~\cite{Zenaiev:2015rfa}, the PROSA Collaboration investigated the impact of heavy-flavour measurements of the LHCb
at $ \sqrt s= $7 TeV on the gluon PDF in a global QCD analysis together with inclusive and heavy-flavour production cross
sections at HERA. Note that their study was performed based on xFitter framework~\cite{Alekhin:2014irh,Botje:2010ay,James:1975dr}, using the Mangano-Nason-Ridolfi (MNR) code~\cite{Mangano:1991jk,MNRcode} 
in a fixed flavour number scheme (FFNS) with $ N_f=3 $
active flavours. They concluded that these data impose additional
constraints on the gluon and the sea-quark distributions at low $ x $ of the proton momentum, down to
$ x\sim 5\times 10^{-6} $. At the same time, the impact of heavy-flavour measurements of the LHCb
at $ \sqrt s= $7 TeV was investigated on the NNPDF3.0~\cite{Ball:2014uwa} gluon PDF 
using the Bayesian reweighting method by the authors of Ref.~\cite{Gauld:2015yia}
where the FONLL predictions were used for the theoretical calculations.
They found that by inclusion of the LHCb measurements, the PDF
uncertainties in the NNPDF3.0 gluon PDF are reduced by more than a factor three.
It has also been shown that the central value at small-$ x $ of the gluon PDF preferred by the LHCb charm data is less
steep than that of the global fit. The analysis of Ref.~\cite{Gauld:2015yia} has recently been continued in Ref.~\cite{Gauld:2016kpd}
where the authors have investigated the impact of forward charm production 
data provided by LHCb for all three different center-of-mass energies: 5 TeV, 7 TeV, and 13 TeV including the cross-section ratios between data were
taken at different center-of-mass values $ R_{13/5} $ and $ R_{13/7} $ on the NNPDF3.0 gluon PDF.
They have demonstrated these data lead to a reduction of the PDF uncertainties of the gluon down to $ x\simeq 10^{-6} $.
As the last point, note that as mentioned before, because of the large theoretical uncertainties of
the NLO calculations, the direct inclusion of absolute heavy-flavour production
cross sections into a PDF fit can not affect the uncertainties of the extracted PDFs.
However, it has been shown that using normalised representations of the LHCb measurements
can improve description of the small-$ x $ gluon~\cite{Zenaiev:2015rfa,Gauld:2015yia}.
A similar outcome is achieved using the ratios of cross sections between
different center-of-mass energies~\cite{Gauld:2016kpd}.

Recently, LHCb Collaboration has presented a new measurement of the $B^{\pm}$ production cross-section in $ pp $ collisions at $\sqrt s =$ 7 and 13 TeV corresponding to 1.0 $fb^{-1}$ and 0.3 $fb^{-1}$, respectively~\cite{Aaij:2017qml}. In this case, the $B^{\pm}$ mesons 
(as well as their charge-conjugate states) have been reconstructed in decays 
mainly containing a $J/{\psi}K^{\pm}$ in final states. Moreover, the cross sections have been measured as a function of the transverse momentum, $p_T$, and rapidity, $y$, of the reconstructed hadrons in the region $ 0< p_\textrm{T}< 40 $ GeV and $ 2.0<y<4.5 $. 
In this work, 
we are going to study the impact of these new data on PDFs by performing some global analyses. In addition, the effect of these data on the charm and beauty mass is also investigated.  The framework of the theoretical calculations for the heavy-flavour production at $ pp $ collisions, as well as the QCD fit, is described in the next section.

%%%%%%%%%%%%%%%%%%%%%%%%%%%%%%%%%%%%%%%%%%%%%%%%%%%%%%%%%%%%%%%%%%%%%%%%%%%%%%%%%
\section{QCD analysis framework}
\label{sec:three}
In this section, we present a brief overview of the theoretical formalism and experimental data which are used in our QCD analysis for investigating the impact of LHCb data on PDFs and the mass of heavy quarks.  
It should be noted that the PDFs extraction is performed using the xFitter package~\cite{Alekhin:2014irh,Botje:2010ay,James:1975dr}, at the NLO approximation. The bases for all PDFs determination are the HERA DIS measurements, so we use the HERA inclusive production cross sections~\cite{Abramowicz:2015mha} in our analysis to obtain essential information on the proton PDFs. In order to put further limits on gluon distribution and the mass of charm and beauty quarks, the heavy-flavour production cross sections~\cite{Abramowicz:2014zub,Abramowicz:1900rp} measured at HERA are also included in the QCD analysis. In Sec.~\ref{sec:four}, within two separate analyses, the LHCb updated beauty production data at 7 TeV and their new data at 13 TeV~\cite{Aaij:2017qml} are respectively included to study the impact of theses data on the extracted PDFs and the mass of heavy quarks. All experimental data sets used in the present study have been listed in the first column of Table~\ref{tab:one}. 
Note that we applied the kinematics cuts $ Q^2 > 3.5 $ GeV$ ^2 $ and $ W^2 > 15 $ GeV$ ^2 $ on the DIS data to avoid the non-perturbative effects. It should also be noted that in the present work, the QCD coupling constant is taken to be equal to 
$\alpha_S(m_Z)^{N_f=3} =0.1059$ in the NLO 3-flavour $\overline{MS}$ scheme ~\cite{Botje:2010ay}. 
For inclusive and heavy flavor production at HERA, the renormalisation and factorisation scales are set to $\mu_r = \mu_f = Q$ and $\mu_r = \mu_f=\sqrt{Q^2+4m^2_Q}$, respectively, and for LHCb data they are set to $\mu_r = \mu_f=\sqrt{p^2_T+m^2_Q}$, where $m_Q$ is the pole mass of $c$ or $b$ quarks~\cite{Mangano:1991jk,MNRcode}.

Since the main objective of the present study is not a comprehensive global analysis of PDFs and we would like to find out the impact of LHCb data on PDFs,
it is satisfying to use of simple flexible parametrization forms for input parton densities such as HERAPDF form~\cite{Chekanov:2005nn}.
However, it should be noted that the optimal parametrization forms for the PDF fit can be
found through a parametrization scan as described in~\cite{Aaron:2009aa}. The central HERAPDF parametrization at the initial scale $Q_0^2=1.4$ GeV$^2$ is: 

\begin{eqnarray}\label{eq:paramf}
xg(x) &=& A_g x^{{B_g}}(1-x)^{C_g}- A_g^\prime x^{{B_g^\prime}}(1-x)^{{C_g^\prime}}, 
 \nonumber \\
xu_{\rm v}(x) &=& A_{u_{\rm v}} x^{B_{u_{\rm v}}}(1-x)^{C_{u_{\rm v}}}
(1+E_{u_{\rm v}} x^2),
\nonumber \\
xd_{\rm v}(x) &=& A_{d_{\rm v}}x^{B_{d_{\rm v}}} (1-x)^{C_{d_{\rm v}}},\nonumber \\
x \bar U(x) &=& A_{\bar U} x^{B_{\bar U}}(1-x)^{C_{\bar U}}(1+D_{{\bar U}}x),
\nonumber \\
x \bar D(x)&=& A_{\bar D} x^{B_{\bar D}}(1 -x)^{C_{\bar D}}.
\end{eqnarray}
which $xu_{v}(x)$ and $ xd_v(x)$ are the valance distribution, and
 $x\bar U(x)$ and $ x\bar D(x)$ are anti-quarks distribution, where $x\bar U(x)=x\bar u(x)$ 
and $x\bar D(x)=x\bar d(x) + x \bar s(x)$,
and finally $ xg(x)$ is the gluon distribution, with more flexible form. Note that according to the MSTW analysis, for the gluon distribution, the $C_g^\prime = 25$ is fixed~\cite{Thorne:2006qt,Martin:2009ad}.

In the HERAPDF approach, $ A_{u_v} $, $ A_{d_v} $ and $ A_g $ are the normalisation parameters which are determined with the help of QCD sum rules.
Moreover, for the case of $ B $ parameters in sea quark PDFs, 
they are considered to be equal as $  B_{\bar{U}}=B_{\bar{D}} $. Furthermore, an additional  
constraint as $  A_{\bar{u}}=A_{\bar{D}}(1-f_s) $ is considered, which ensures
the same behaviour of  the $ x\bar{u} $ 
and $ x\bar{d} $  as $ x\rightarrow 0 $.
The contribution of the strange quark density is taken to be proportional
to $ x\bar{s}=f_s x\bar{D} $. It has been shown that a value of 0.31 is a good estimation
for factor $ f_s $~\cite{Martin:2009ad,Mason:2007zz,Aad:2012sb}. 
After these simplifying assumptions, the number of unknown parameters, which should be 
determined by the fit, will be 16.

The framework we use in the present study for making the theoretical calculations of the heavy-flavour production at $ pp $ collisions
is similar to the study performed in Ref.~\cite{Zenaiev:2015rfa}. To be more precise, 
we use the massive NLO calculations in the FFNS~\cite{Nason:1987xz,Nason:1989zy,Beenakker:1990maa}
with the number of flavours $ N_f=3 $ that is implemented in xFitter by using original routines from the
MNR calculations~\cite{Mangano:1991jk,MNRcode}. 
The MNR calculations have also been used recently by the LHCb Collaboration to describe the production of prompt $D^0$~mesons 
in proton-lead and lead-proton collisions at the LHC~\cite{Aaij:2017gcy}.
One of the advantages of such calculations
is to include the fragmentation of the heavy quark into a particular final-state hadron.
Actually, for taking into account the transition of the heavy quark 
into the observed heavy-flavoured hadron, one can multiply the cross section with the appropriate branching fraction.
However, using a suitable fragmentation function describing the hadronisation of the heavy quark
can give us everything we are looking for. In the present analysis, following the PROSA Collaboration~\cite{Zenaiev:2015rfa},
we use the parametrisations obtained by Kartvelishvili {\it et al.}~\cite{Kartvelishvili:1977pi}.
The fragmentation function uncertainties are assigned to the measurements and are treated as correlated. However, it should be noted that in addition to using new data, there is another important difference between the present study and the analysis performed before by the PROSA Collaboration.
Actually, in our analysis the newest HERA combined data~\cite{Abramowicz:2015mha} are used which are related to HERA II period, while the PROSA Collaboration used those data were taken during the HERA I period.

%%%%%%%%%%%%%%%%%%%%%%%%%%%%%%%%%%%%%%%%%%%%%%%%%%%%%%%%%%%%%%%%%%%%%%%%%%%%%%%%%%%
\section{Global analysis of PDFs including the LHCb beauty production data}
\label{sec:four}
As mentioned, the main objective of the present paper is studying the impact of LHCb new measurements of the $B^{\pm}$ production cross-section in $ pp $ collisions at $\sqrt{s} =$ 7 and 13 TeV on the flavour composition of quarks within the proton and its gluon content, as well as their uncertainties by performing QCD global analysis. Although the potential constraint of the older LHCb beauty production data at $\sqrt{s} =$ 7 TeV on the PDFs at smaller values of $ x $ has been investigated by PROSA collaboration~\cite{Zenaiev:2015rfa}, it is also of interest to quantify the impact of new LHCb data at $ \sqrt s= $13 TeV (and also at $ \sqrt s= $7 TeV with better luminosity) compared with the older data.

In the previous section, the phenomenological framework of our analysis was introduced. Now we are in a position to perform such an analysis and study in details the impact of LHCb data 
on the behaviour of PDFs. For this aim, in this section we perform three analyses as follows. 
In the first analysis, we just consider the combined HERA I+II inclusive cross-section measurements, as well as the measurements of charm and beauty production at HERA, which are necessary to constrain PDFs. In the second analysis, we include also the updated LHCb measurement of $B^{\pm}$ production cross-section data at center-of-mass energy of $ \sqrt s= $7 TeV to assess the impact of these data on the PDFs. It should be noted that, in order to reduce the scale dependence of the theoretical prediction, we use the normalised cross-section,  $\frac{{\rm d}\sigma}{{\rm d}y} / \frac{{\rm d}\sigma}{{\rm d}y_0}$, for $B^{\pm}$ production which are obtained from the absolute measurements published by LHCb, with $\frac{{\rm d}\sigma}{{\rm d}y_0}$ being the cross section in the center rapidity bin, $3 < y < 3.5$, though the impact of absolute measurements on PDFs is also studied in the following. In this way, $N_{\rm dat}=108$ new data points will be added to the first analysis.
Finally, in the third analysis, we substitute the normalised $B^{\pm}$ production cross-section data at $ \sqrt s= $7 TeV with the same measurements at $ \sqrt s= $13 TeV, to
study also the effects of these new data on the PDFs. 

It is well known now that the $c$-quark mass is about
1.5 GeV, while this value is about 4.5 GeV for the case of $b$-quark. %Then, comparing with the QCD scale value $\Lambda_{QCD} \sim$ 0.25 GeV, the mass of charm and beauty quarks can be treated as a hard scale in pQCD. 
 In the present analysis, the effect of LHCb data on the mass of heavy quarks is also investigated. Considering the masses of the charm and beauty quarks as free parameters, their optimal values are determined for each of three QCD analyses described above. 

As mentioned in the previous section, for the theoretical calculations, we use the massive NLO calculations~\cite{Nason:1987xz,Nason:1989zy,Beenakker:1990maa} in the FFNS. 
The fully exclusive parton cross sections for heavy-quark production is presented in the MNR calculations~\cite{Mangano:1991jk}. 
The UA1 Collaboration were used successfully these calculations to describe its measurement of the beauty production cross-section in $p \bar p$ 
collisions \cite{Albajar:1995fz} and the Tevatron \cite{Abulencia:2006ps}. One of the important ingredients of such calculations is using the suitable values for heavy quark fragmentation fractions.
In order to use the LHCb inclusive $B^{\pm}$ 
hadron production data in our analyses, we use the fragmentation fractions for $b$-flavoured hadrons from~\cite{Aaij:2013noa}.

The list of experimental data and also the values of the $\chi ^2$ and number of data points for each data sets, and the total $\chi ^2$ for each analyses
have been illustrated in Table~\ref{tab:one}. The second column of the table shows the results of the base analysis which dose not include the LHCb data. The columns
labeled by ``LHCb 7 norm'' and ``LHCb 13 norm'' contain the results of the second and third analyses 
that include also the LHCb normalised data at $ \sqrt s= $7 and 13 TeV, respectively. 
The values of the total $\chi ^2$ divided by the number of degrees 
of freedom, have been given in the last row of the table. 
As can be seen, it is equal to 1.187, 1.186 and 1.164 for the base fit using the HERA combined data, the analyses including the LHCb data at 7 and 13 TeV, respectively. At first glance, it means that the simultaneous inclusion of the LHCb $B^{\pm}$ 
hadron production data and HERA DIS combined data can lead to a satisfying fit on these data. Actually, according to the results obtained, it can be clearly seen that the $\chi ^2$ per points for the LHCb data at both $ \sqrt s= $7 and 13 TeV is perfect. However, a slight tension between
the LHCb data and HERA combined data can also be elicited by comparing the $\chi ^2$ per points for
HERA1+2 NCep 920 and HERA1+2 CCep data before and after including the LHCb data in the analysis.
This tension in more visible for the analysis with LHCb data at 7 TeV in analogy to the analysis containing LHCb data at 13 TeV.

The optimal values of the input PDF parameters of Eq.~\ref{eq:paramf} at the initial scale $Q^{2}$= 1.4 GeV$^{2}$, as well as the charm and bottom quark masses extracted from the fit have been given in Table~\ref{tab:two} (see the columns included label ``norm" for the analyses containing the LHCb normalised data), for all three global analyses described above. Note that we have used the blue color to distinguish between the parameters excluded from the fit and considered to be fixed
with the free parameters. Another important point should be mentioned is that for the case of base analysis without considering the LHCb data, we have exactly used the parametrisation of HERAPDF2.0 analysis~\cite{Abramowicz:2015mha} with an extra $ D $ parameter for $ \bar U $ distribution (see Eq.~\ref{eq:paramf}), while for the analyses including the LHCb data at 7 and 13 TeV, this parameter has been considered to be zero. According to the results obtained, one of the interesting points can be achieved from Table~\ref{tab:two} is that the inclusion of the LHCb data in the analysis leads to a decrease in values of the heavy quark masses, especially for the analysis containing the LHCb data at 7 TeV.

\begin{figure}
\center
\includegraphics[width=0.32\textwidth]{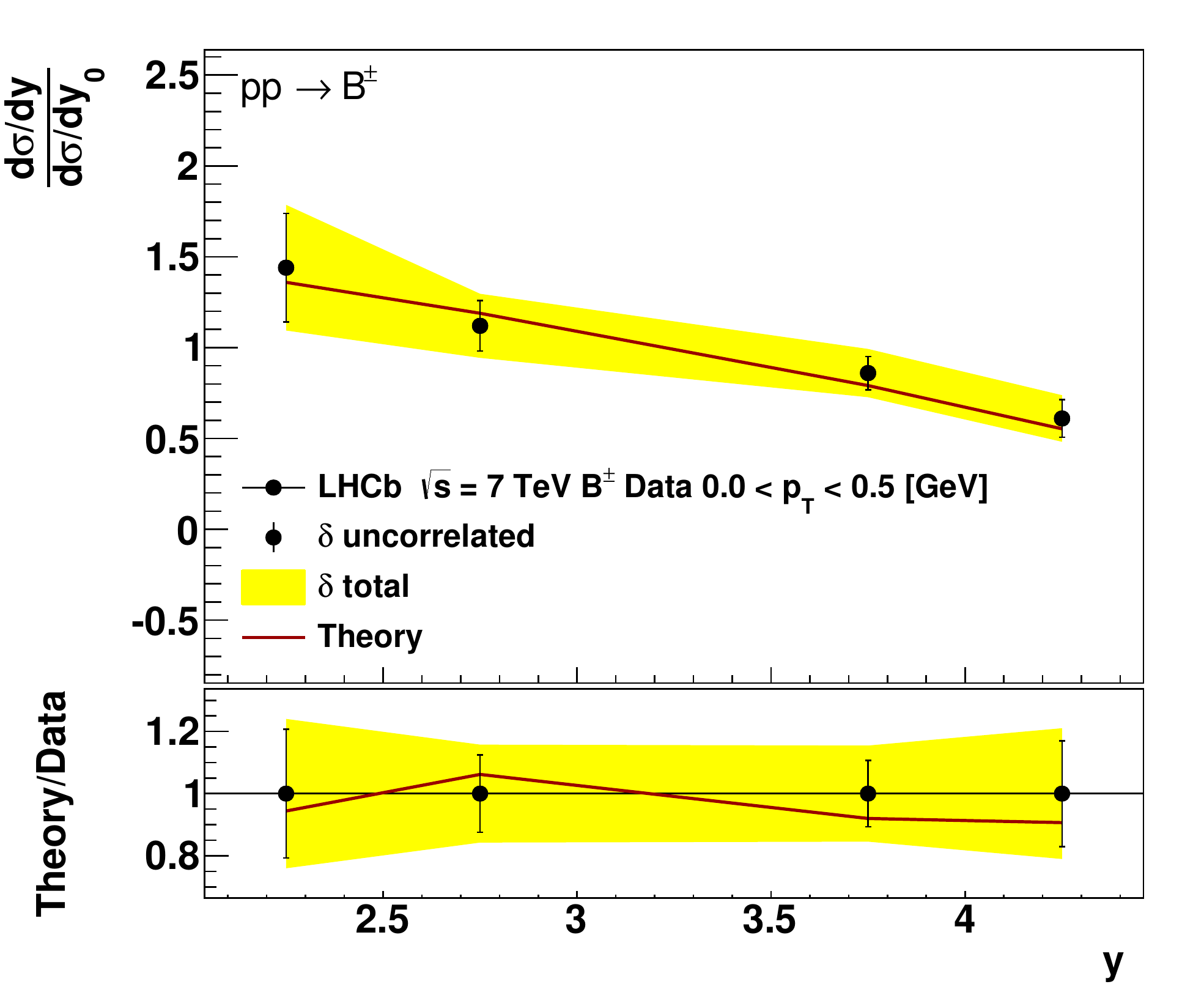}
\includegraphics[width=0.32\textwidth]{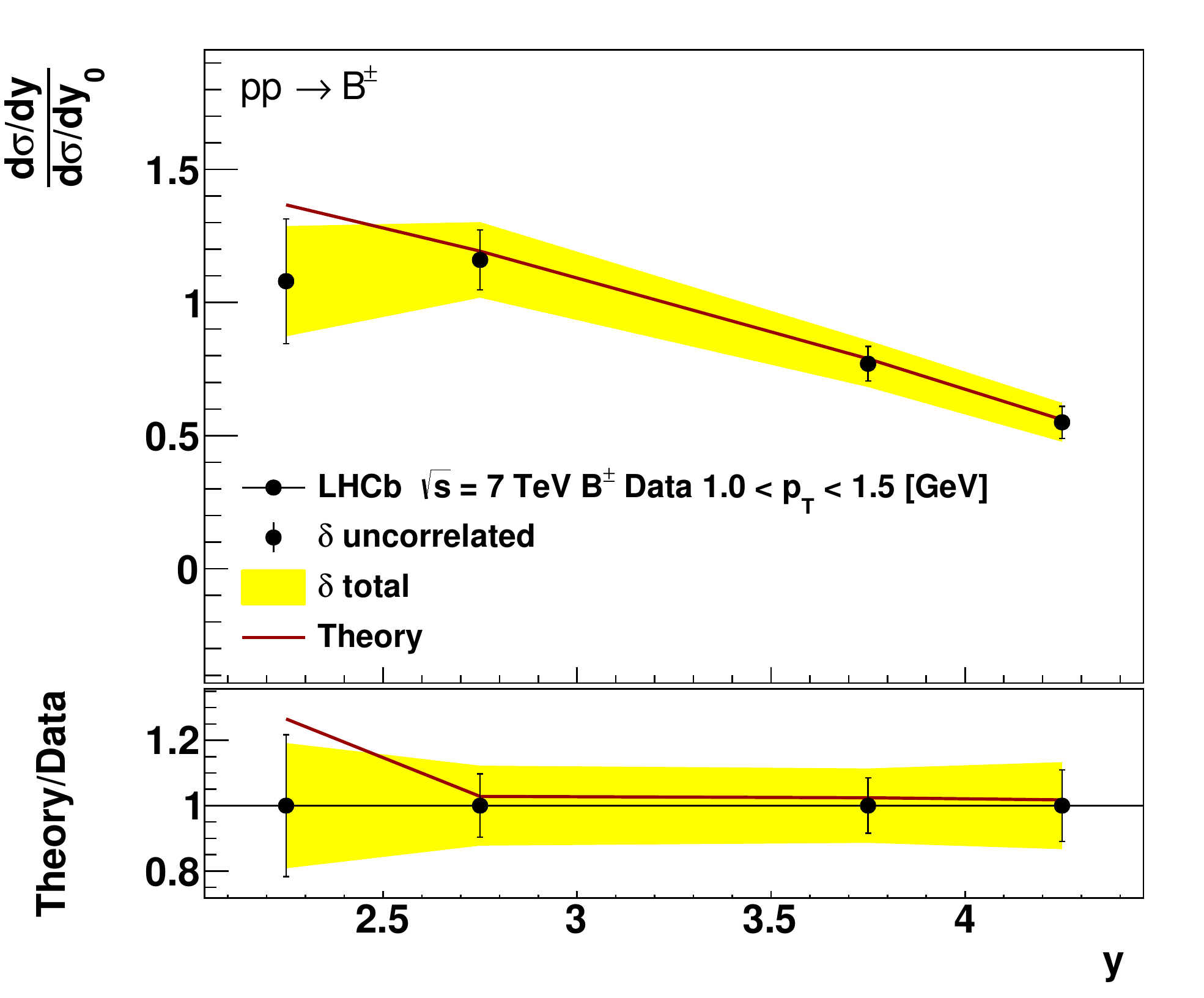}
\includegraphics[width=0.32\textwidth]{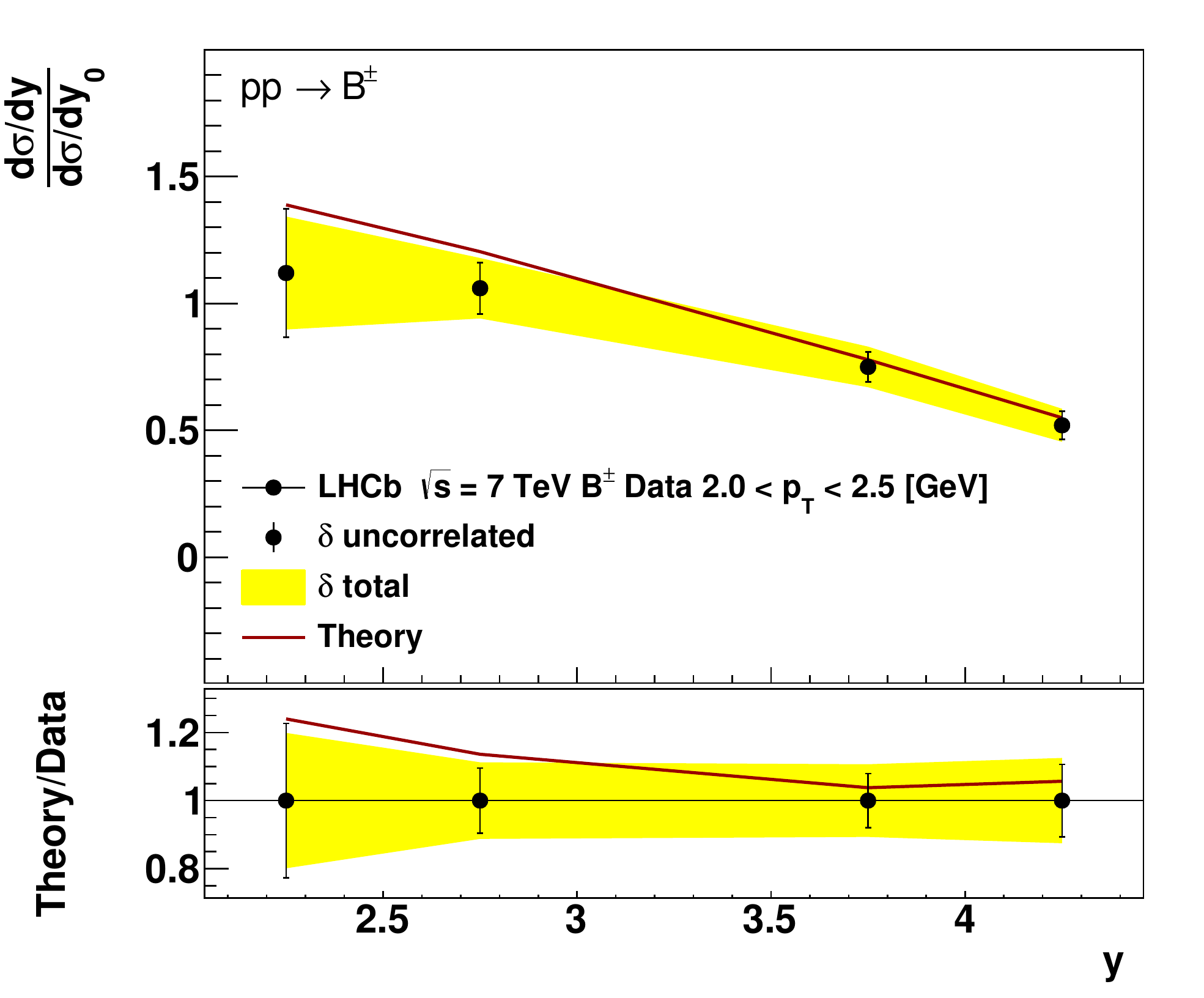}\\
\includegraphics[width=0.32\textwidth]{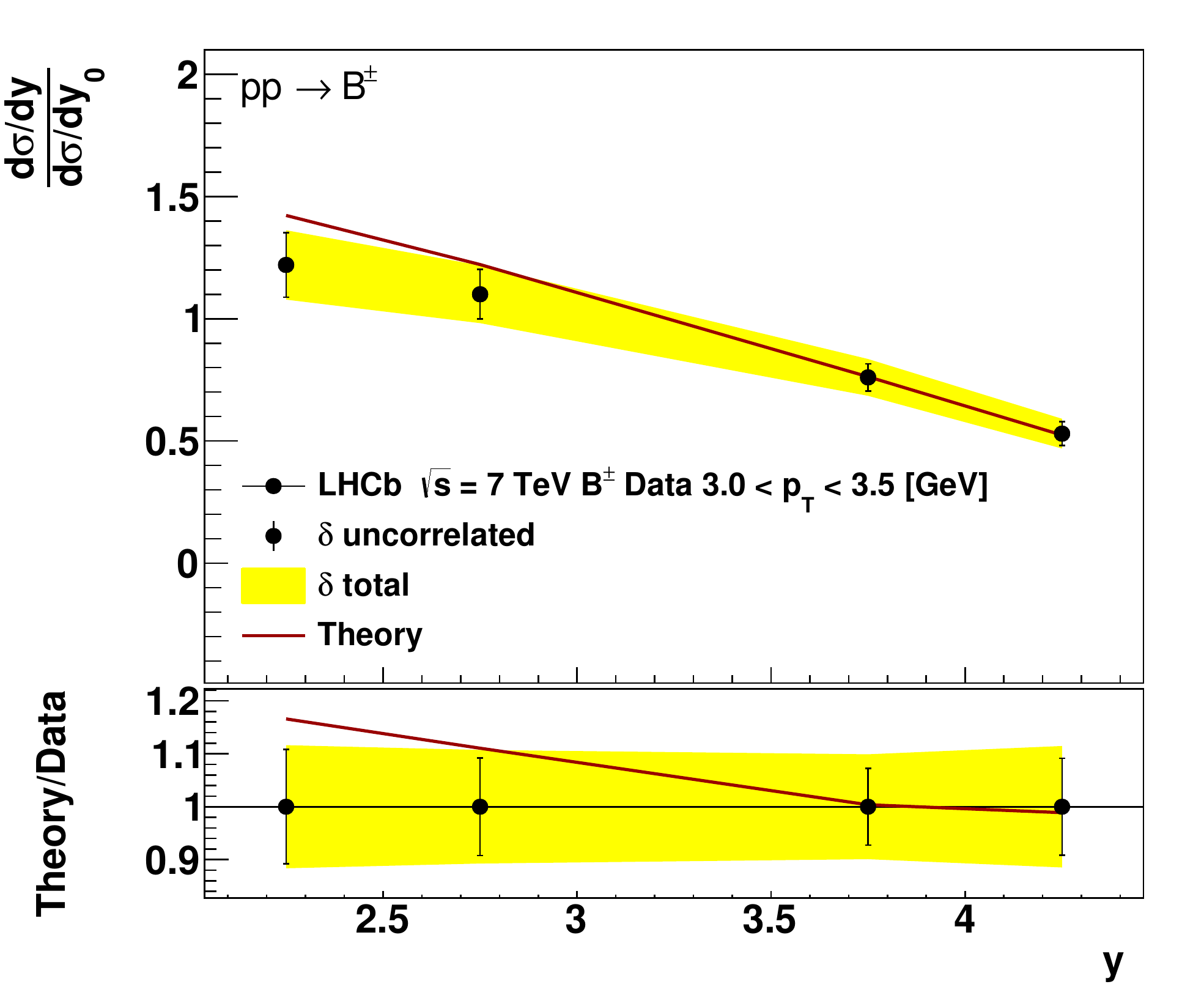}
\includegraphics[width=0.32\textwidth]{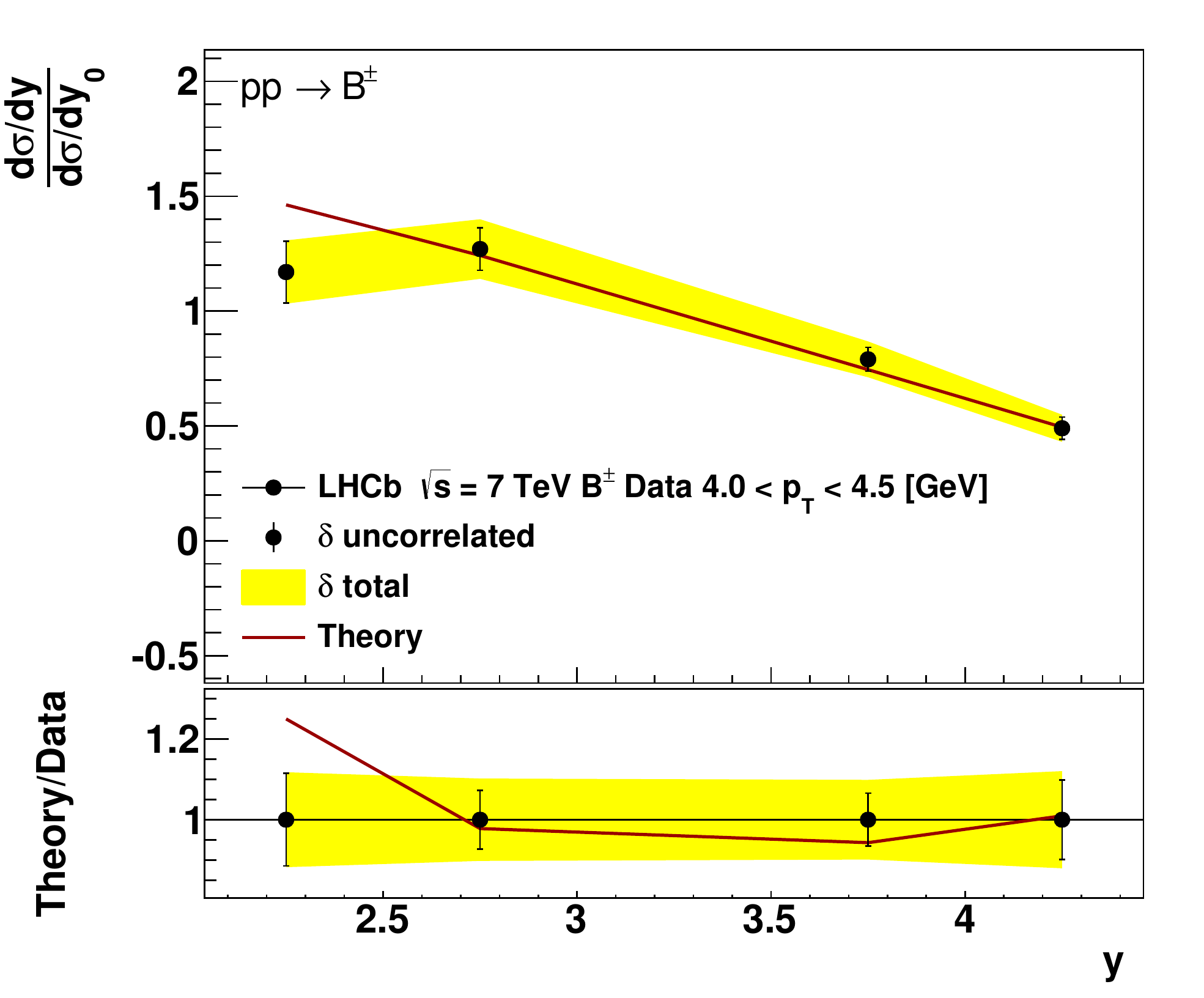}
\includegraphics[width=0.32\textwidth]{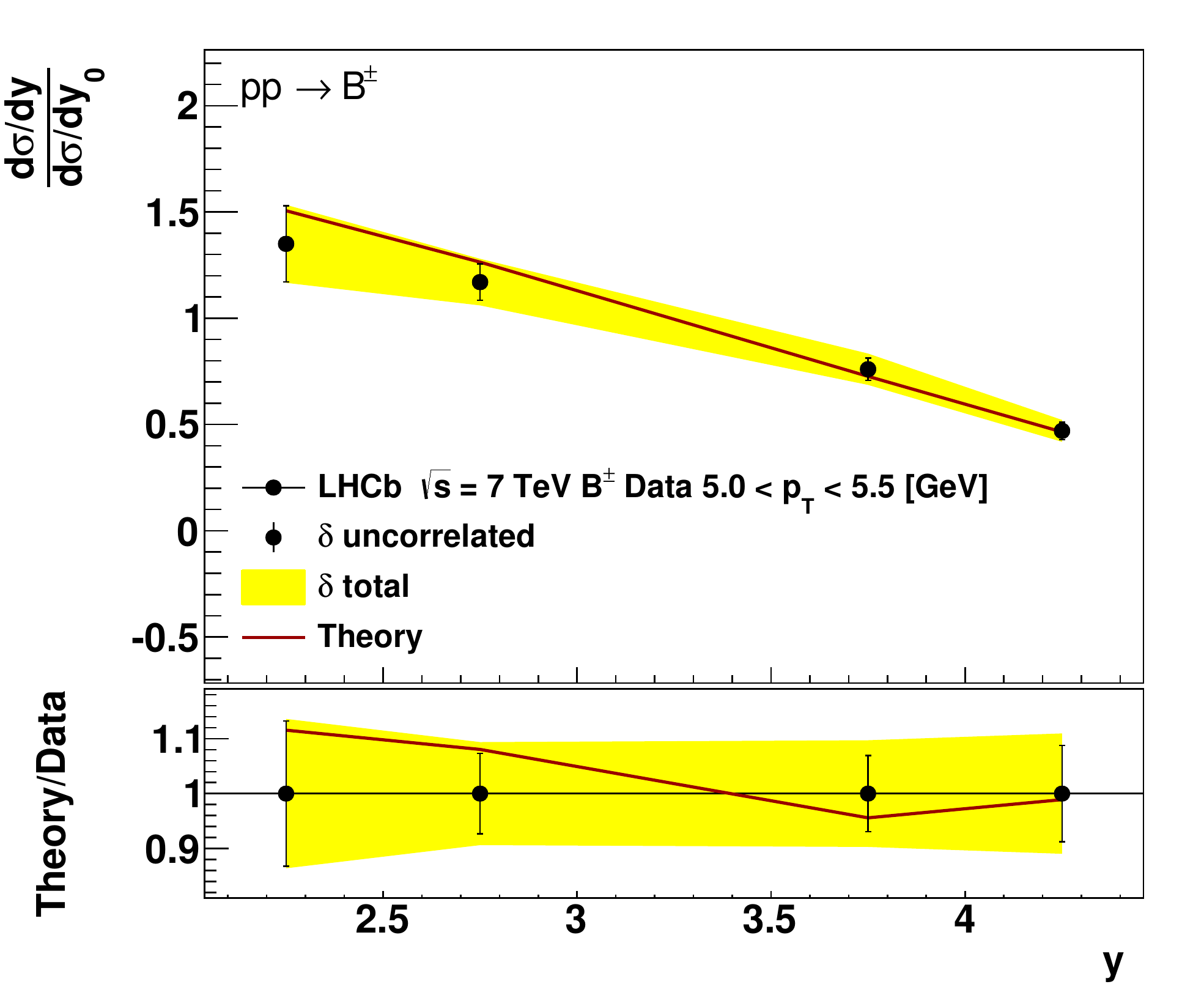}\\
\includegraphics[width=0.32\textwidth]{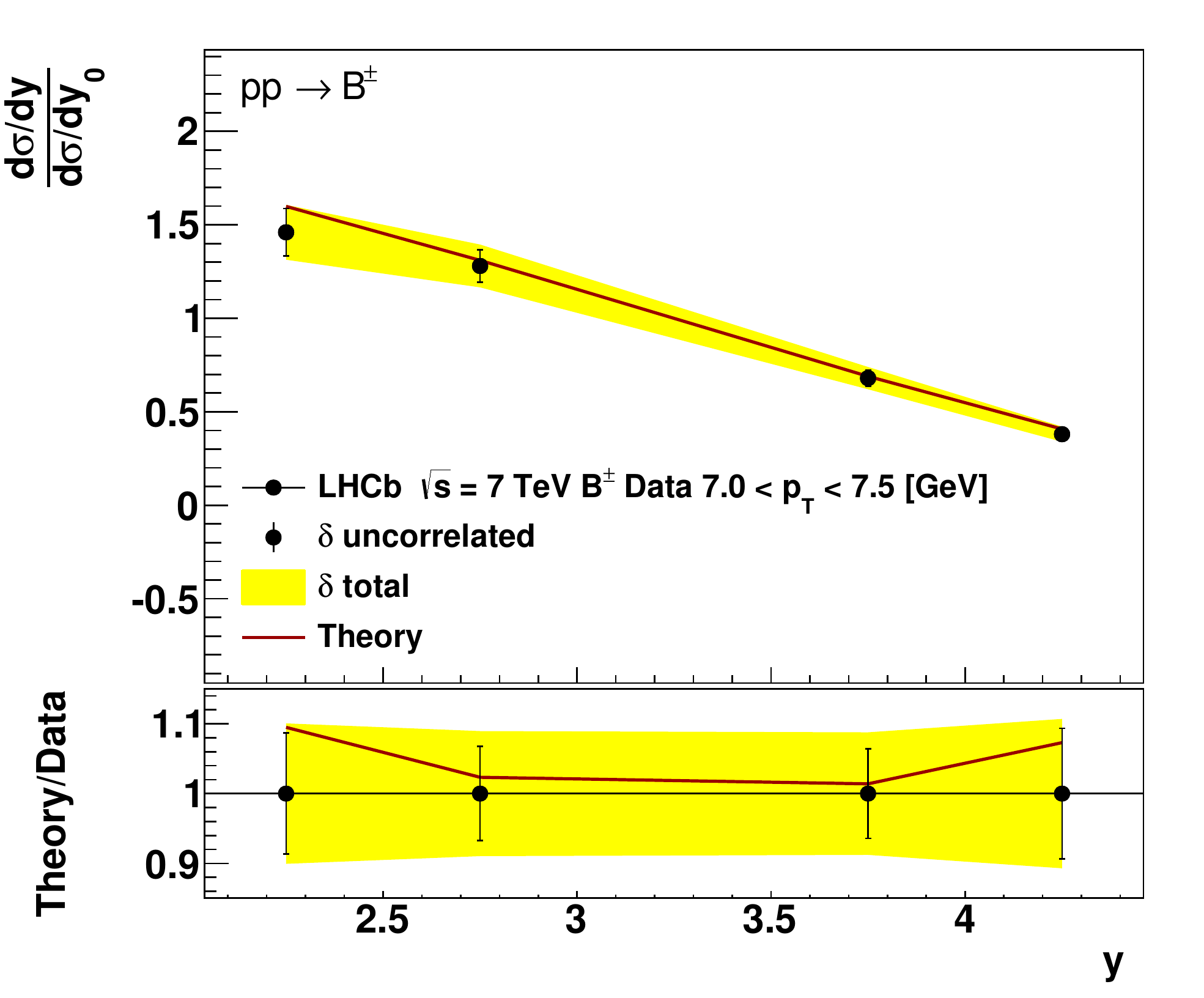}
\includegraphics[width=0.32\textwidth]{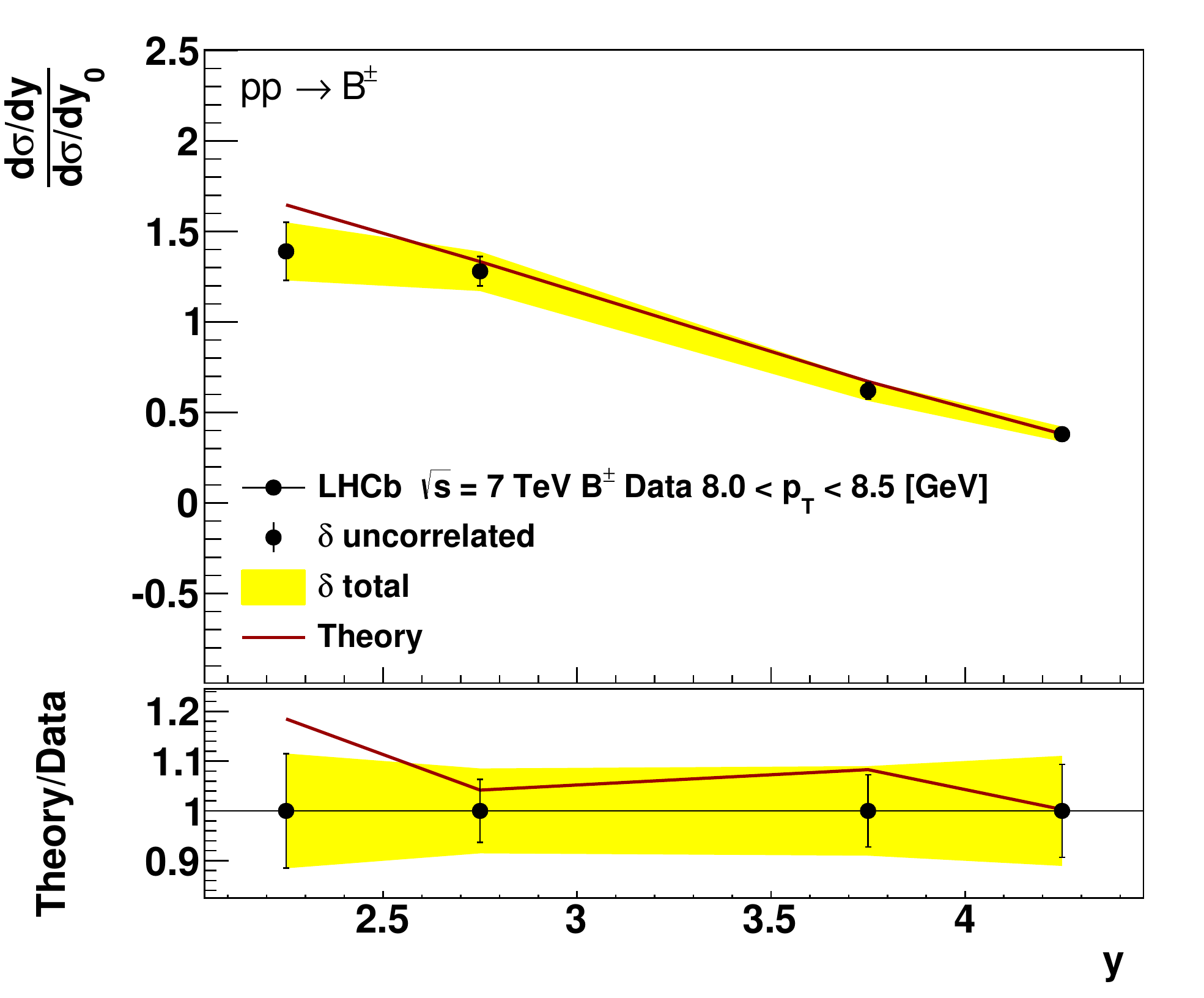}
\includegraphics[width=0.32\textwidth]{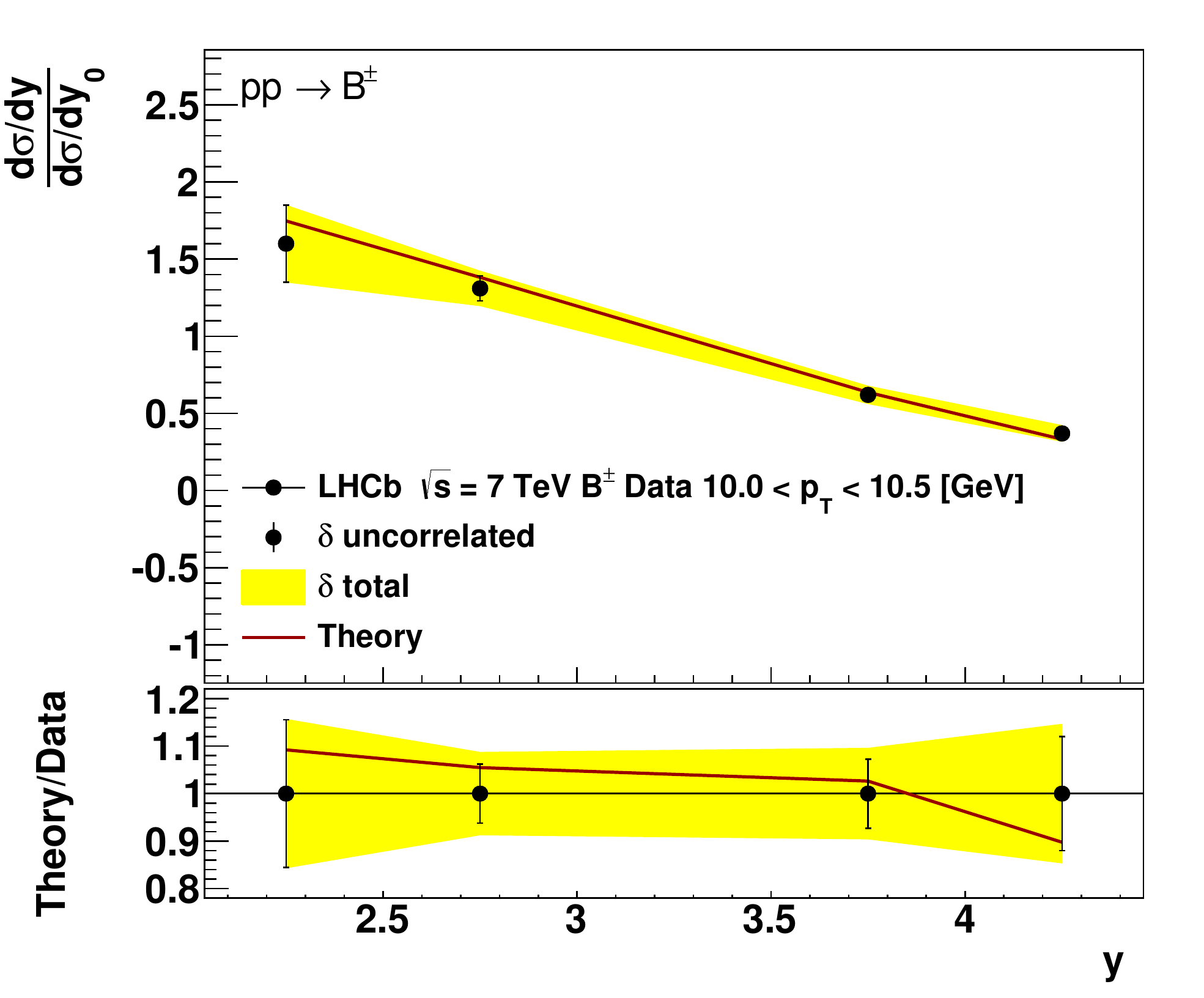}\\
\includegraphics[width=0.32\textwidth]{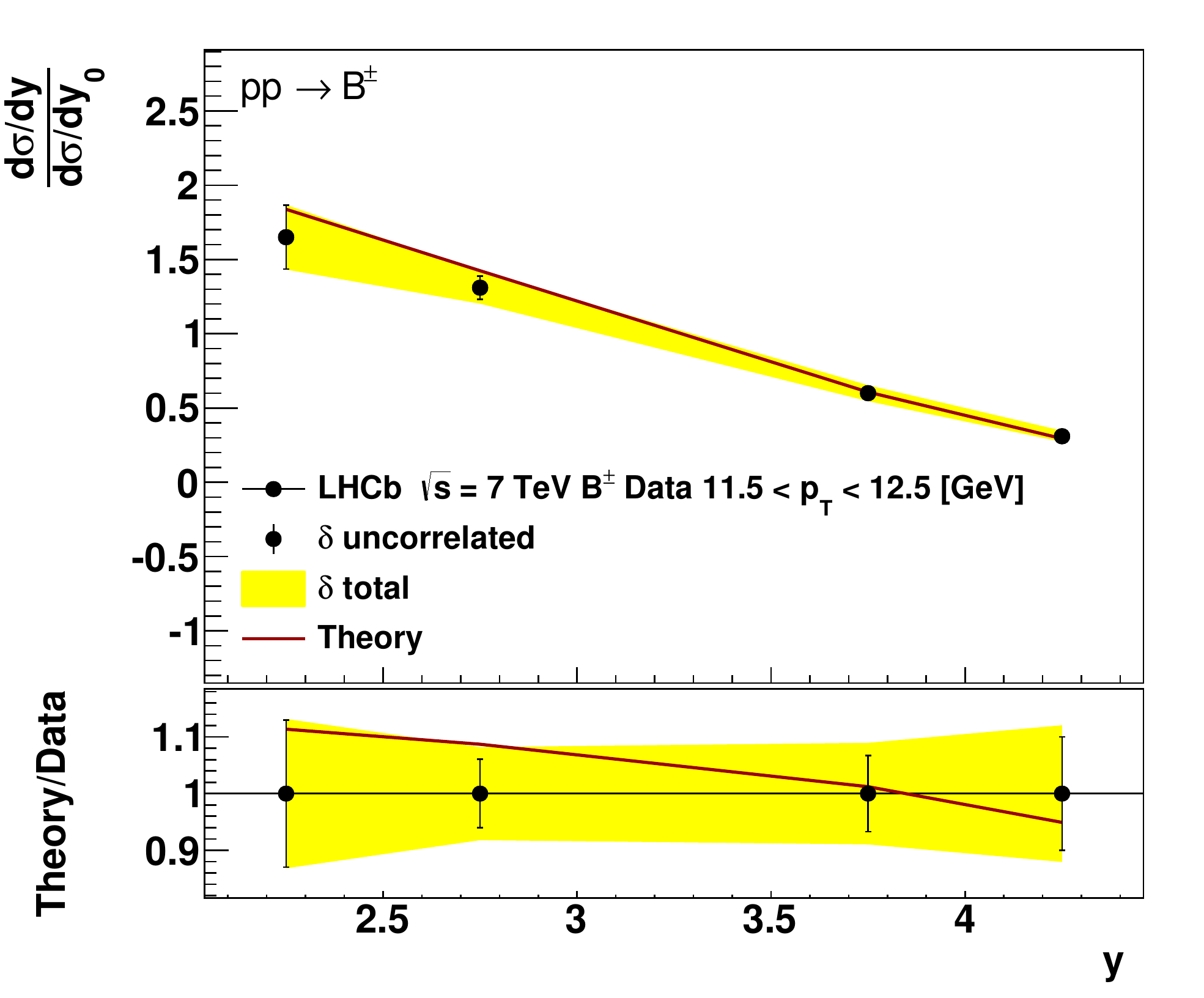}
\includegraphics[width=0.32\textwidth]{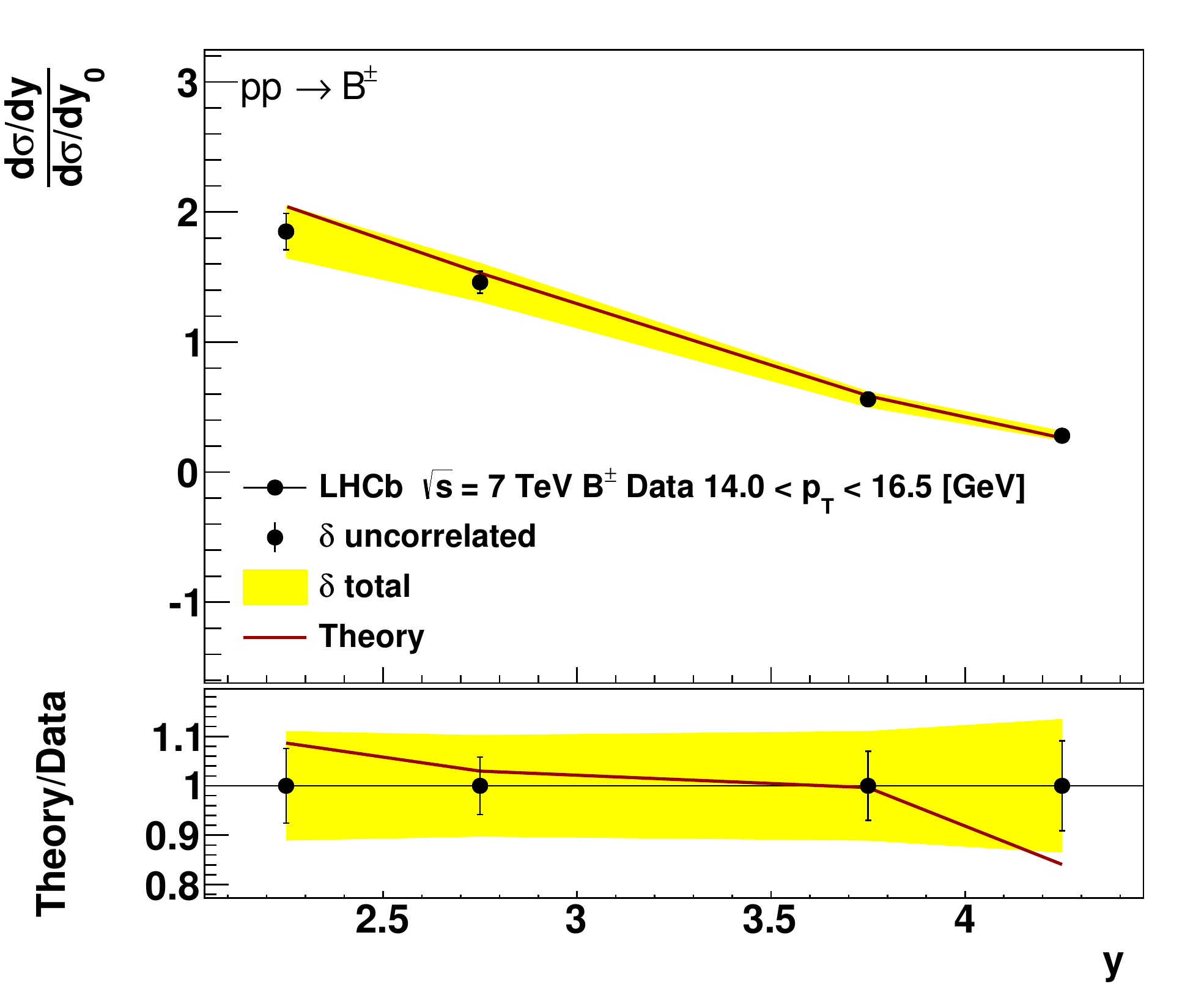}
\includegraphics[width=0.32\textwidth]{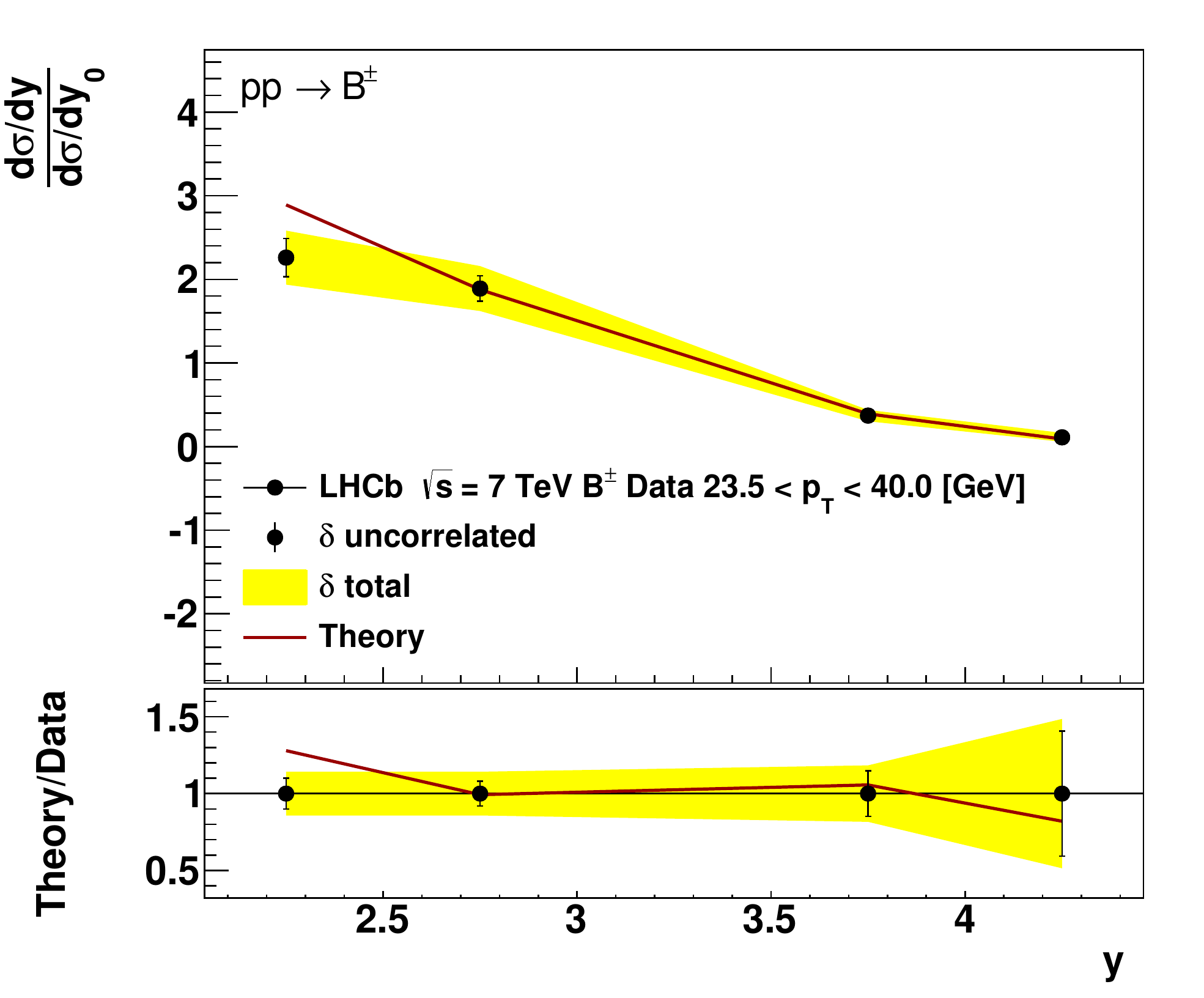}\\
\caption{ The fit results for a representative subset of the LHCb 7 TeV normalised cross sections~\cite{Aaij:2017qml}. From up to down, for production of $B^{\pm}$ mesons for $0.0 < p_T < 0.5$~GeV, up to $23.5 < p_T < 40.0$~GeV. 
 In the bottom panels the ratios theory/data are shown. } \label{fig:one}
\end{figure}

In Fig.~\ref{fig:one}, the LHCb normalised cross sections for production of $B^{\pm}$ mesons at 7 TeV~\cite{Aaij:2017qml} have been compared to the theory predictions based on the PDFs extracted from the fit. This figure contains a comprehensive comparison between the theory and experiment in wide range of the transverse momentum from $0.0 < p_T < 0.5$~GeV up to $23.5 < p_T < 40.0$~GeV as a function of rapidity $ y $. As expected from the value of $\chi ^2$ per points presented in Tab.~\ref{tab:one}, there is an excellent agreement between the LHCb data with QCD theory.  
 
\begin{figure}
\center
\resizebox{0.32\textwidth}{!}{%
  \includegraphics{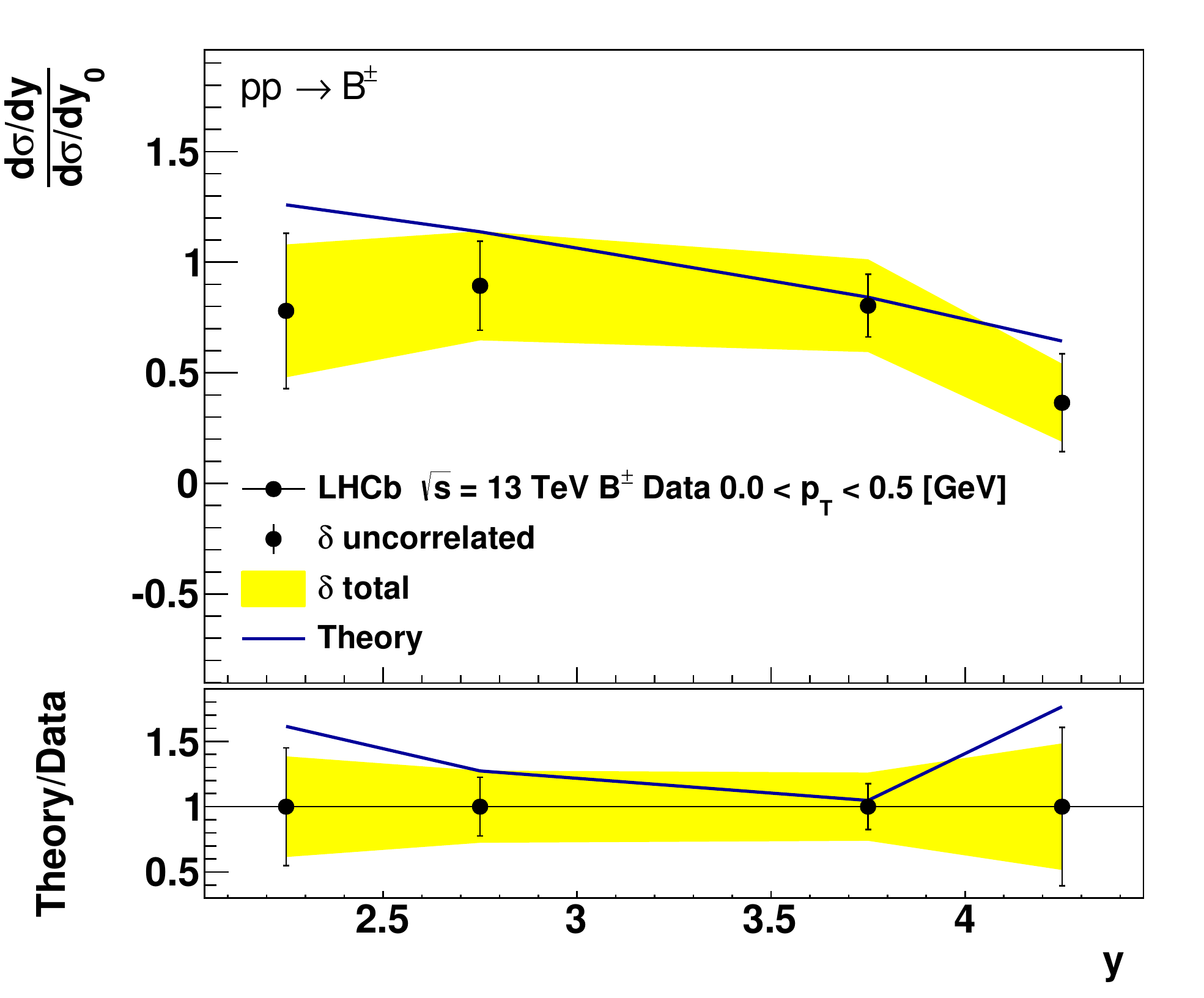}}
\resizebox{0.32\textwidth}{!}{%
  \includegraphics{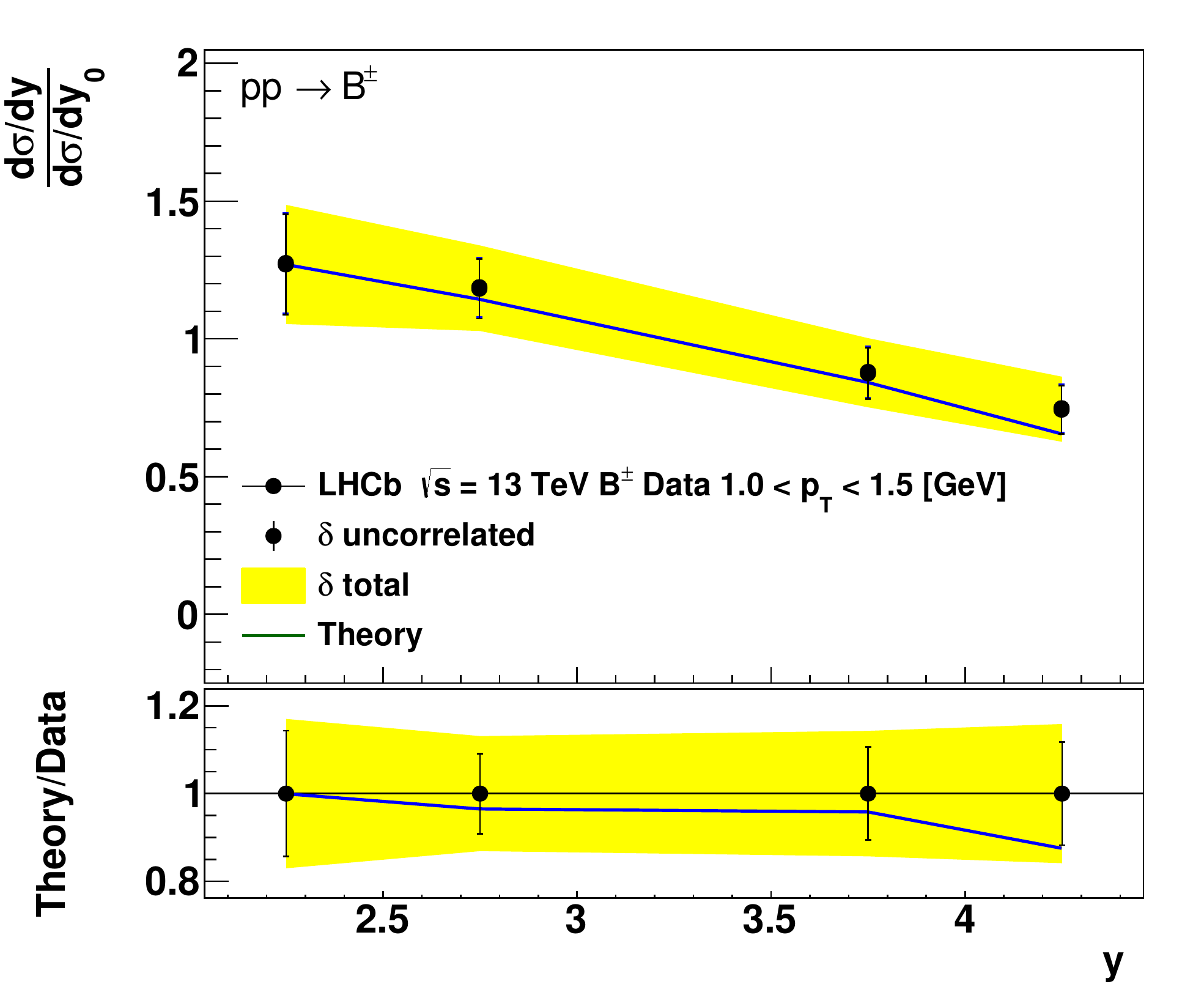}}
\resizebox{0.32\textwidth}{!}{%
  \includegraphics{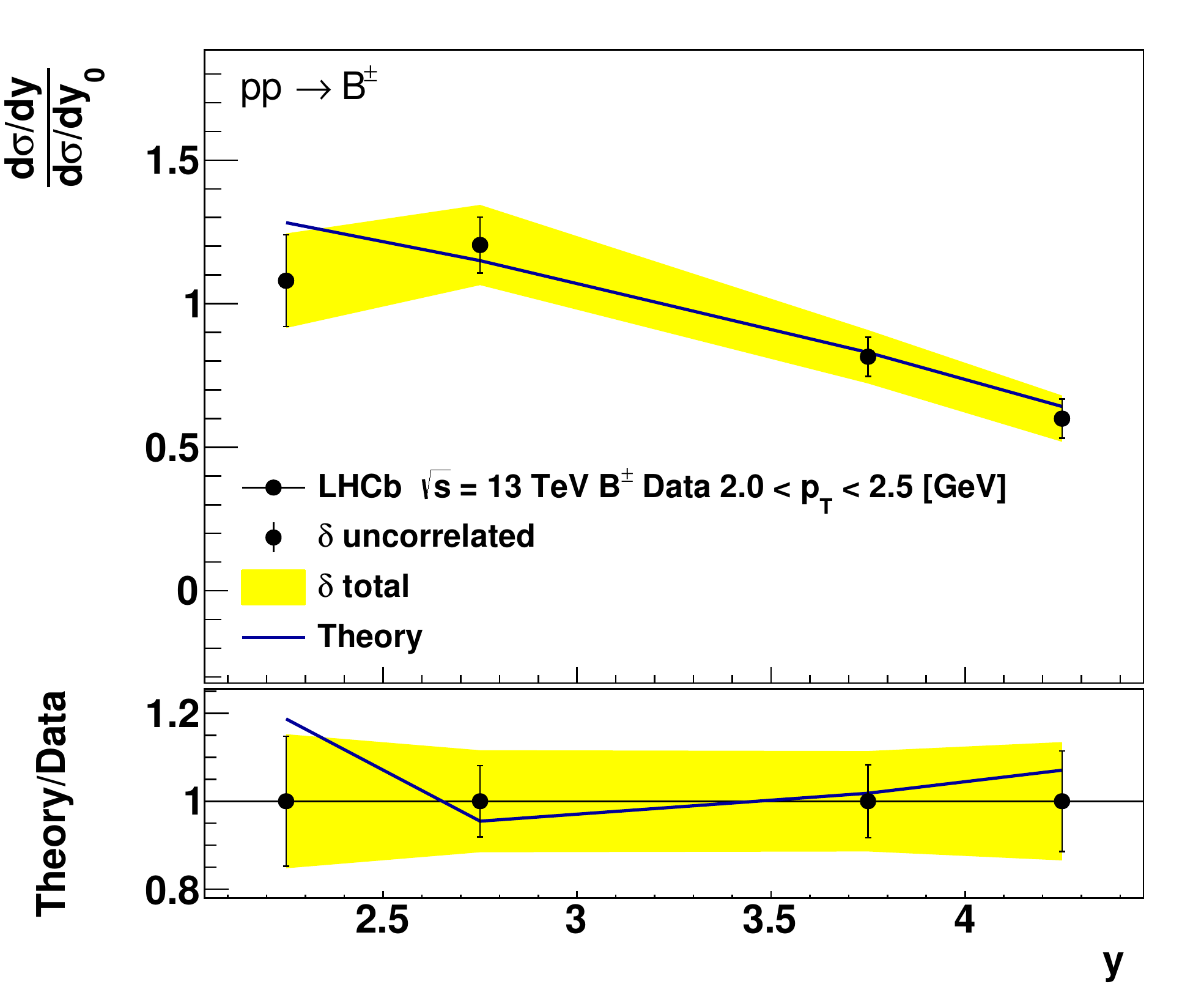}}\\
\resizebox{0.32\textwidth}{!}{%
  \includegraphics{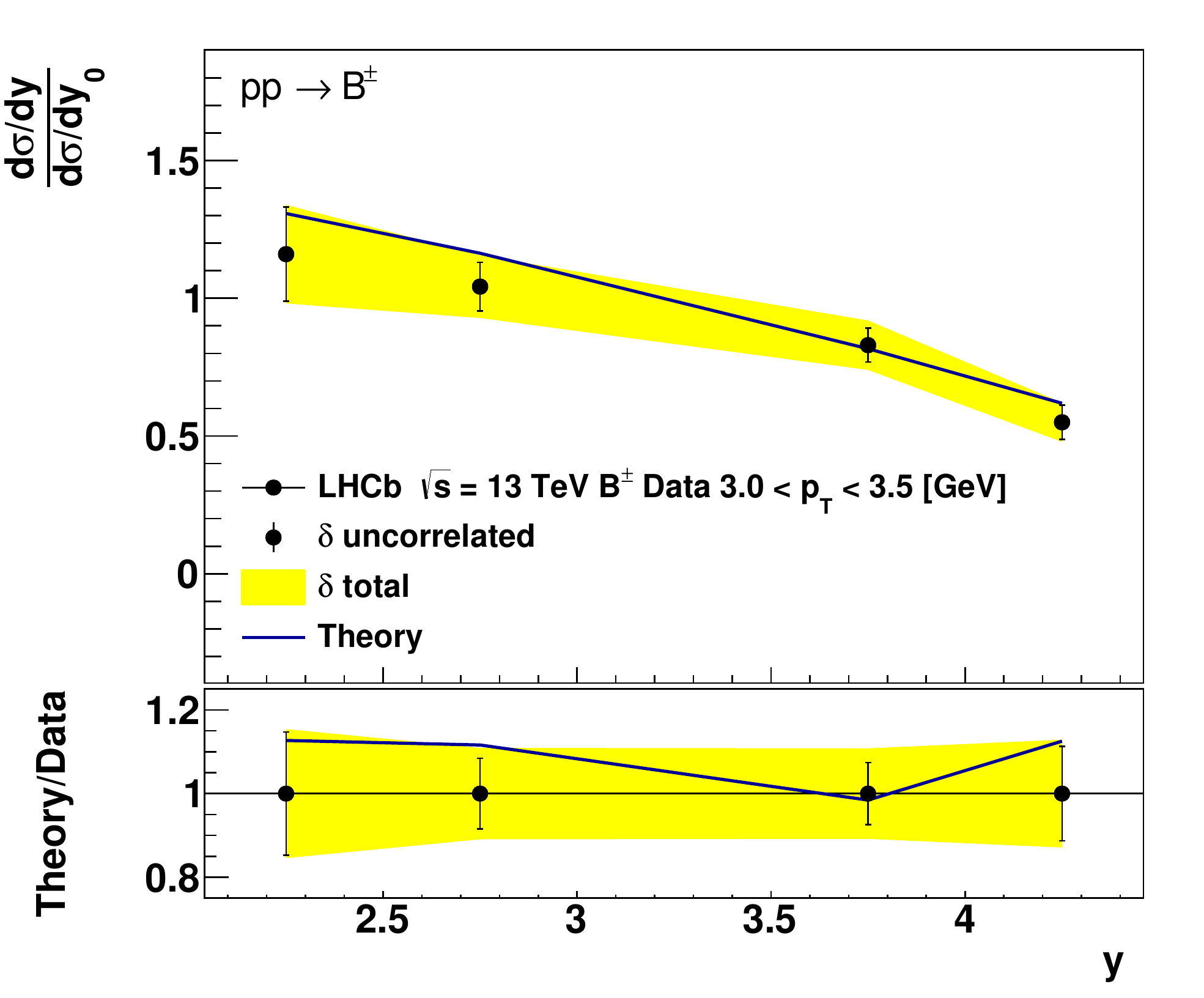}}
\resizebox{0.32\textwidth}{!}{%
  \includegraphics{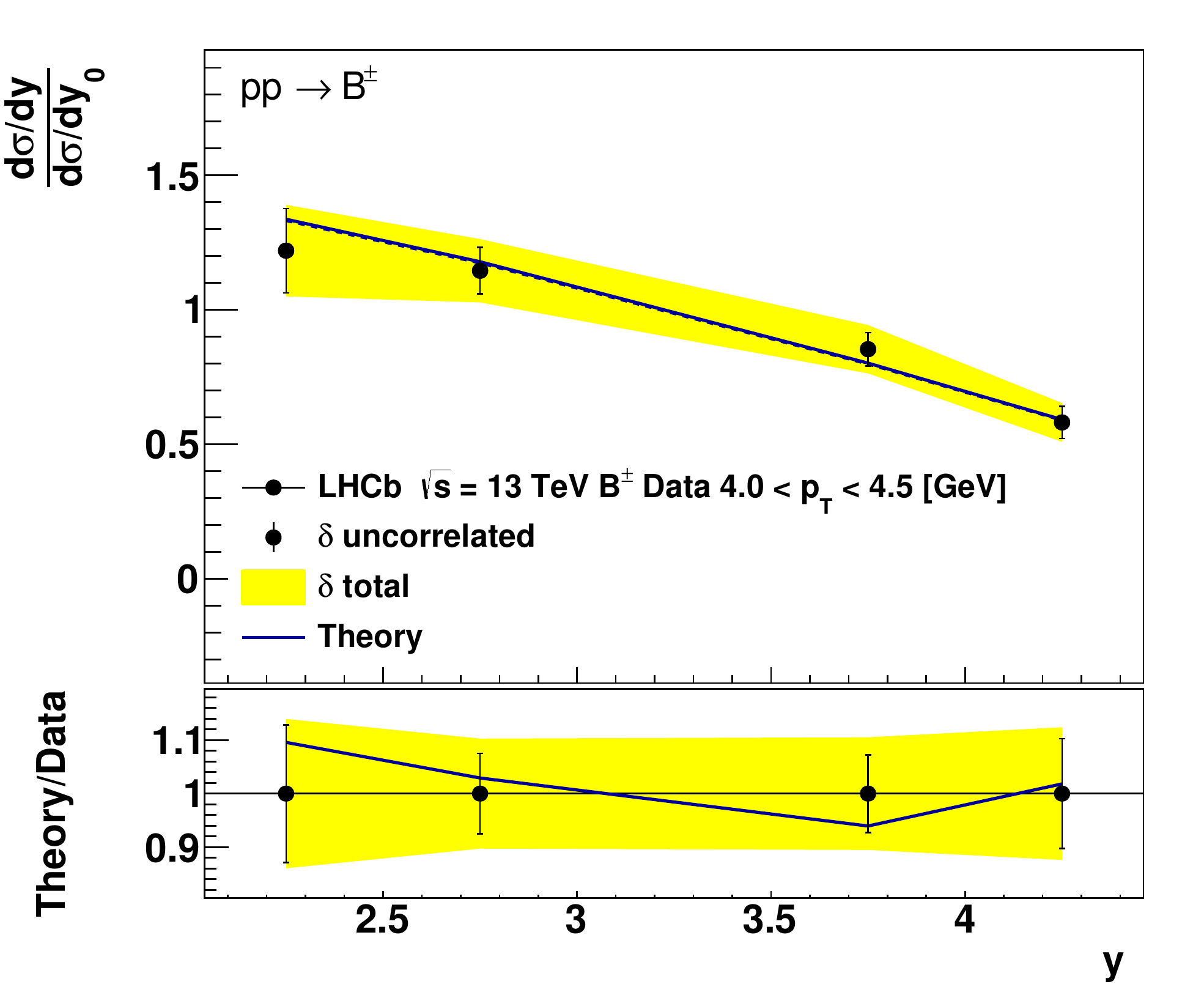}}
  \resizebox{0.32\textwidth}{!}{%
  \includegraphics{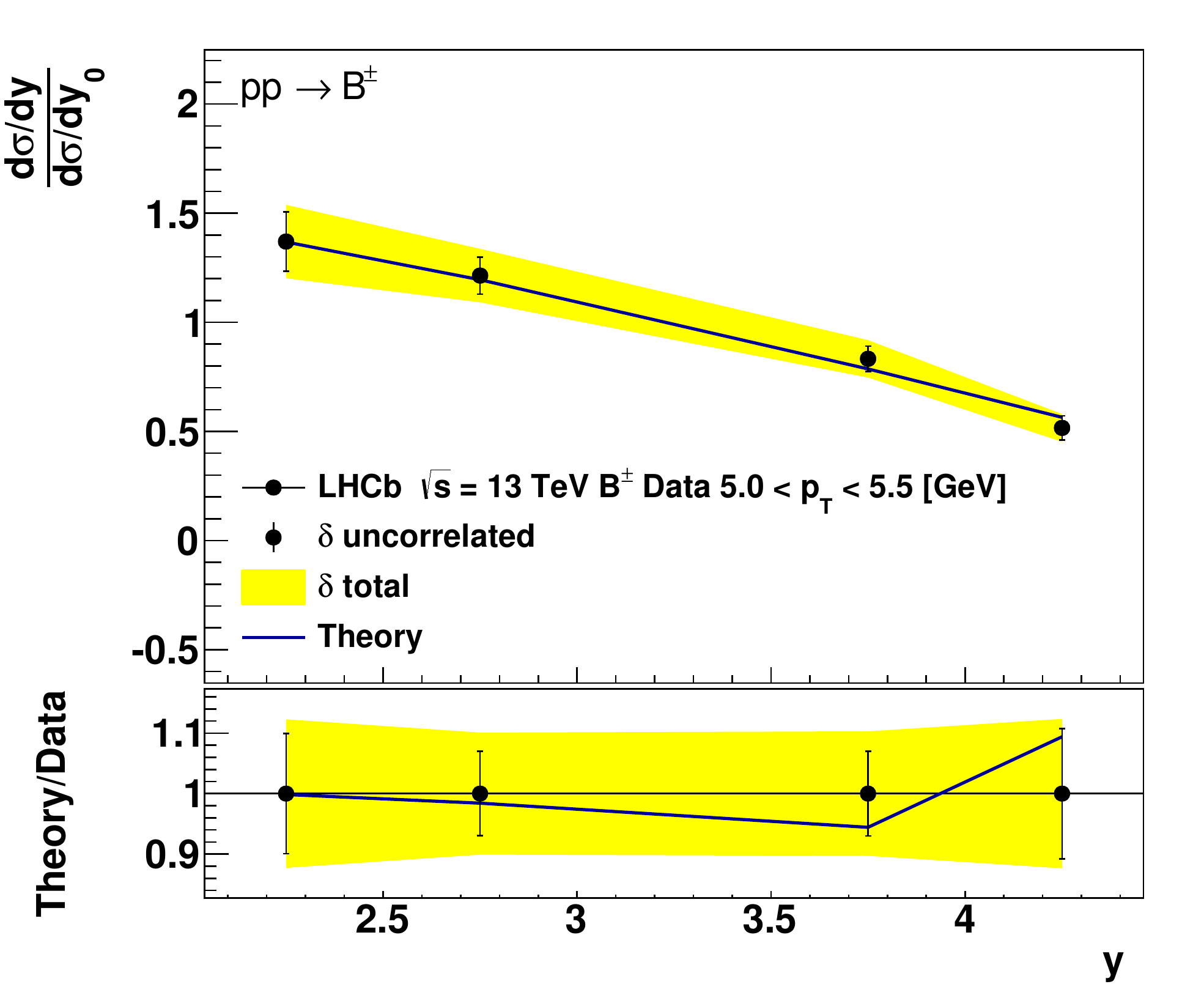}}\\
\resizebox{0.32\textwidth}{!}{%
  \includegraphics{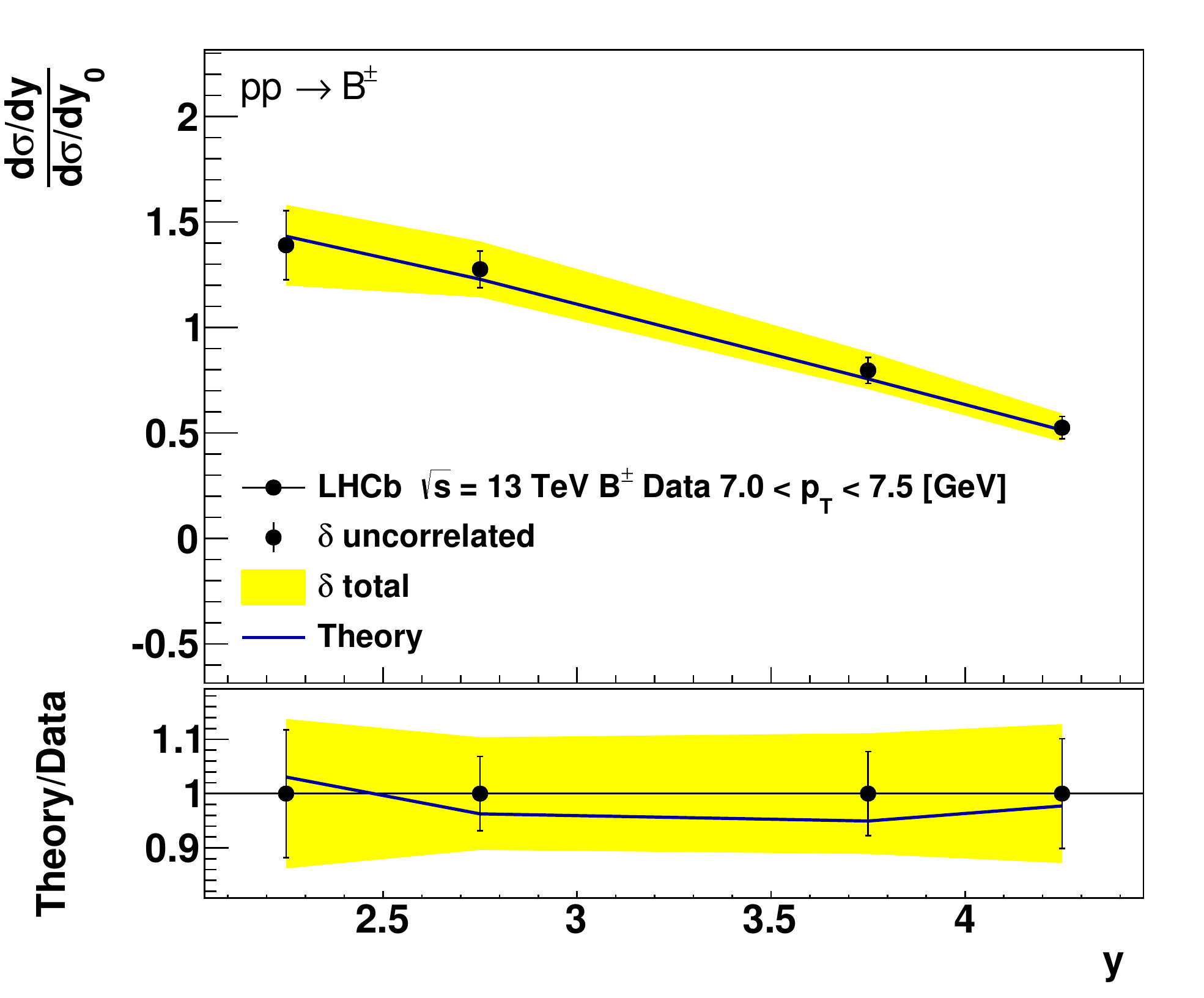}}
\resizebox{0.32\textwidth}{!}{%
  \includegraphics{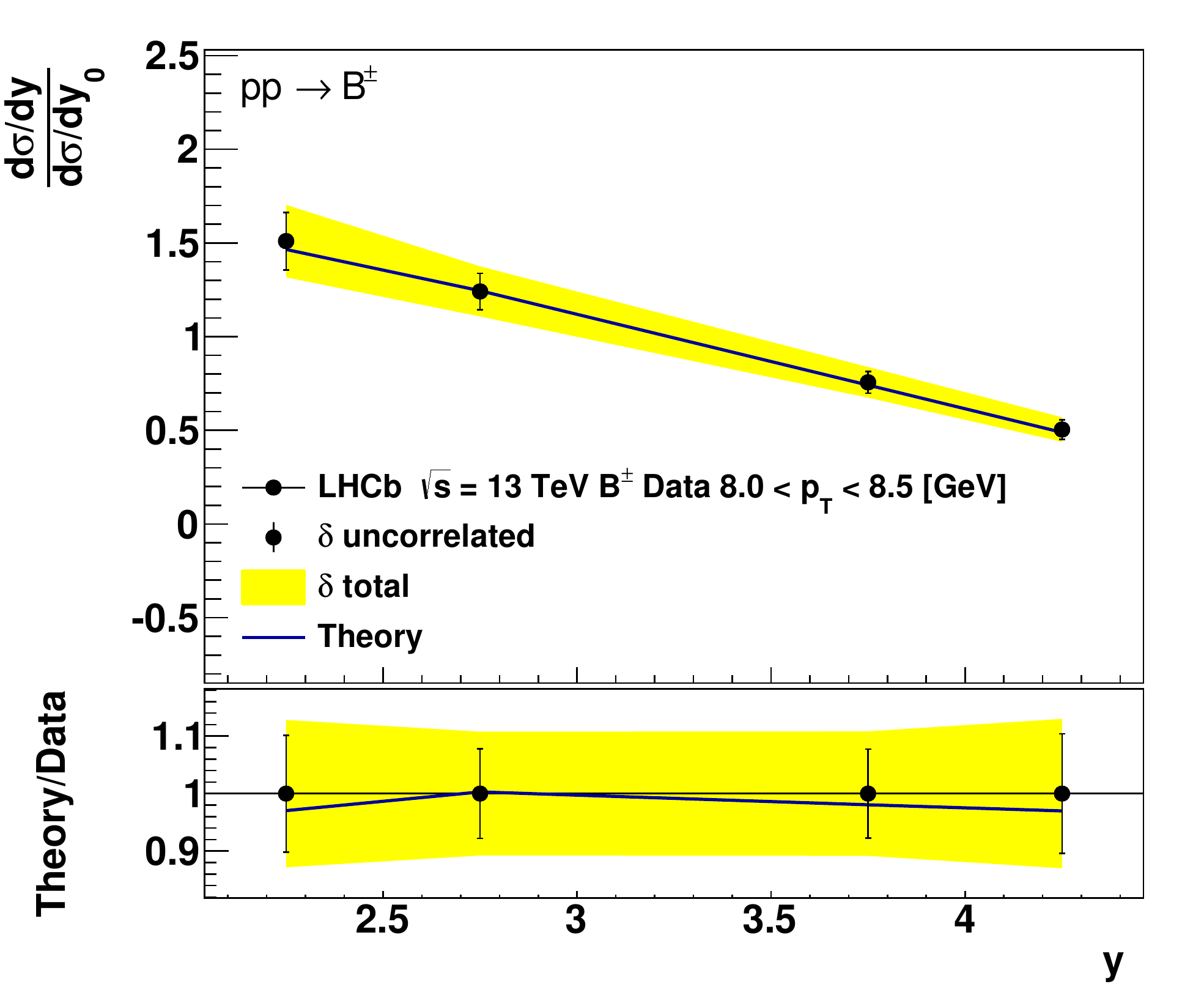}}
\resizebox{0.32\textwidth}{!}{%
  \includegraphics{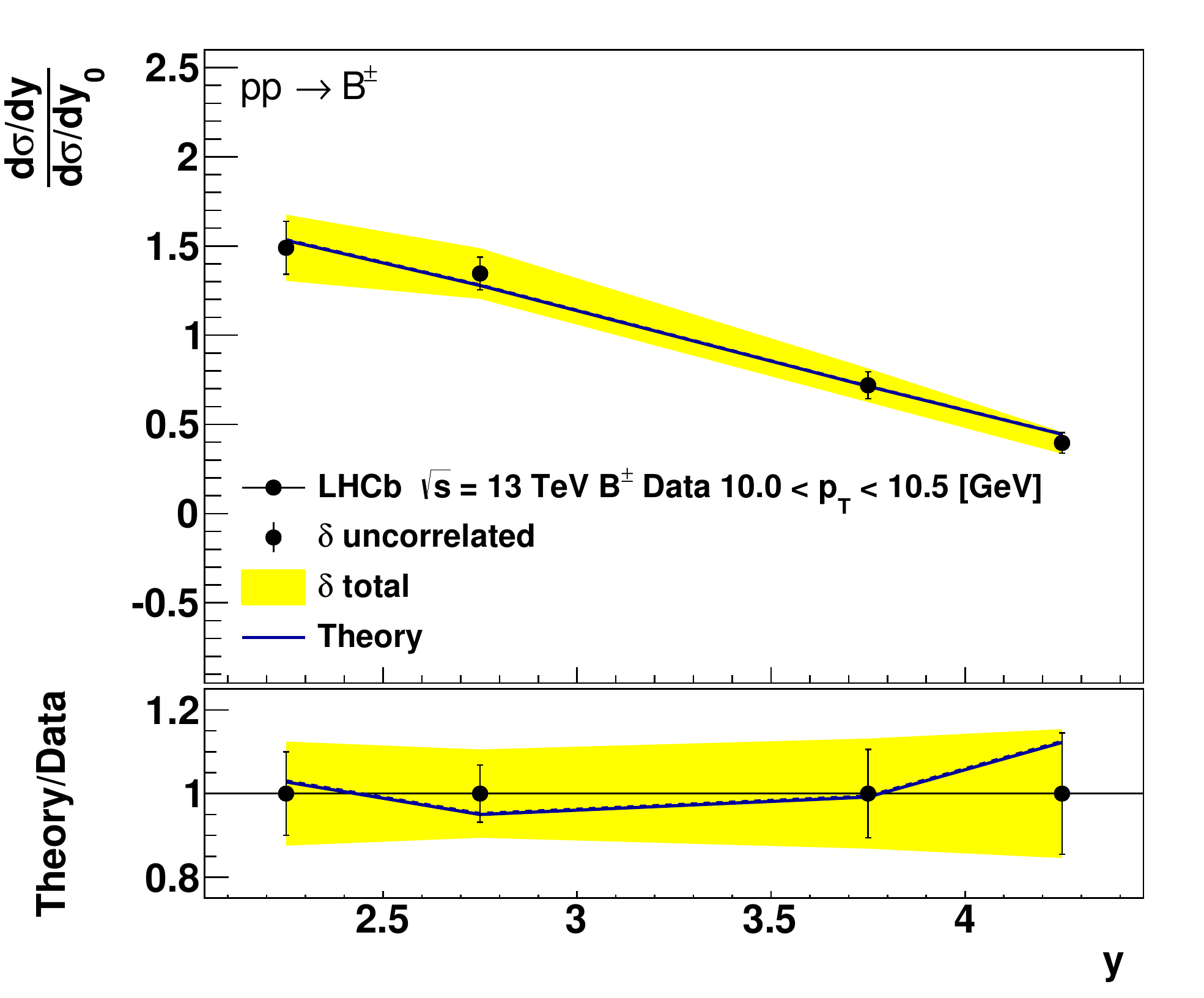}}\\
\resizebox{0.32\textwidth}{!}{%
  \includegraphics{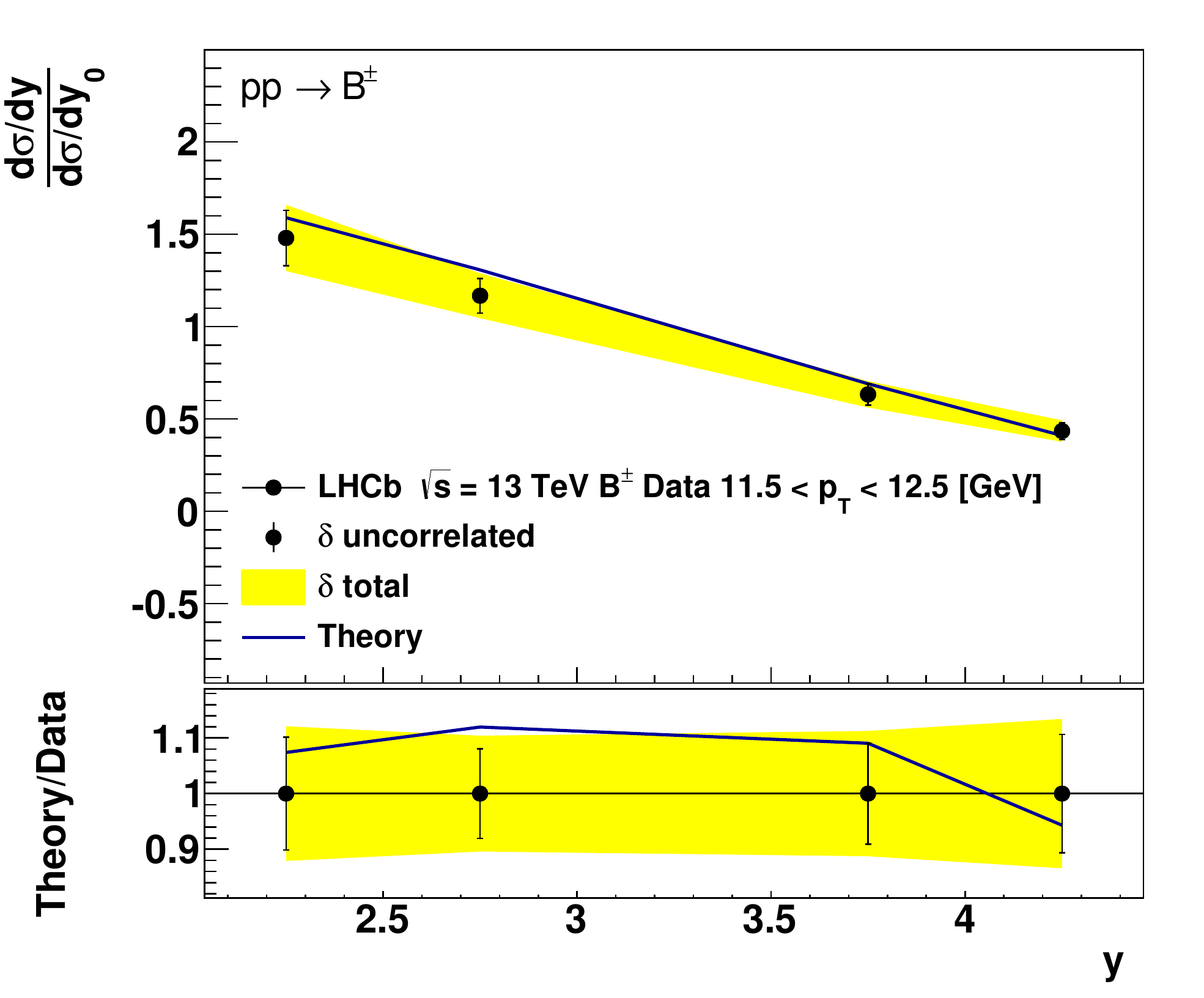}}
  \resizebox{0.32\textwidth}{!}{%
  \includegraphics{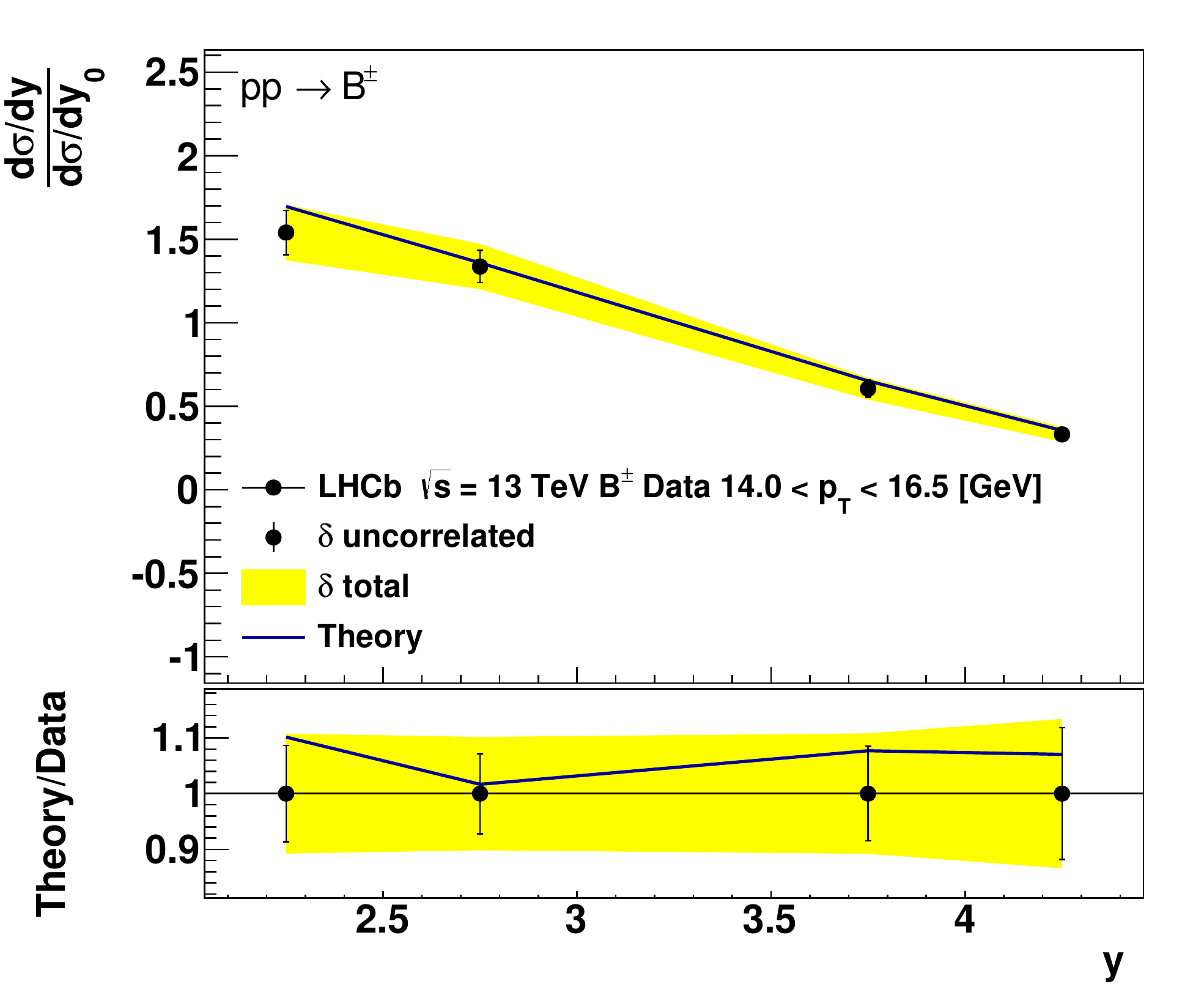}}
\resizebox{0.32\textwidth}{!}{%
  \includegraphics{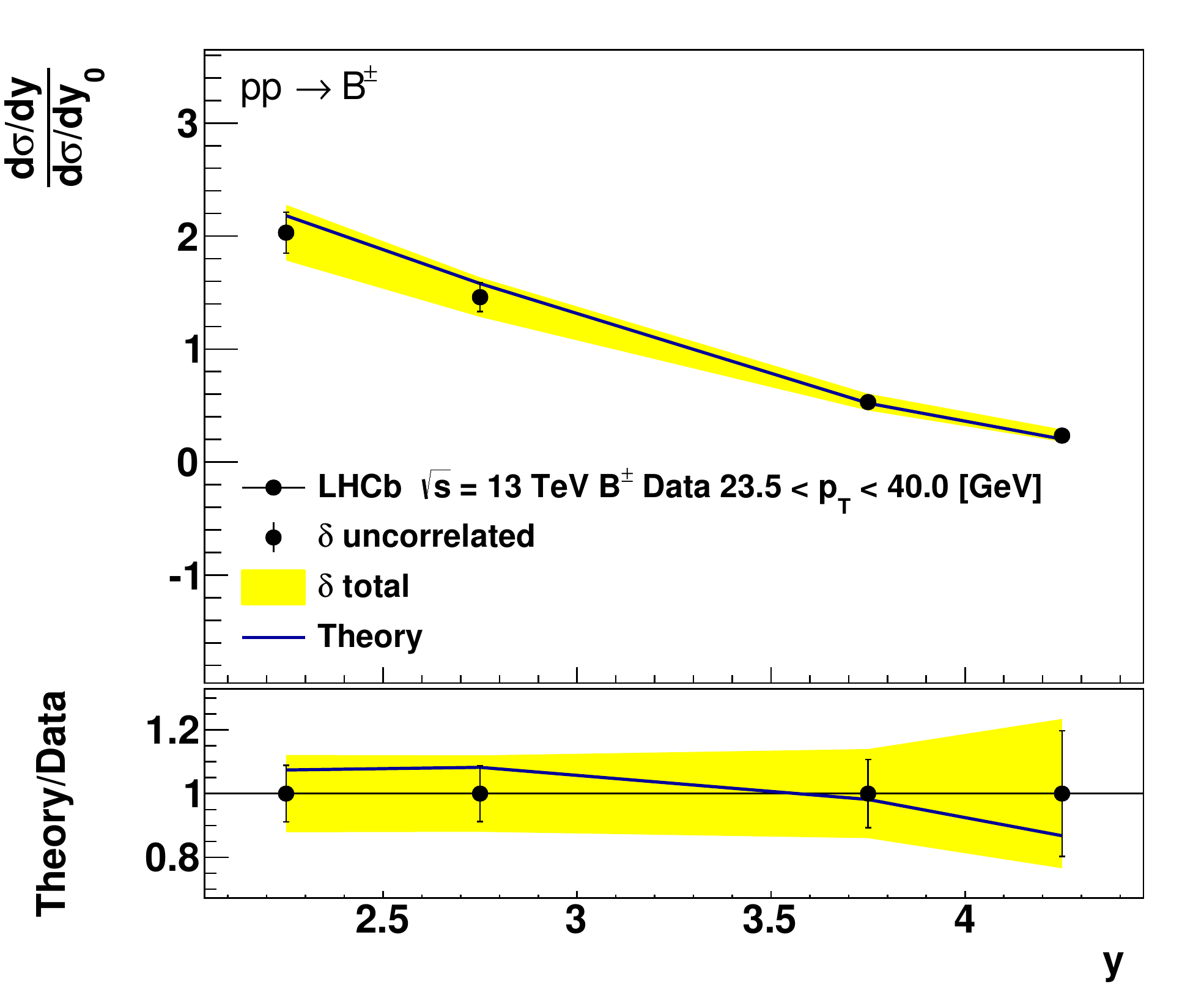}}\\
\caption{ The fit results for a representative subset of the LHCb 13 TeV normalised cross sections~\cite{Aaij:2017qml}. From up to down, for production of $B^{\pm}$ mesons for $0.0 < p_T < 0.5$~GeV, up to $23.5 < p_T < 40.0$~GeV. 
 In the bottom panels the ratios theory/data are shown.} \label{fig:two}
\end{figure}

Fig.~\ref{fig:two} shows the same results as Fig.~\ref{fig:one}, but for the case of LHCb 13 TeV normalised cross sections. As can be seen, the agreement between the theoretical predictions using the PDFs obtained and the experimental data is again excellent. There is also a slight deviation for the lowest transverse momentum range $0.0 < p_T < 0.5$~GeV.
 
\begin{figure}
\center
\includegraphics[width=0.4\textwidth]{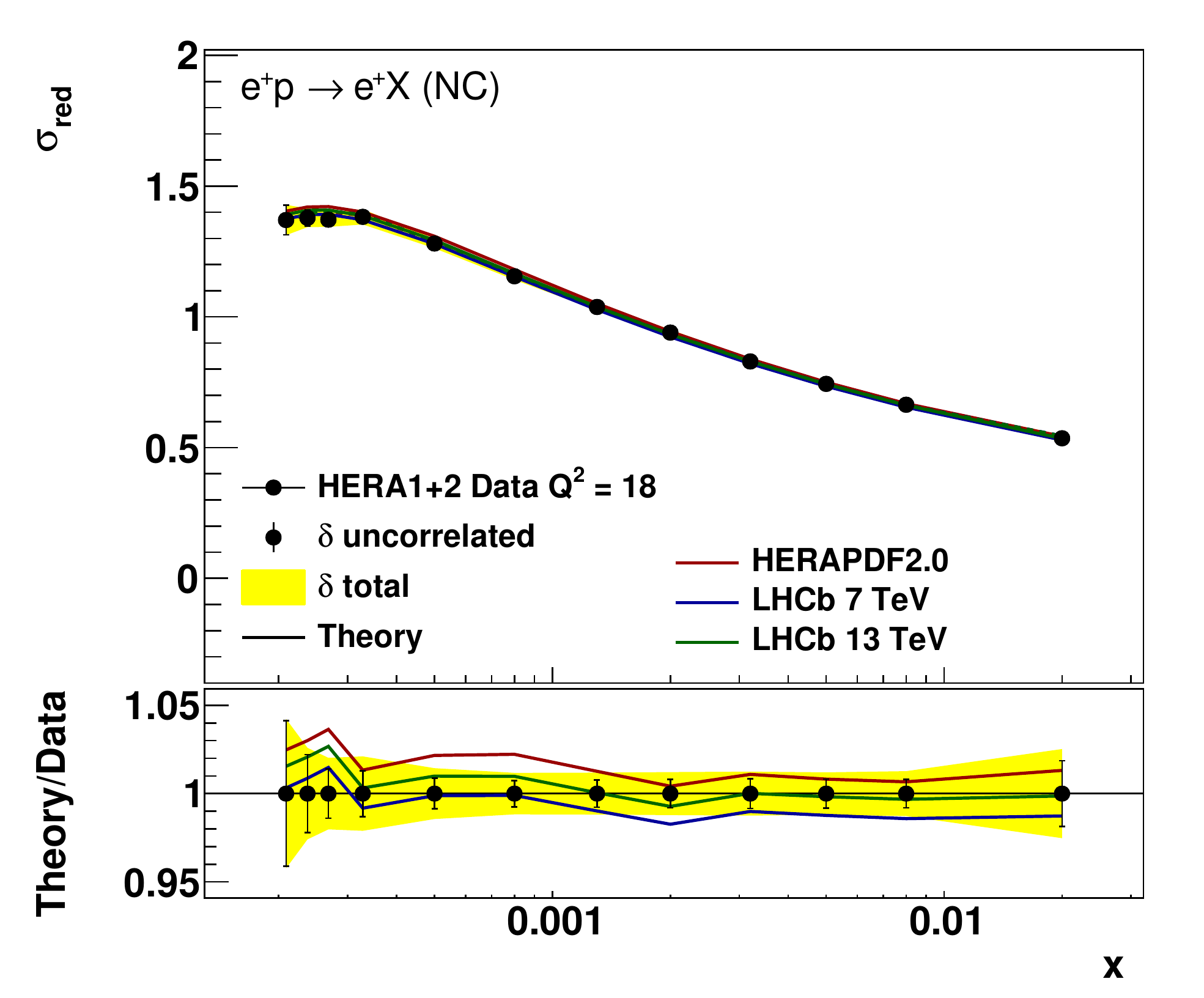}
\includegraphics[width=0.4\textwidth]{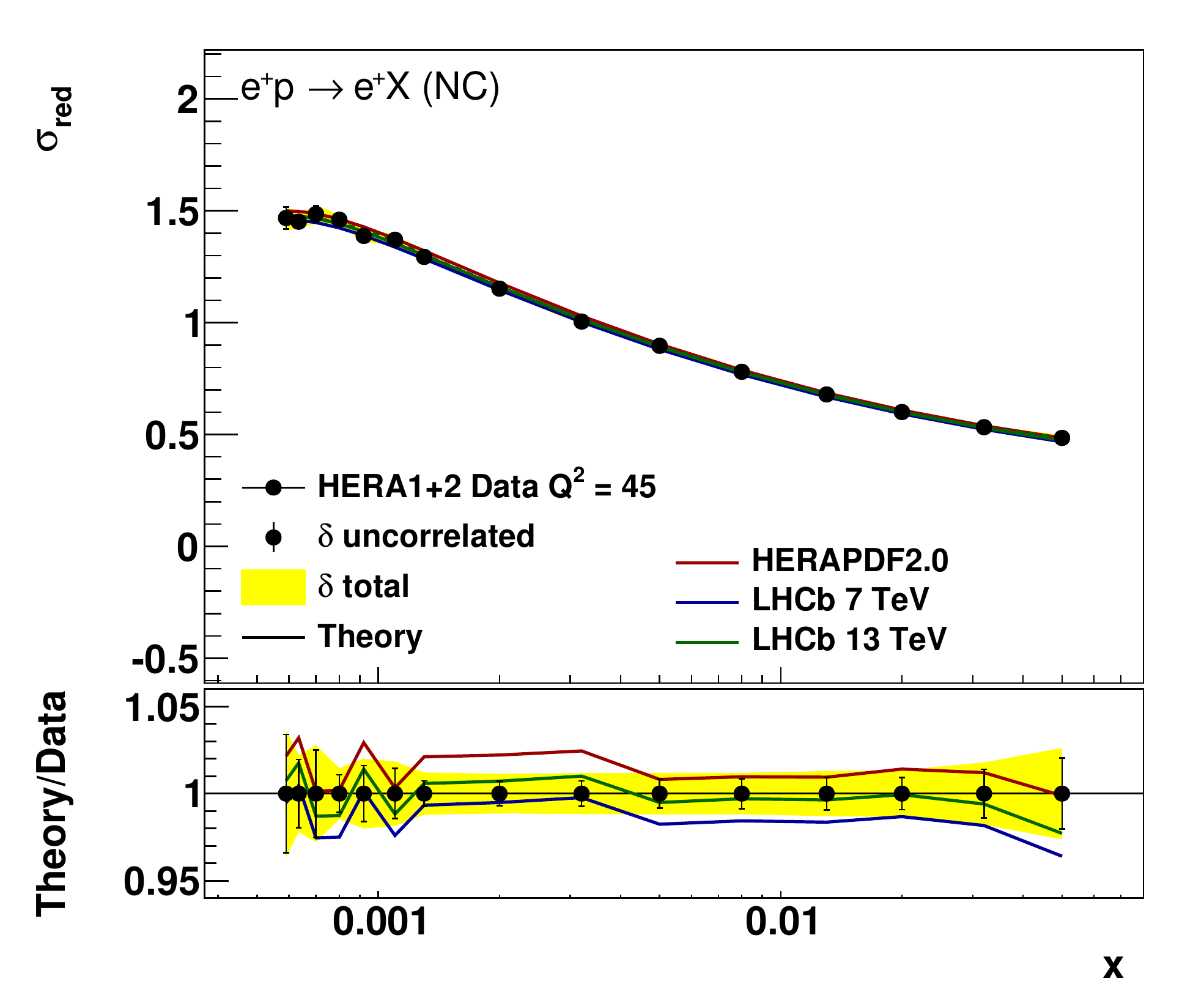}\\
\includegraphics[width=0.4\textwidth]{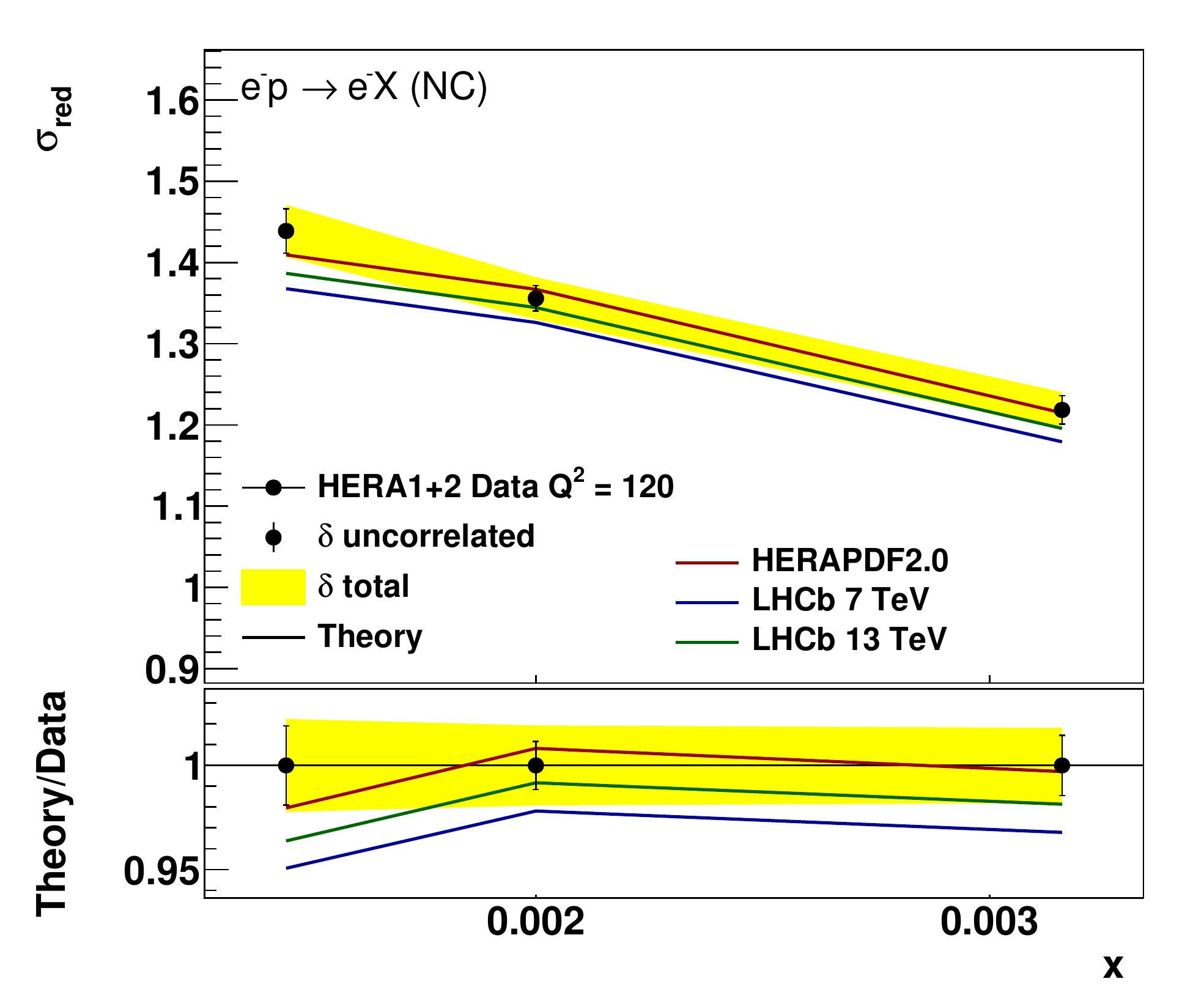}
\includegraphics[width=0.4\textwidth]{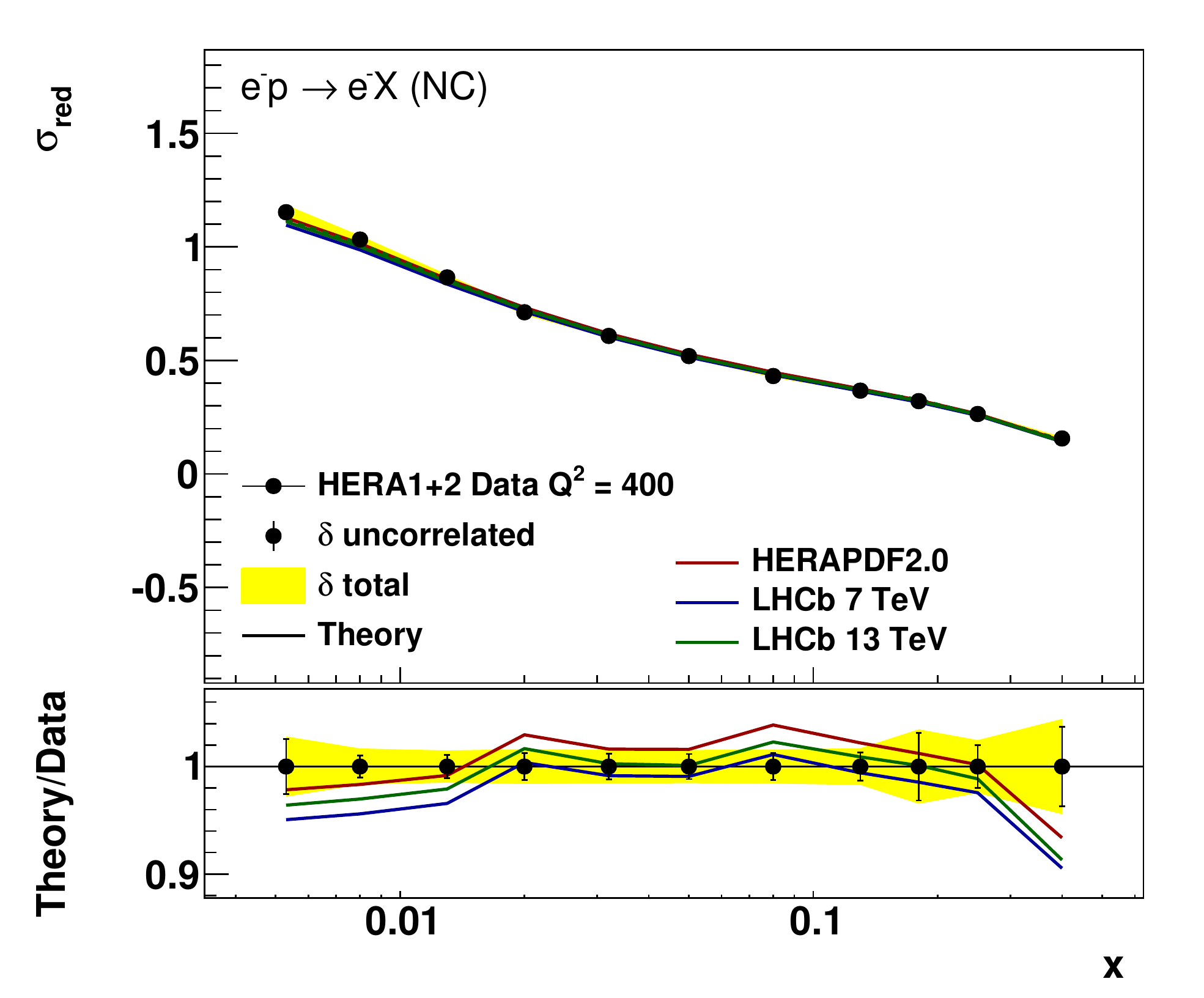}\\
\includegraphics[width=0.4\textwidth]{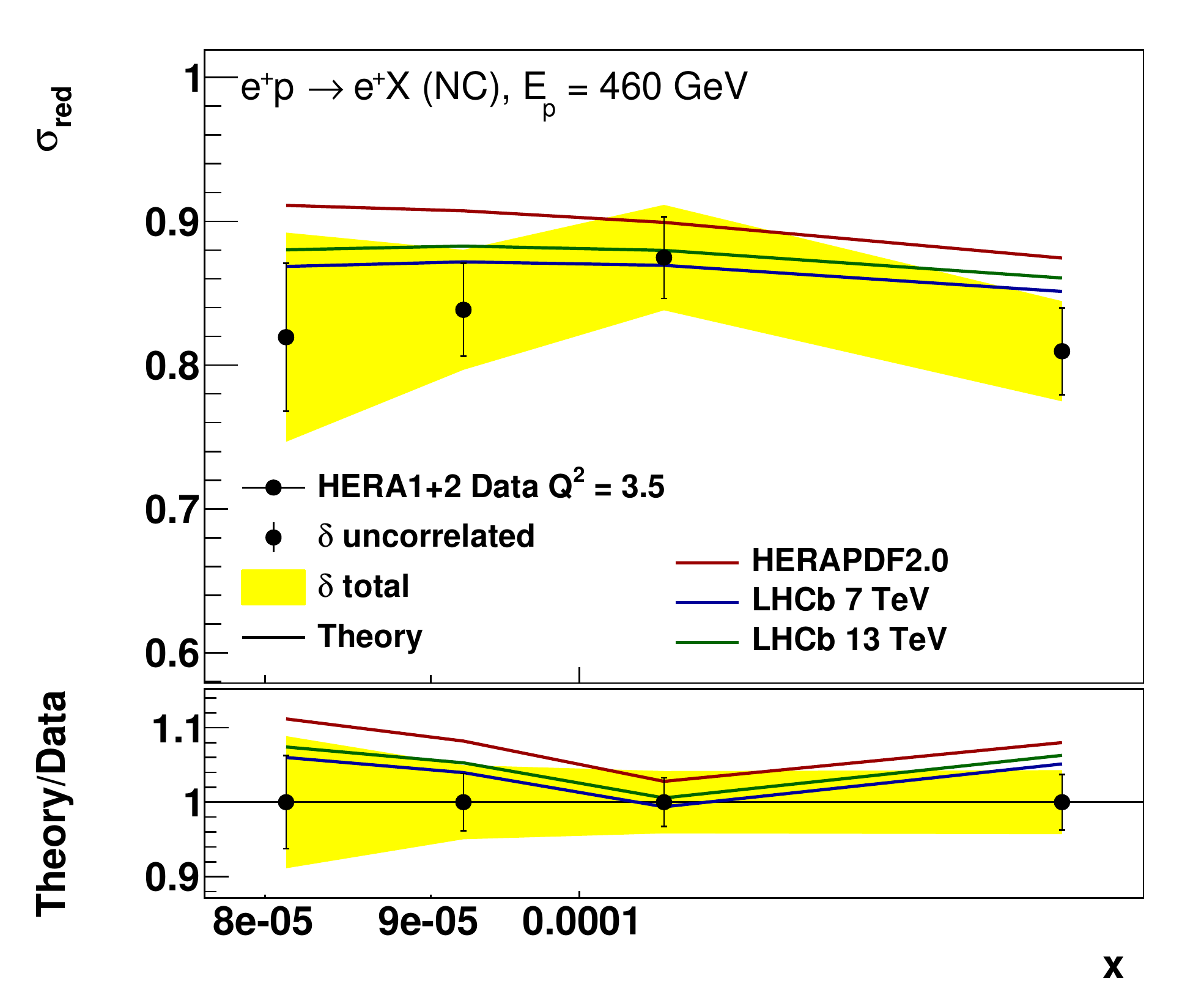}
\includegraphics[width=0.4\textwidth]{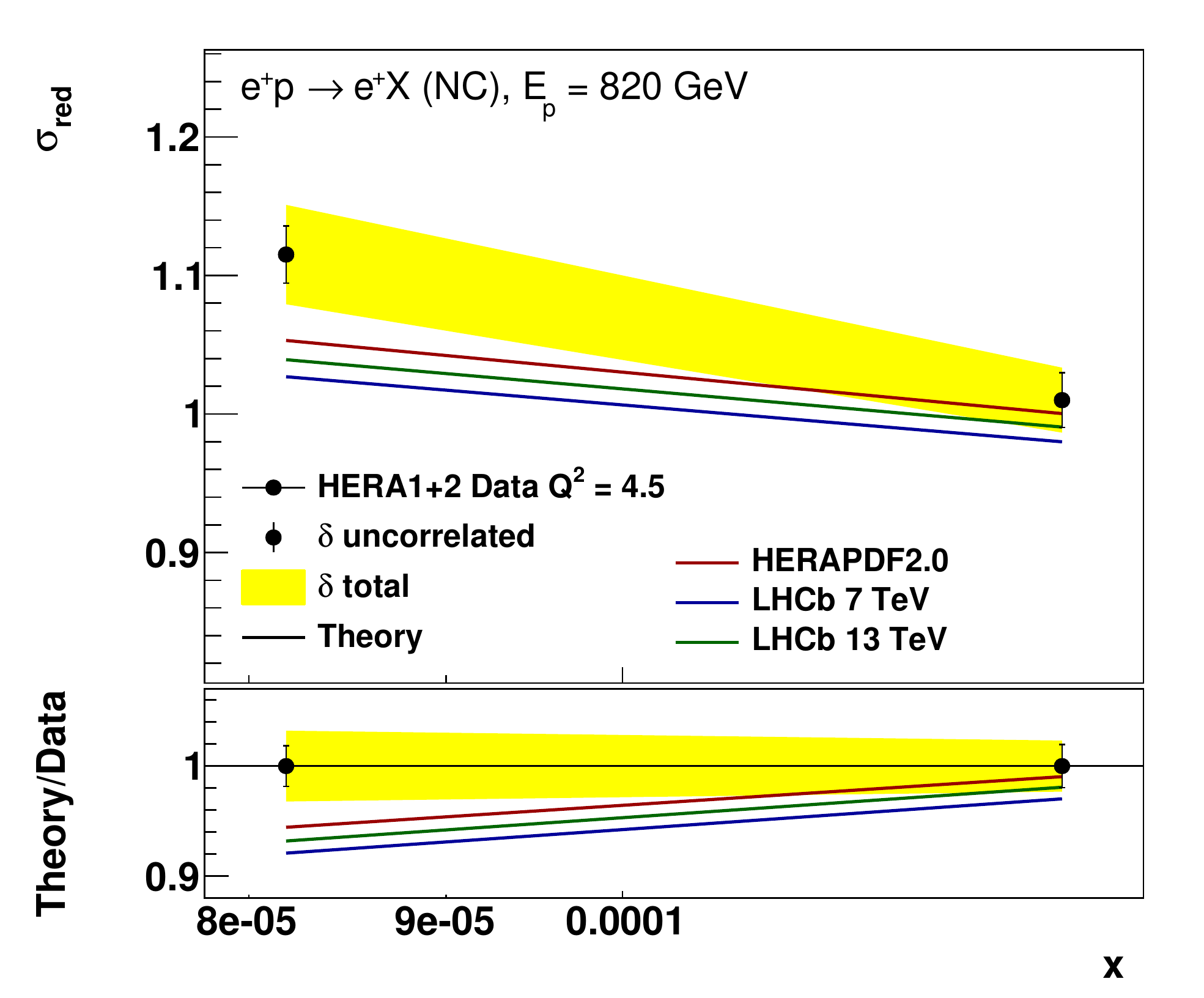}\\
\includegraphics[width=0.4\textwidth]{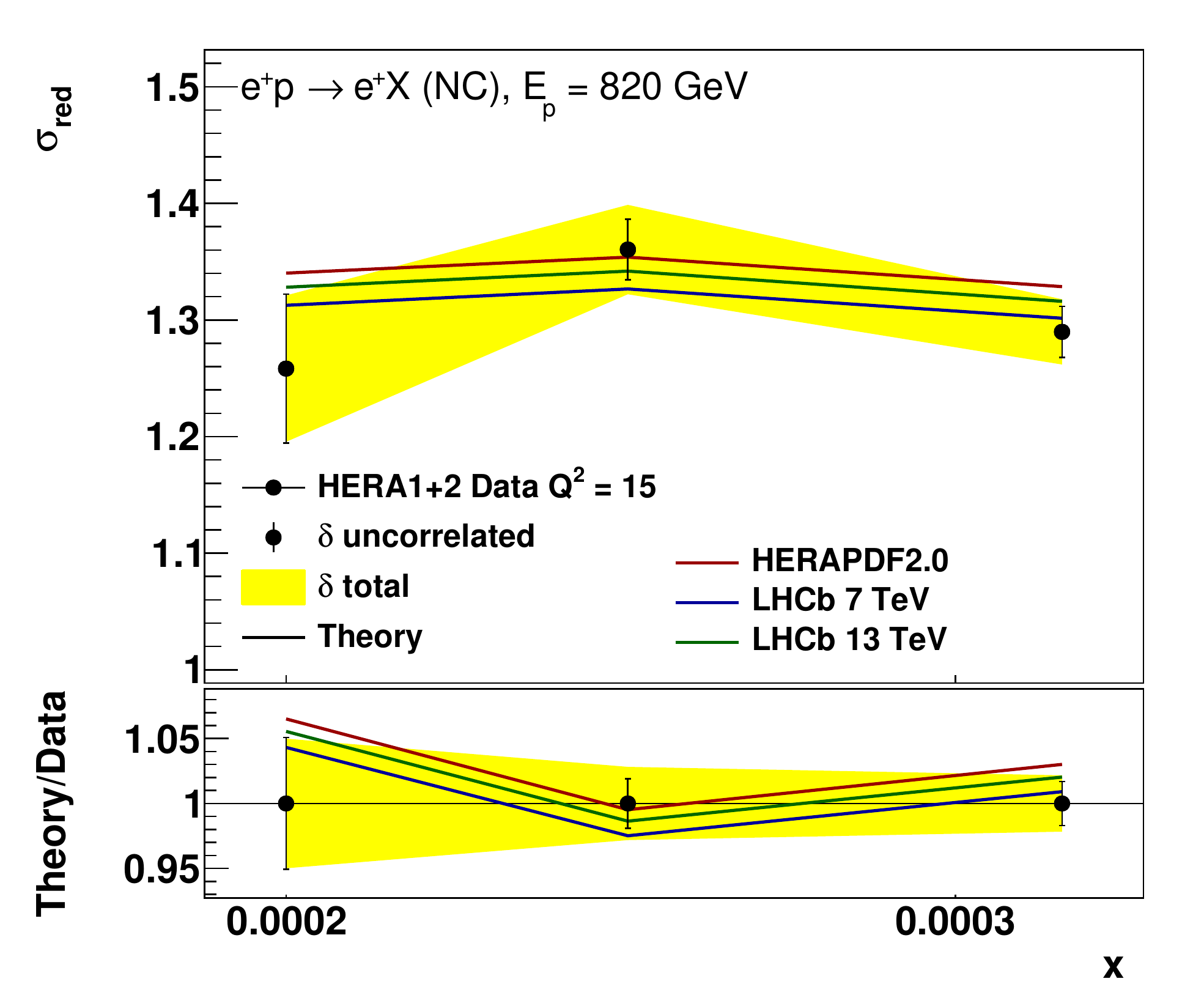}
\includegraphics[width=0.4\textwidth]{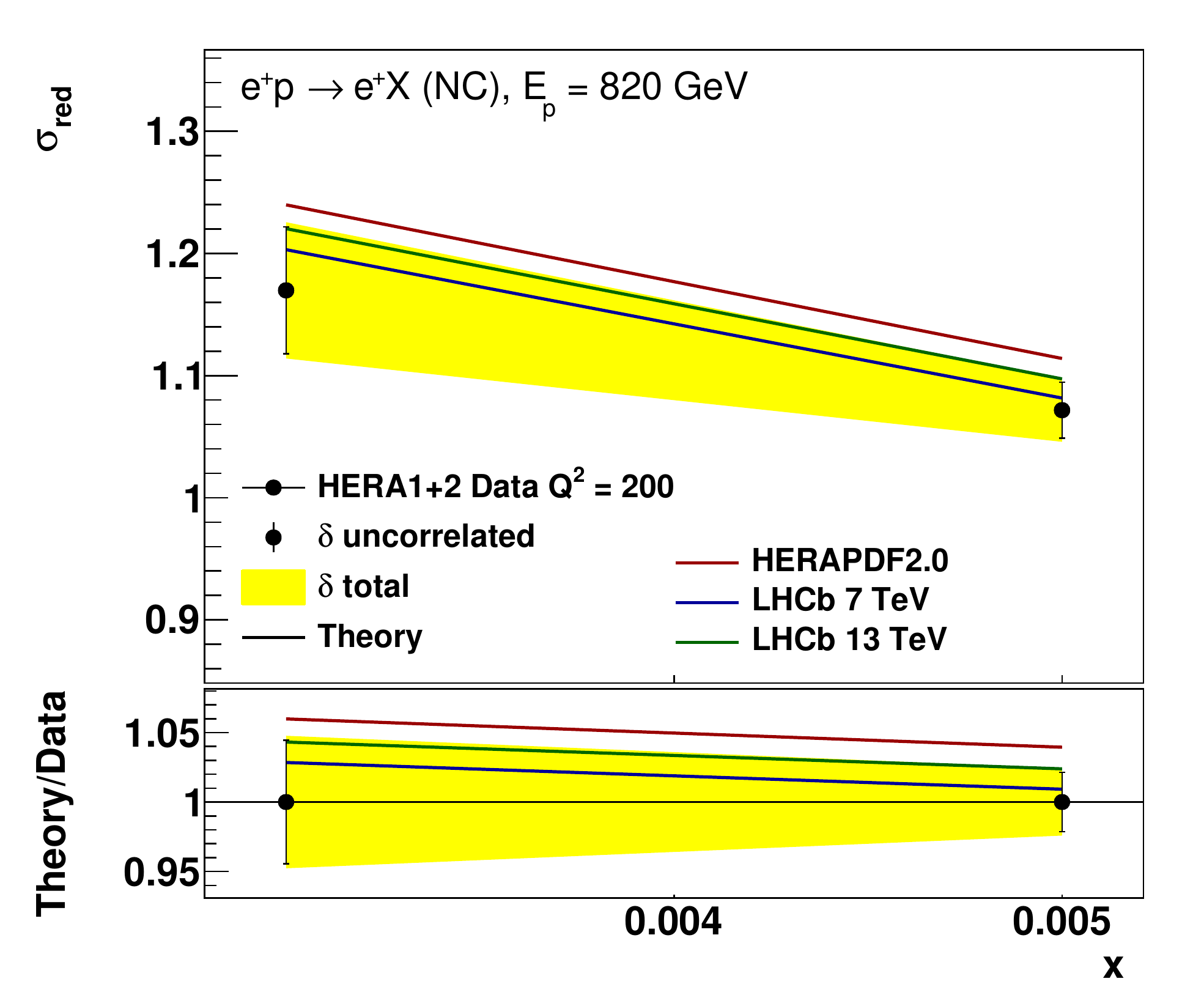}\\
\caption{The fit results for a representative subset of the HERA DIS combined data~\cite{Abramowicz:2015mha}, for the fits which are included only the HERA data (red),  considering the LHCb 7 TeV data sets (blue), and considering LHCb 13 TeV data sets (green) in substitution of 7 TeV data sets~\cite{Aaij:2017qml}.  } \label{fig:three}
\end{figure}

Fig.~\ref{fig:three} shows the fit results for a representative subset of the HERA DIS combined data~\cite{Abramowicz:2015mha}, for all three analyses performed in this paper. The panels have been chosen in such a way that represent both the goodness of fit and the existing tension between the HERA and LHCb data. According to the results obtained, some interesting points can be concluded. Firstly, note that the inclusion of the LHCb data in the analysis leads to a better fit of the HERA data at some kinematic regions. For example, see the panels containing the HERA1+2 NCep data with $ E_p=820 $ GeV at larger scales $ Q^2=15 $ and 200 GeV$ ^2 $. On the other hand, due to some existing tensions between the HERA and LHCb data, the agreement between the theory and experiment gets worse by inclusion the LHCb data in the analysis. This fact can be clearly seen from the panel corresponding to the HERA1+2 NCep data with $ E_p=820 $ and 460 GeV at low scales $ Q^2=4.5 $ and 3.5 GeV$ ^2 $, respectively. However, there are some kinematic regions in which the changes in agreement between the theory and experiment are not considerable after including the LHb data.  
 
\begin{figure}
\center
\includegraphics[width=0.40\textwidth]{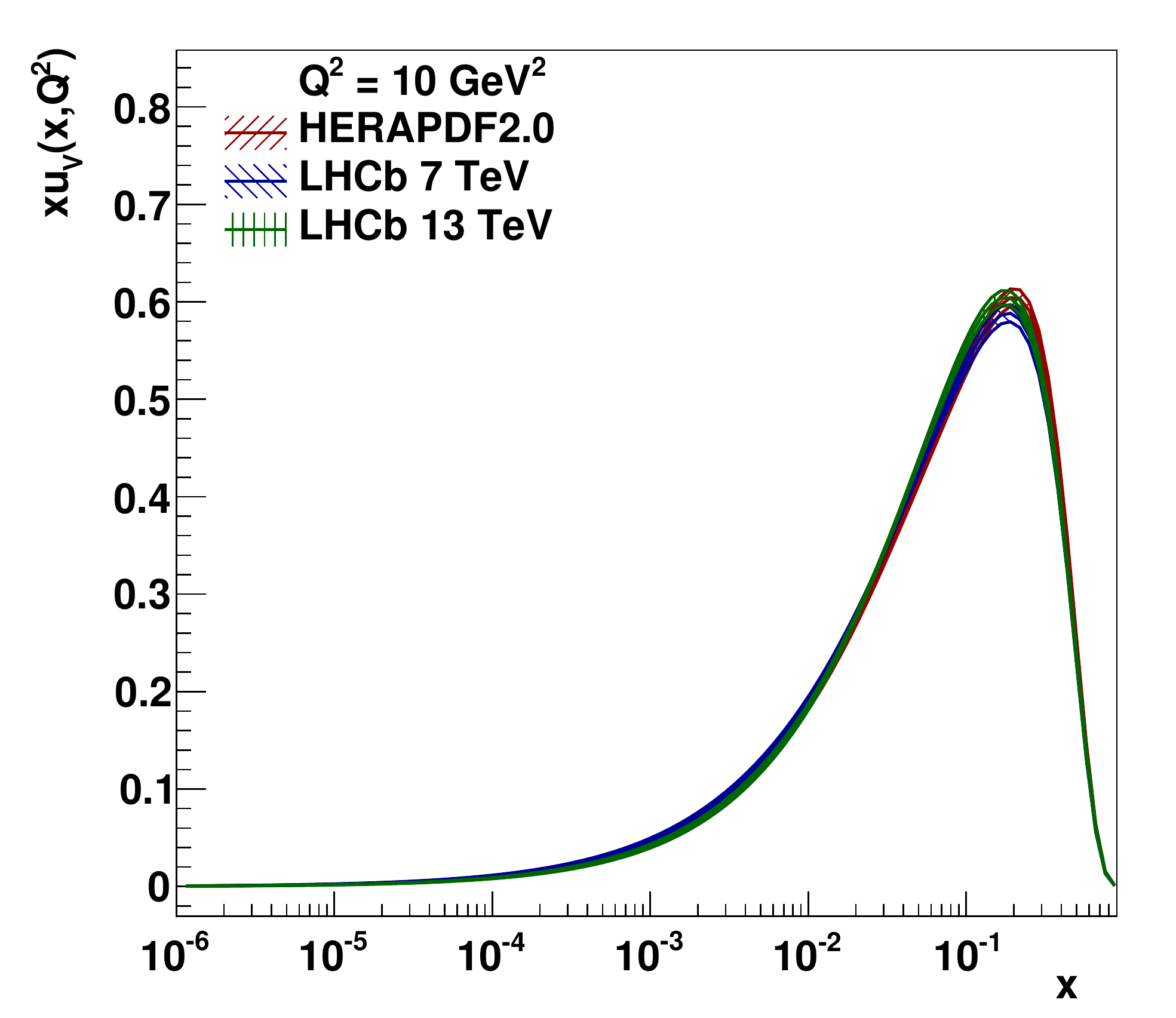}
\includegraphics[width=0.4\textwidth]{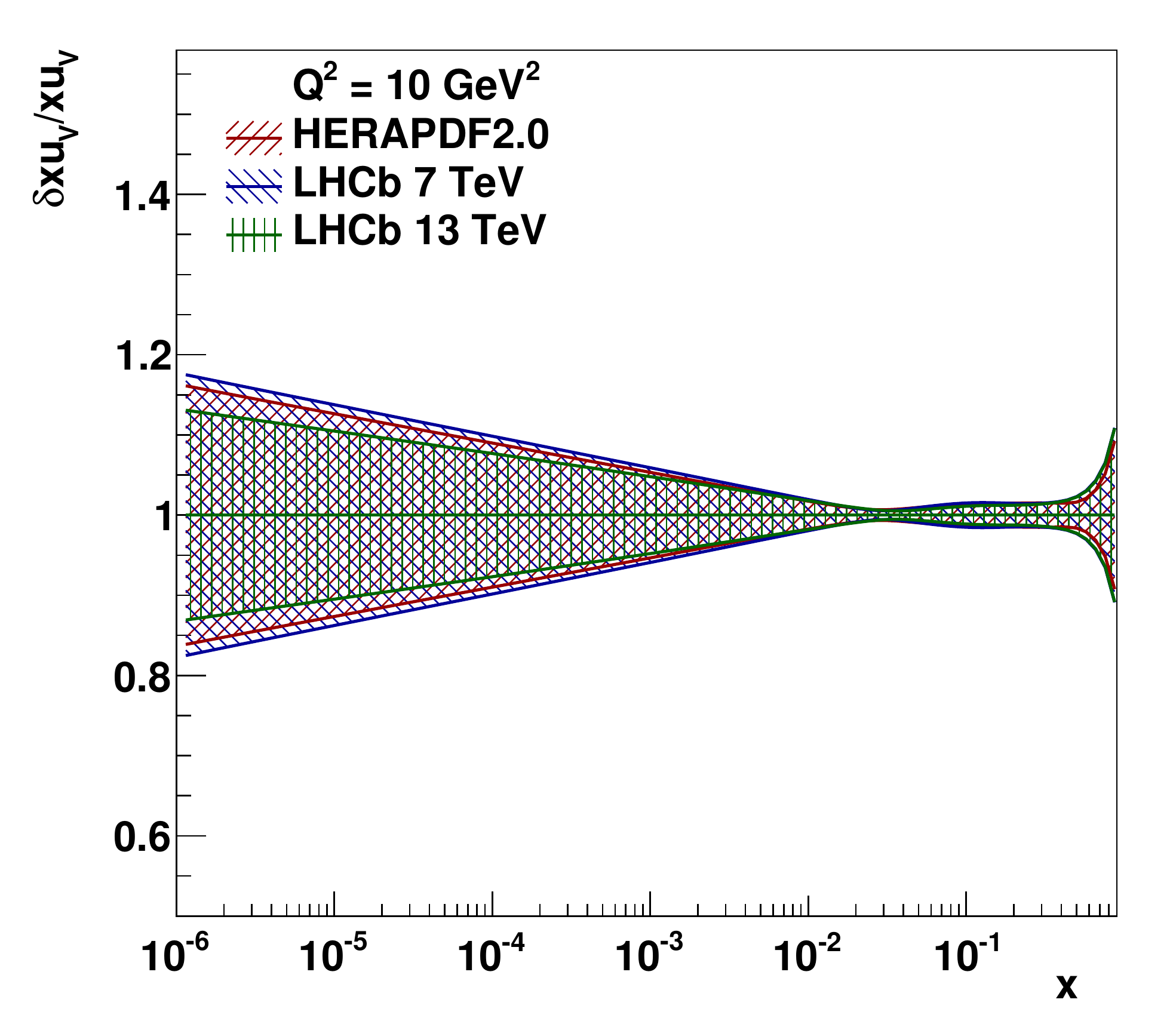}\\
\includegraphics[width=0.40\textwidth]{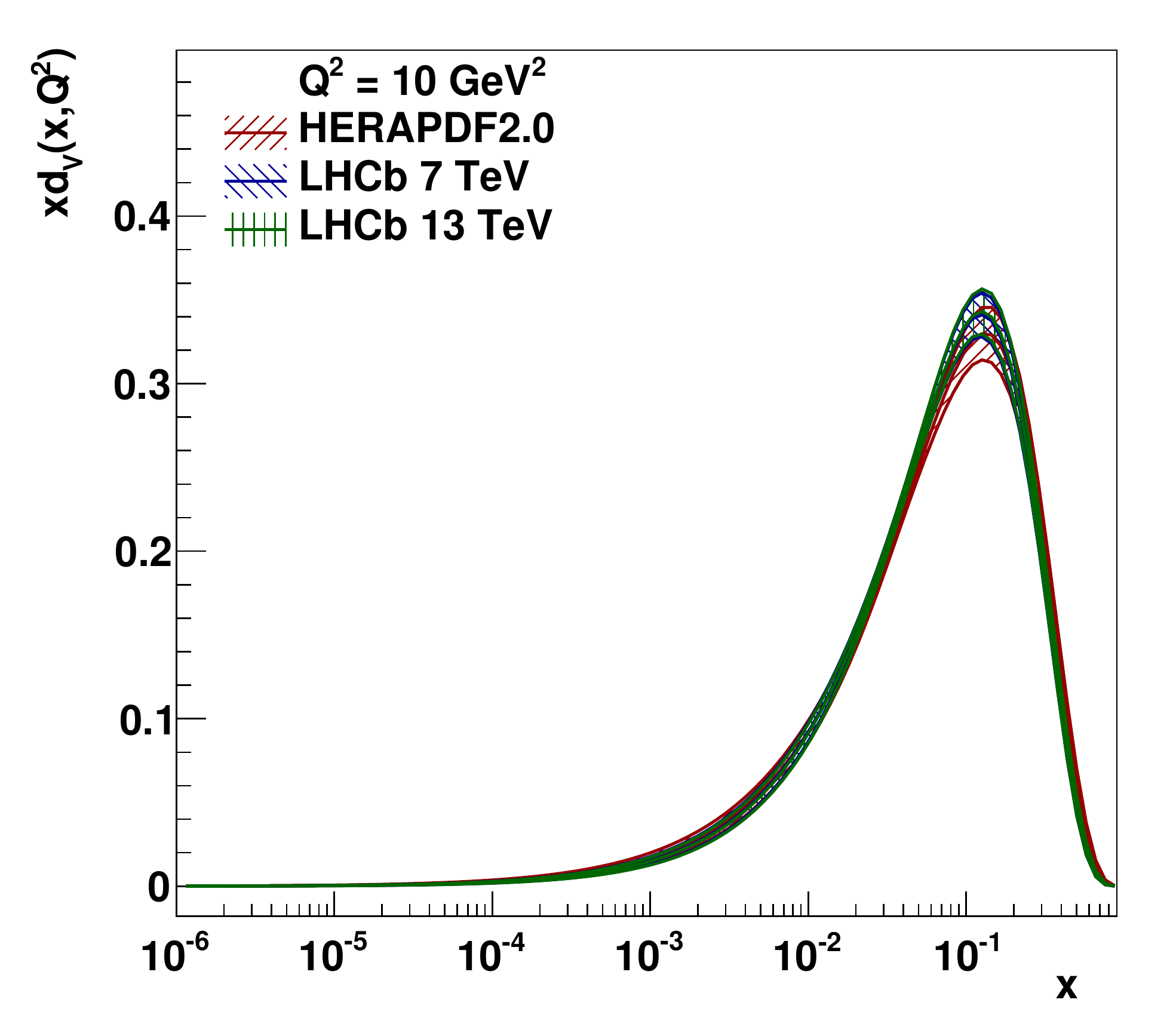}
\includegraphics[width=0.4\textwidth]{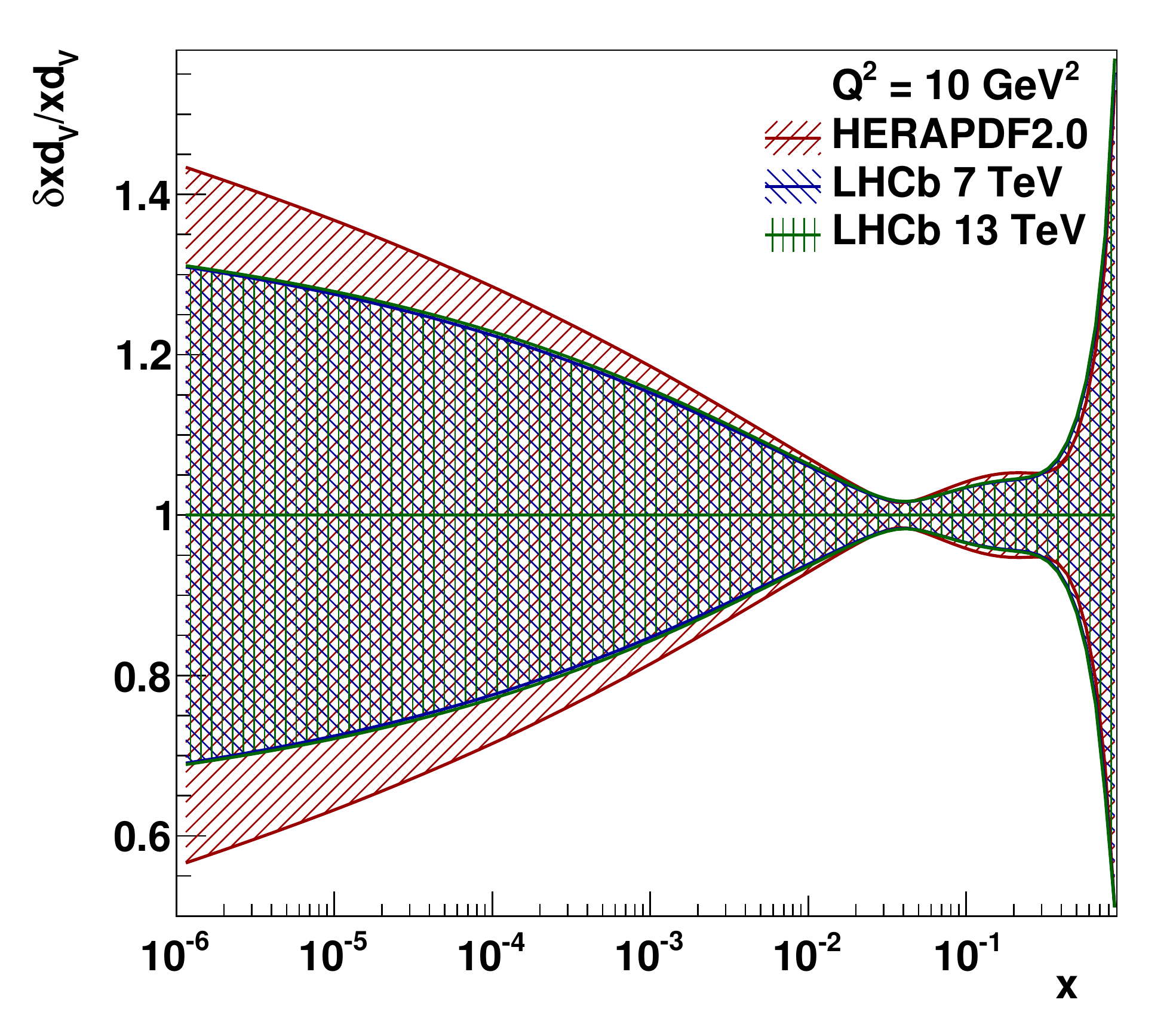}\\
\includegraphics[width=0.40\textwidth]{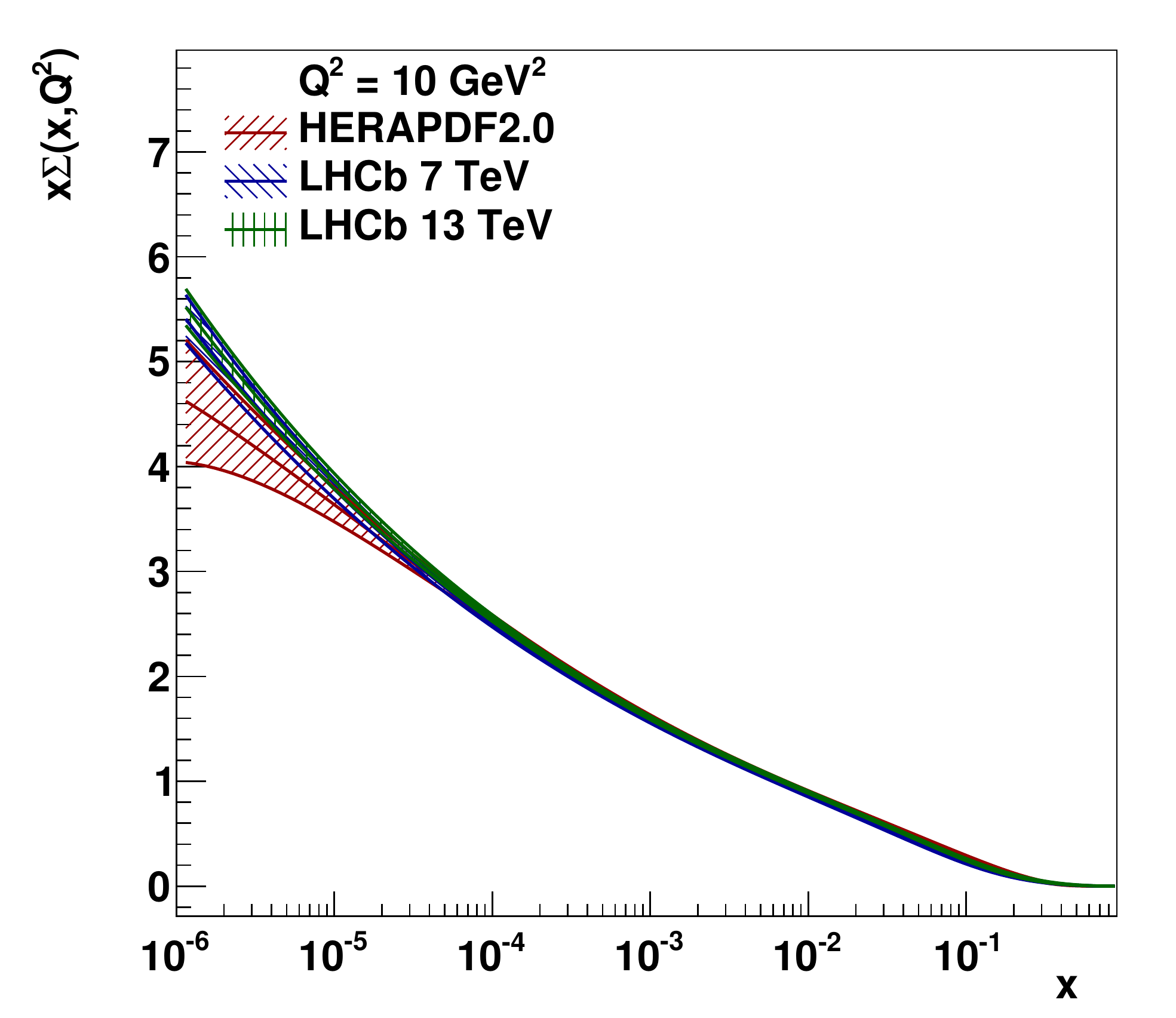}
\includegraphics[width=0.4\textwidth]{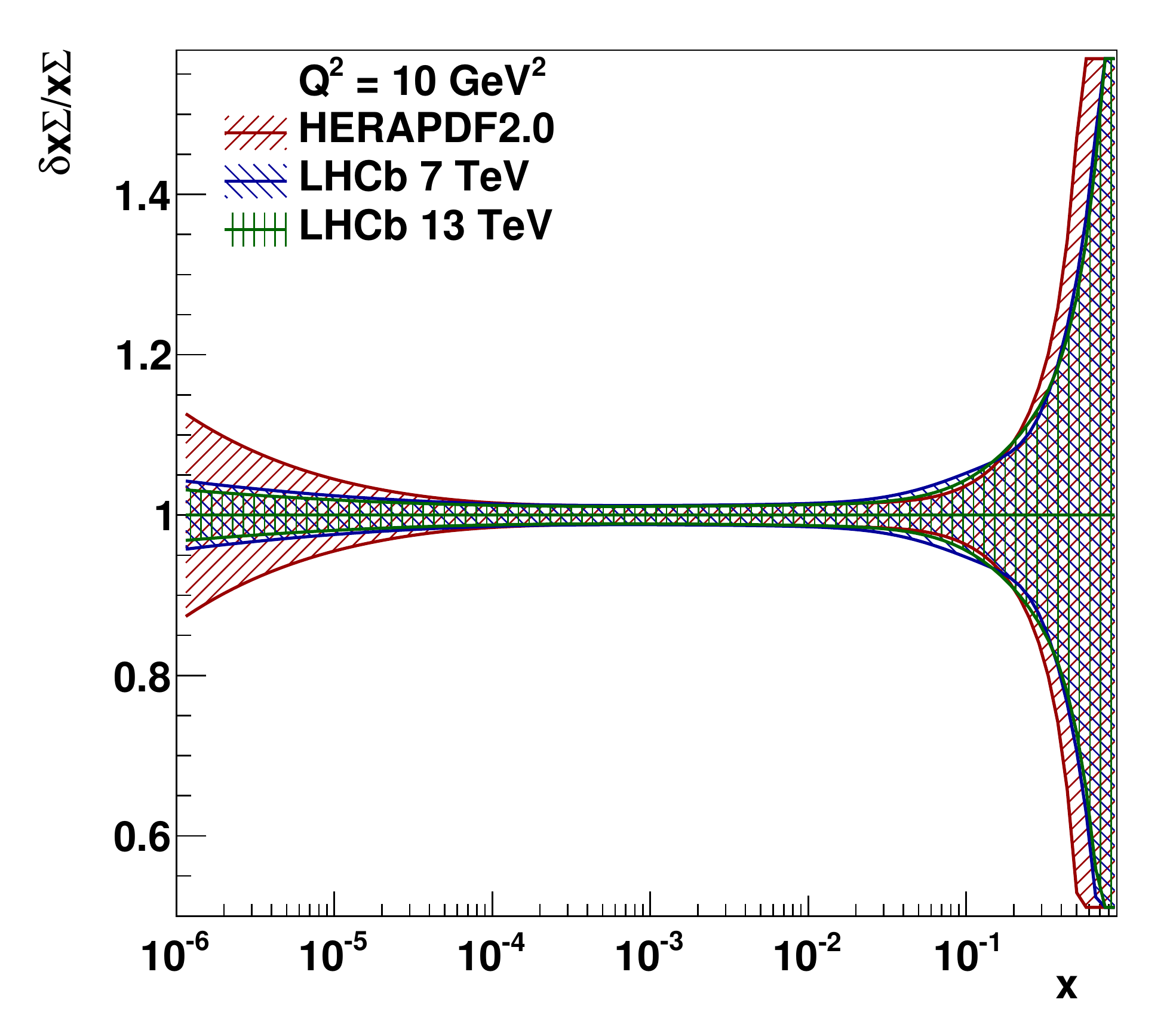}\\
\includegraphics[width=0.40\textwidth]{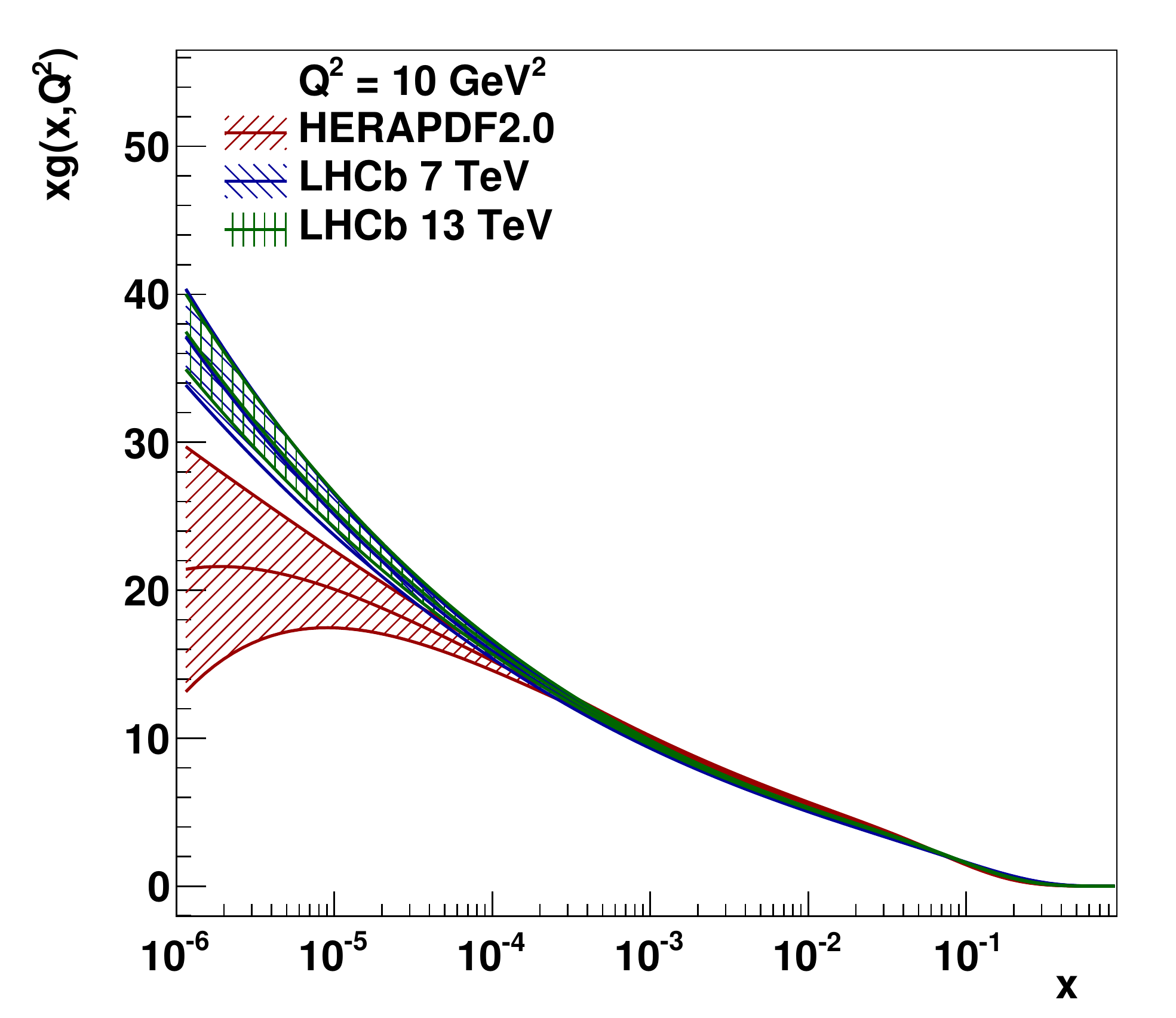}
\includegraphics[width=0.4\textwidth]{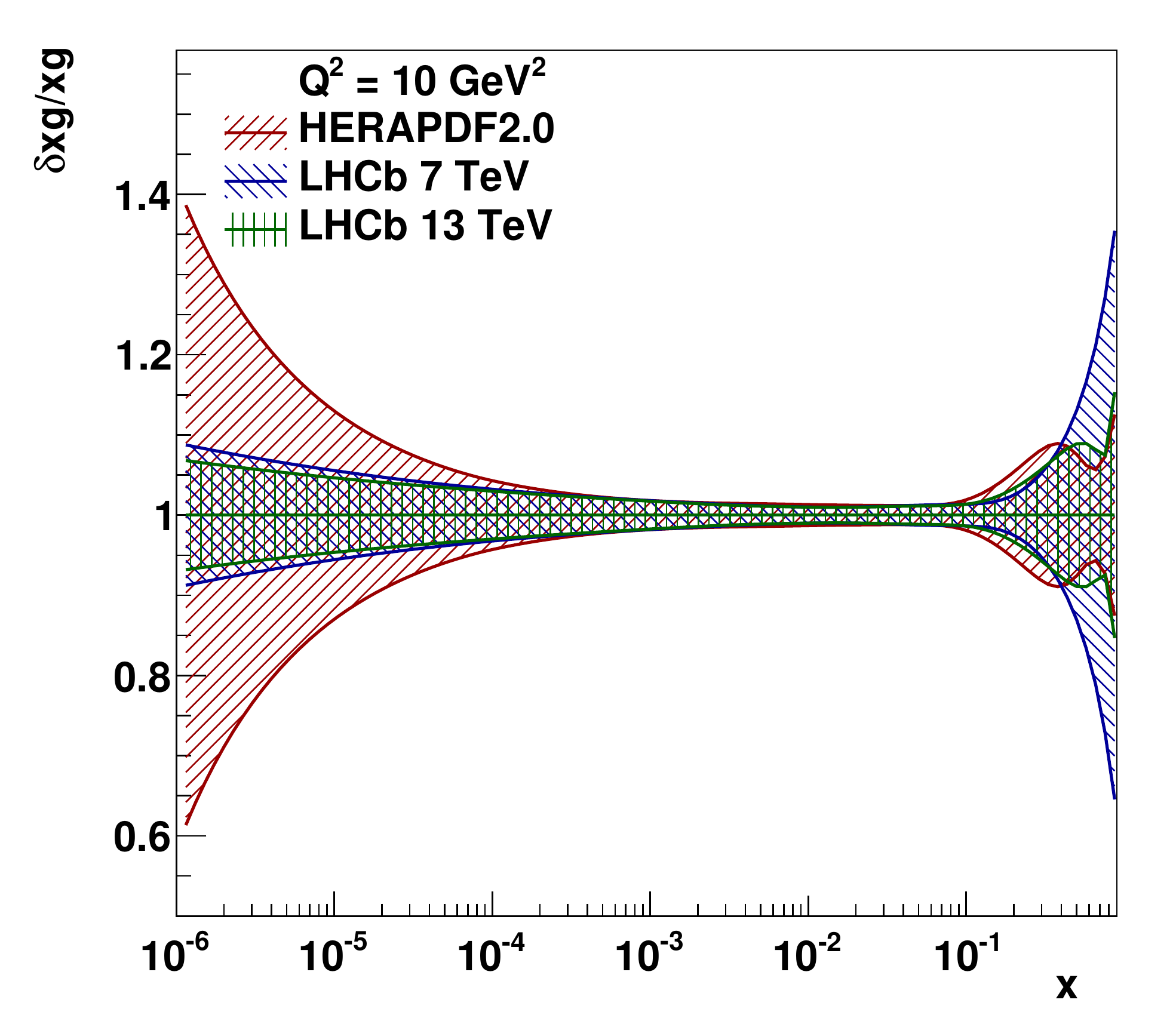}\\
\caption{  The extracted PDFs (left) and its relative PDF uncertainties (right) at the scale $Q^{2}$= 10 GeV$^{2}$, as a function of $x$ for $u_{v}$,  $d_{v}$, $g$, and $sea$, determined
with a fit to the HERA DIS data (red), adding with normalised LHCb 7 TeV data (blue)~\cite{Aaij:2017qml}, and finally adding with normalised 13 TeV (green) data~\cite{Aaij:2017qml}.  The widths of the bands represent the experimental uncertainties, and the sea-quark distribution is defined as $\Sigma=2 \cdot (\bar{u}+\bar{d}+\bar{s})$.} \label{fig:four}
\end{figure}

Fig.~\ref{fig:four} shows a comparison between the extracted PDFs (left) and their relative uncertainties (right) for $ u_v $ and $ d_v $ valence-quarks, gluon $ g $, and sum of sea-quarks $ \Sigma $ with their total  uncertainties at the scale $Q^{2}$= 10 GeV$^{2}$, as a function of $x$, for three different analyses. It should be noted that, the total uncertainty in this figure involves only the experimental uncertainties, and the model and parametrisation uncertainties have not been taken into account (they are studied separately at the end of this section). As can be seen from this figure, the inclusion of the LHCb data at both 7 and 13 TeV leads to a high impact on gluon and sea distributions at the low-$x$ kinematic ranges. In general, we can say that by inclusion the LHCb data at 7 TeV and 13 TeV, PDFs are almost affected similarly. In comparison to the base fit results which have not been contained the LHCb data, the most changes are appeared at small values of $ x $ as expected, where the LHCb data are dominant. 
%Moreover, it is observed  that the gluon and sea distribution at large value of $x$ are more  affected with inclusion of LHCb data at 7 TeV in comparison with the LHCb data at 13 TeV. This discrepancy may be due to the difference in the kinematic region which covered by LHCb data at 7 and 13 TeV. According to the leading order relation of $ x=\frac{2p_T}{\sqrt{s}}e^{\pm y} $, at the same rapidity and transverse momentum,  .
Another important thing can be pointed out from the results obtained is the reduction in uncertainties of almost all distributions especially at smaller values of $ x $ by inclusion the LHCb data both for 7 and 13 TeV in the analysis. To be more precise, although the reduction in uncertainties of valence-quark distributions, especially of $ u_v $ is not significant, but it is indeed significant for the sea quarks and gluon distributions. In fact, the reduction in uncertainties of gluon up to 20\% at
$x=10^{-6}$ is one of the most important results of the present analysis, because of its crucial rule in many areas of high energy physics investigations.
Another important result can be concluded from Fig.~\ref{fig:four} is that the inclusion of the LHCb beauty production data at $ \sqrt s= $13 TeV can put somewhat more tighter constraints on the gluon and sea distributions rather than the data at $ \sqrt s= $7 TeV.
%For the case of gluon, although both LHCb data sets at 7 a 13 TeV lead to similar changes in low $ x $, their impacts on the gluon density are different in larger values of $ x $. To be more precise, the impact of 7 TeV data on the gluon distribution at larger $ x $ is more intense rather than the impact of 13 TeV data. 
% Another interesting result obtained through our analyses is the impact of LHCb data on the valence-quark distributions, especially of $ u_v $. In fact, although both LHCb data sets at 7 a 13 TeV lead to similar changes in $ d_v $ distribution in all ranges of $ x $, their impacts on $ u_v $ distribution are completely different at smaller values of $ x $. This finding can be attributed to the further generation of the $ \bar u $ quarks than $ \bar d $ in the proton sea.

For more understanding of how PDFs have affected by including the LHCb data in the analysis according to Fig.~\ref{fig:four}, the sensitivity of data to the PDFs should be examined more accurately. For example, with the
naive LO $2 \rightarrow 2$ kinematics, the momentum fractions in the PDFs of target and projectile typically probed by $B^{\pm}$ production at the LHC can be estimated using the values of proton beam energy $ E_p $, transverse momentum $ p_\textrm{T} $, heavy quark mass $ m_Q $ and
rapidity $ y $ as:
\begin{equation}
x_{1,2}\approx \frac{\sqrt{p_T^2+m_Q^2}}{E_p} e^{\pm y}.
\label{eq2}
\end{equation}
For the case of LHCb data at 7 TeV, such a sensitivity has been investigated in Fig. 1 of Ref.~\cite{Zenaiev:2015rfa}, which shows the LHCb beauty data cover the smaller and larger values of $ x $ than the HERA data, specially for the case of larger rapidities, while the HERA data are dominant at $ 10^{-4} \lesssim x \lesssim 10^{-1} $. Therefore, we expect that by inclusion of the LHCb data, the most changes in the behaviour of PDFs are seen in very small values of $ x $ where the HERA data are not dominant. Another point should be considered is that for the case of data with higher value of center-of-mass energy, namely $ \sqrt{s}=13 $ TeV, the typical $x$ coverage of this data are shifted even to more smaller $ x $. 
This is exactly what has been achieved in Fig.~\ref{fig:four}, where the PDFs have changed in smaller values of $ x $, while remained almost unchanged at other values. 

\begin{figure}
\center
%\includegraphics[width=0.55\textwidth]{pic/d_v.eps}\\
%\vspace{0.5cm}
%\includegraphics[width=0.55\textwidth]{pic/u_v.eps}\\
%\vspace{0.5cm}
\includegraphics[width=0.55\textwidth]{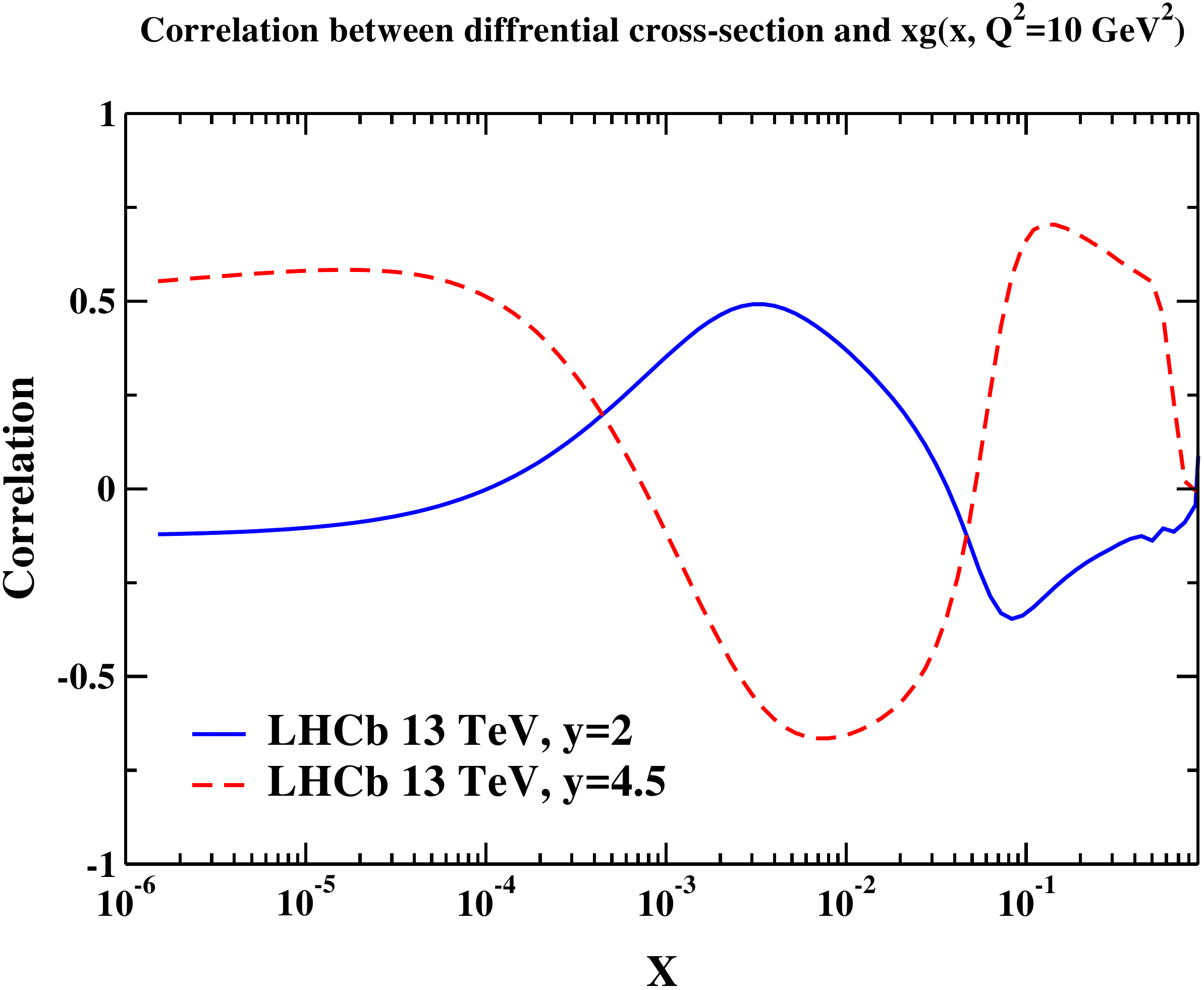}
\caption{Correlations $\cos\varphi$ between the differential cross section of the LHCb $B^{\pm}$ production at $ \sqrt{s}=13 $ TeV for two different rapidity values $ y=2 $ and 4.5, and $ xg(x) $ distributions at $Q^2=10 $ GeV$ ^{2} $.} \label{fig:correlations}
\end{figure}

Nevertheless, investigating the correlations between data and PDFs can be very instructive to explore the sensitivity of data to the PDFs. In this regard, we study such correlations, for instance, between differential cross section of $B^{\pm}$ production measured by LHCb at $ \sqrt{s}=13 $ TeV, and $ xg(x) $ distribution for two different rapidity values $ y=2 $ and 4.5. Generally, correlations can be computed for any variable 
$X(\vec{a})$, where $\vec{a}$ forms a vector in an
$N$-dimensional PDF parameter space, with $N$ being the number of
free parameters in the global analysis that determines these PDFs. 
The correlation between two variables $X(\vec{a})$ and
$Y(\vec{a})$ is computed by~\cite{Nadolsky:2008zw}: 

\begin{eqnarray}
\cos\varphi=\frac{1}{4\Delta X\Delta Y}\sum_{i=1}^{N}(X_{i}^{(+)}-X_{i}^{(-)})(Y_{i}^{(+)}-Y_{i}^{(-)}),
\label{cosphi}
\end{eqnarray}
where $X_{i}^{(+)}$ and $X_{i}^{(-)}$ are the values of $X$ computed
from the two sets of PDFs along the ($\pm$) direction of the $i$-th
eigenvector and the $\Delta X$ is given as:
\begin{equation}
\Delta X=\left\vert \vec{\nabla}X\right\vert =\frac{1}{2}\sqrt{\sum_{i=1}^{N}\left(X_{i}^{(+)}-X_{i}^{(-)}\right)^{2}}.\label{masterDX}\end{equation}
The quantity $\cos\varphi$ characterizes whether $X$ and $Y$ are correlated ($\cos\varphi\approx1$),
anti-correlated ($\cos\varphi\approx-1$), or uncorrelated ($\cos\varphi\approx0$).
Fig.~\ref{fig:correlations}, for instance, demonstrates the correlations $\cos\varphi$ between the differential cross section of the LHCb $B^{\pm}$ production at $ \sqrt{s}=13 $ TeV for two different rapidity values $ y=2 $ and 4.5, and $ xg(x) $ distributions at $Q^2=10 $ GeV$^{2}$. Before discussing the correlations and their compatibility with the results obtained for the PDFs in Fig.\ref{fig:four}, we have to keep in mind one point and that is, according to the relation \ref{eq2}, the data with $ y=2 $
can affect generally PDFs at $ 10^{-4} \lesssim x \lesssim 0.2 $, while the data with higher value $ y=4.5 $ can affect PDFs at even smaller and larger values of $ x $.

Focusing on the results obtained for gluon distribution in Fig.~\ref{fig:correlations},
we see that there are two effective correlation areas at very small and large values of $ x $ for $ y=4.5 $ which can affect the gluon density at these regions. At $ 0.001 \lesssim x \lesssim 0.01 $, the gluon density indicates an anti-correlation and a correlation with differential cross section with $ y=4.5 $ and $ y=2 $, respectively. But, the first one is not effective and on the other hand, the gluon density is well constrained by HERA DIS data in this region, so that the LHCb data cannot change the behaviour of gluon distribution considerably. As can be seen, the results obtained in Fig.~\ref{fig:correlations} are consistent with the results shown in Fig.~\ref{fig:four} for gluon distribution.

%Furthermore, the effect of the LHCb data on the PDF uncertainties can be investigated much more accurately by comparison the relative uncertainty of the three PDFs, which are shown in Fig.~\ref{fig:five} . A significant reduction of the PDF uncertainties is clearly observed in this figure by including the LHCb measurements. The only exception is observed for the case of $ u_v $ distribution belongs to the analysis containing the LHCb data at 7 TeV. It should be also noted that the most changes in uncertainties are totally happened in the small values of $ x $, especially for the case of gluon PDF. Another point to consider is that the behavior of the changes in uncertainties at larger values of $ x $, for the case of gluon distribution, differs from the analysis containing the LHCb 7 TeV data to the ones performed by including 13 TeV data.

\begin{figure}
\center
\includegraphics[width=0.4\textwidth]{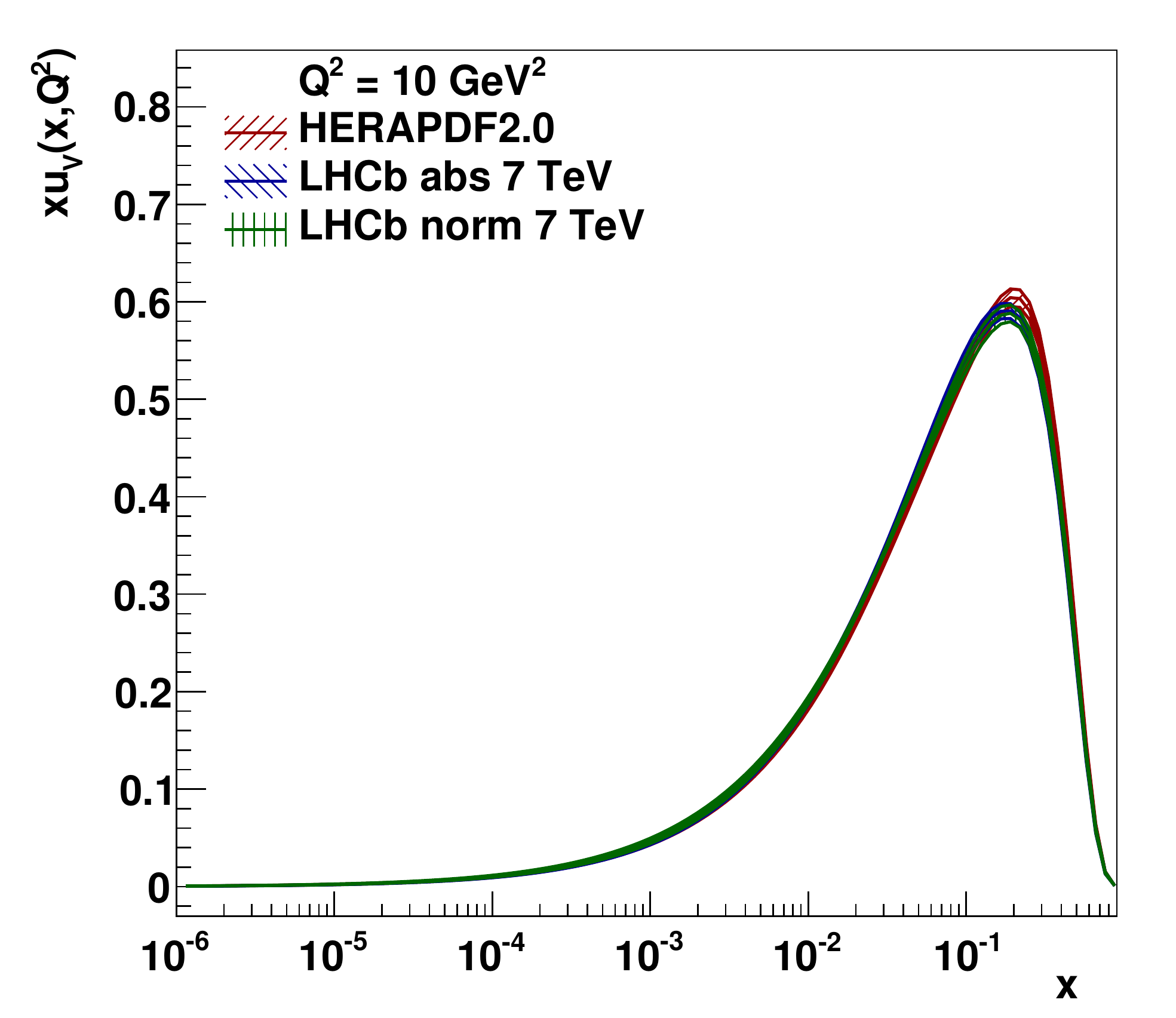}
\includegraphics[width=0.4\textwidth]{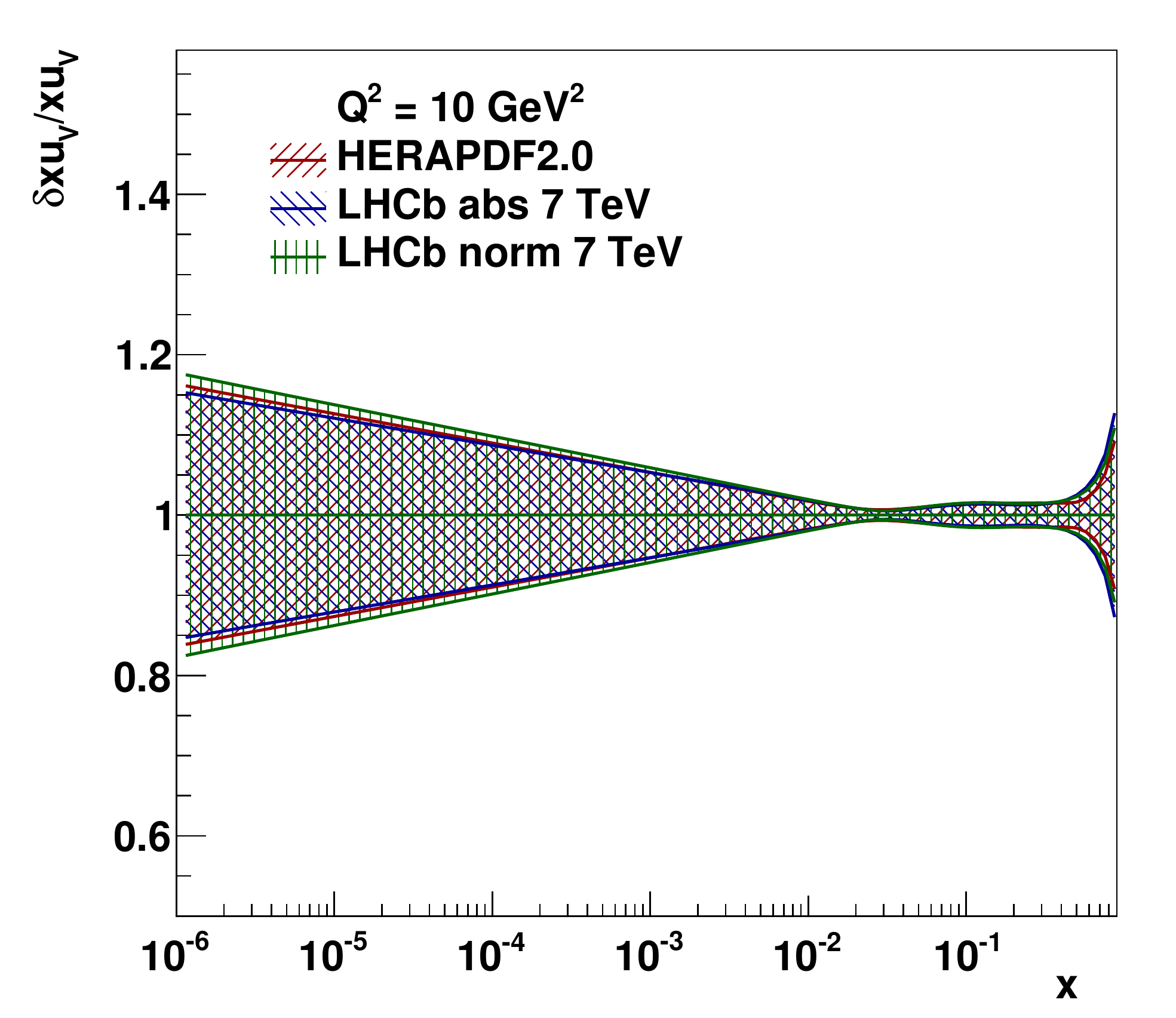}\\
\includegraphics[width=0.4\textwidth]{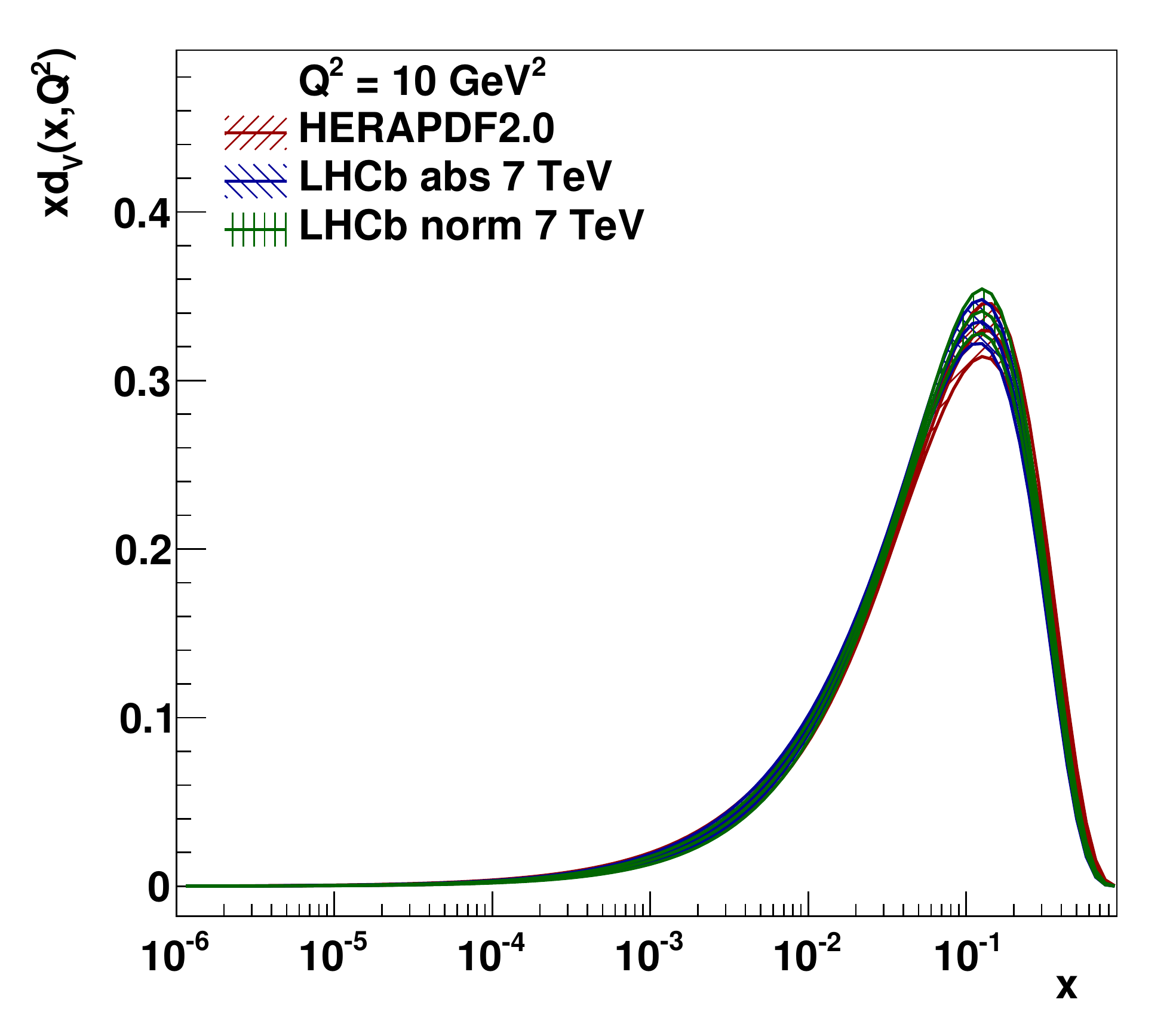}
\includegraphics[width=0.4\textwidth]{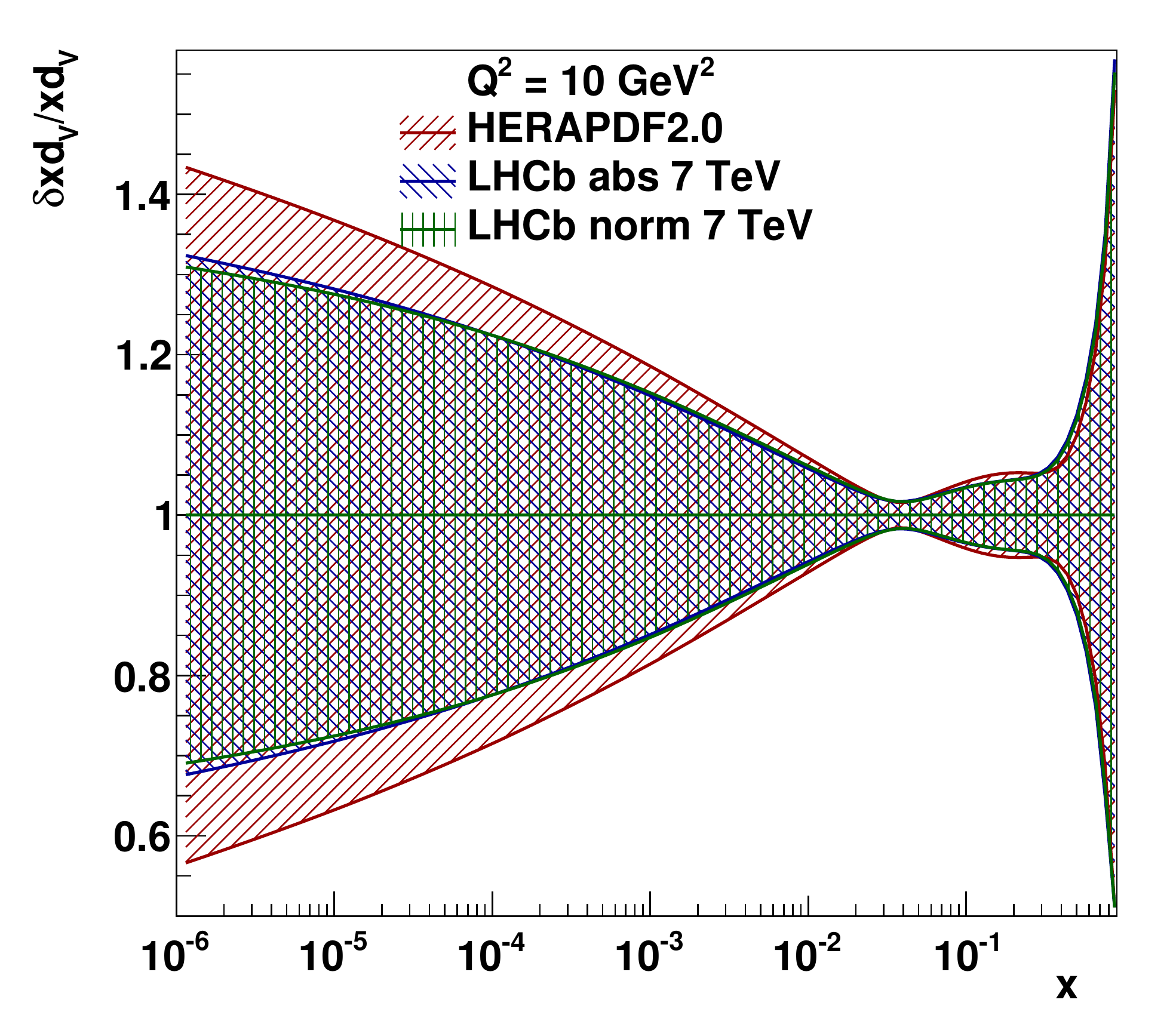}\\
\includegraphics[width=0.4\textwidth]{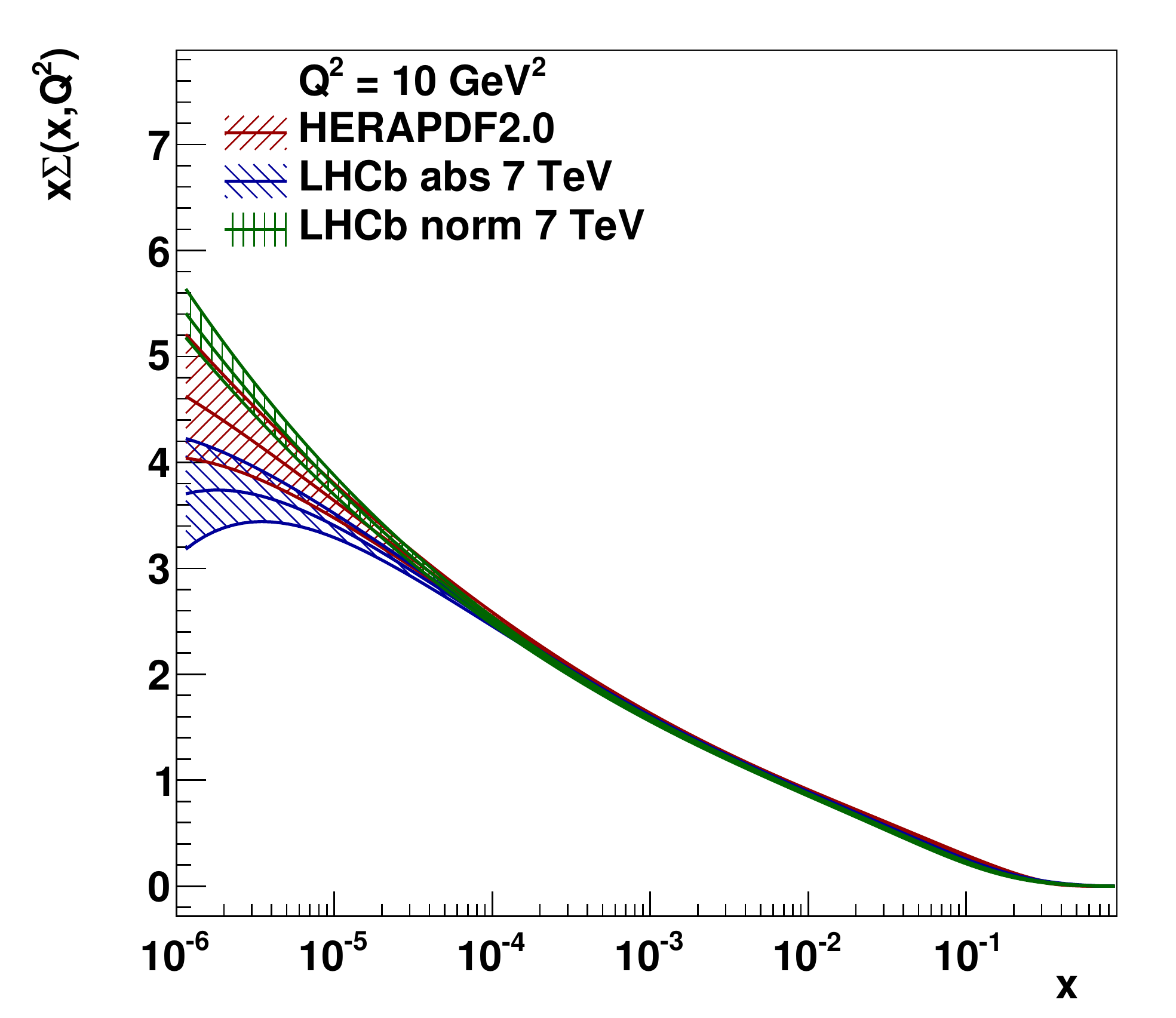}
\includegraphics[width=0.4\textwidth]{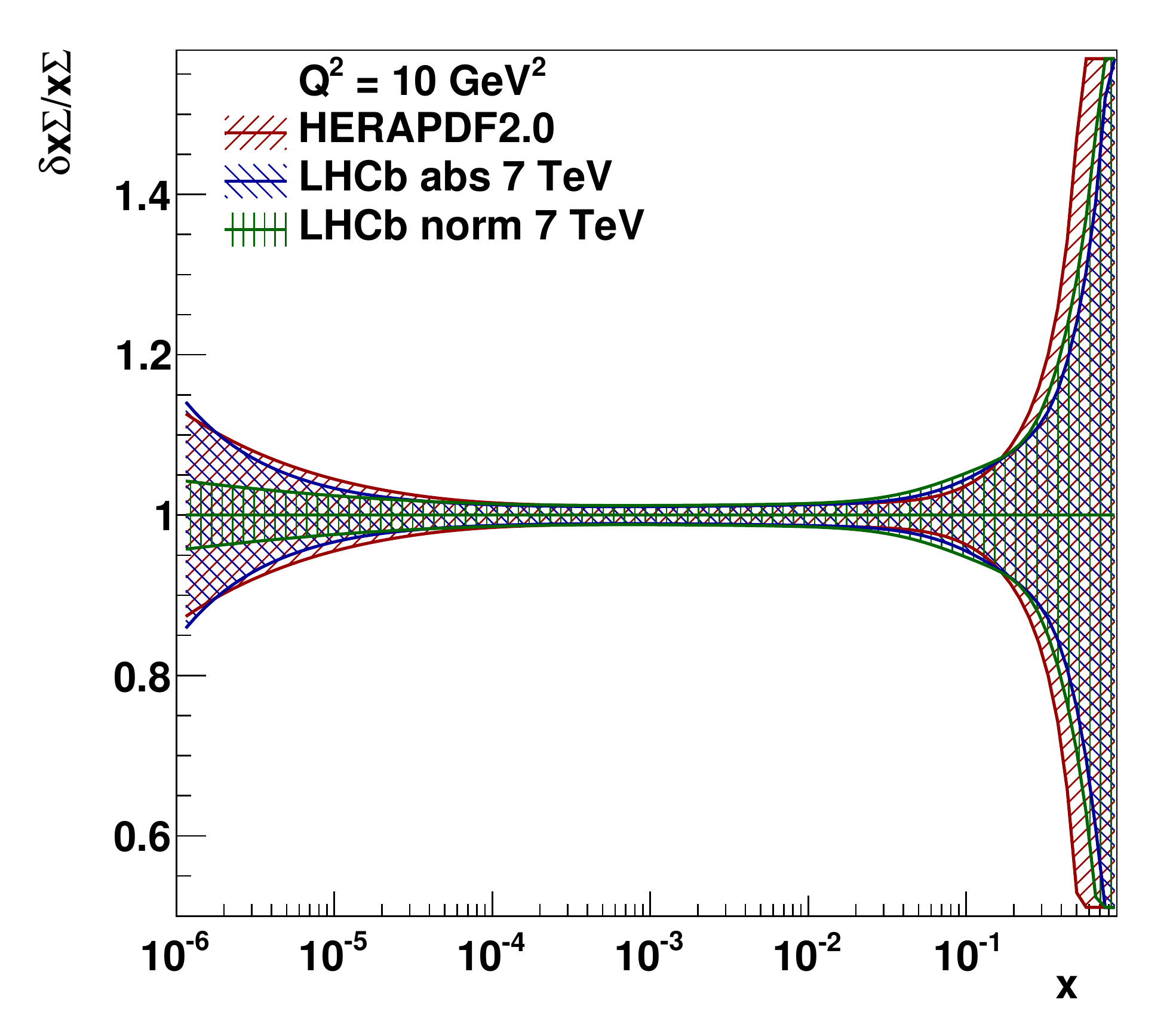}\\
\includegraphics[width=0.4\textwidth]{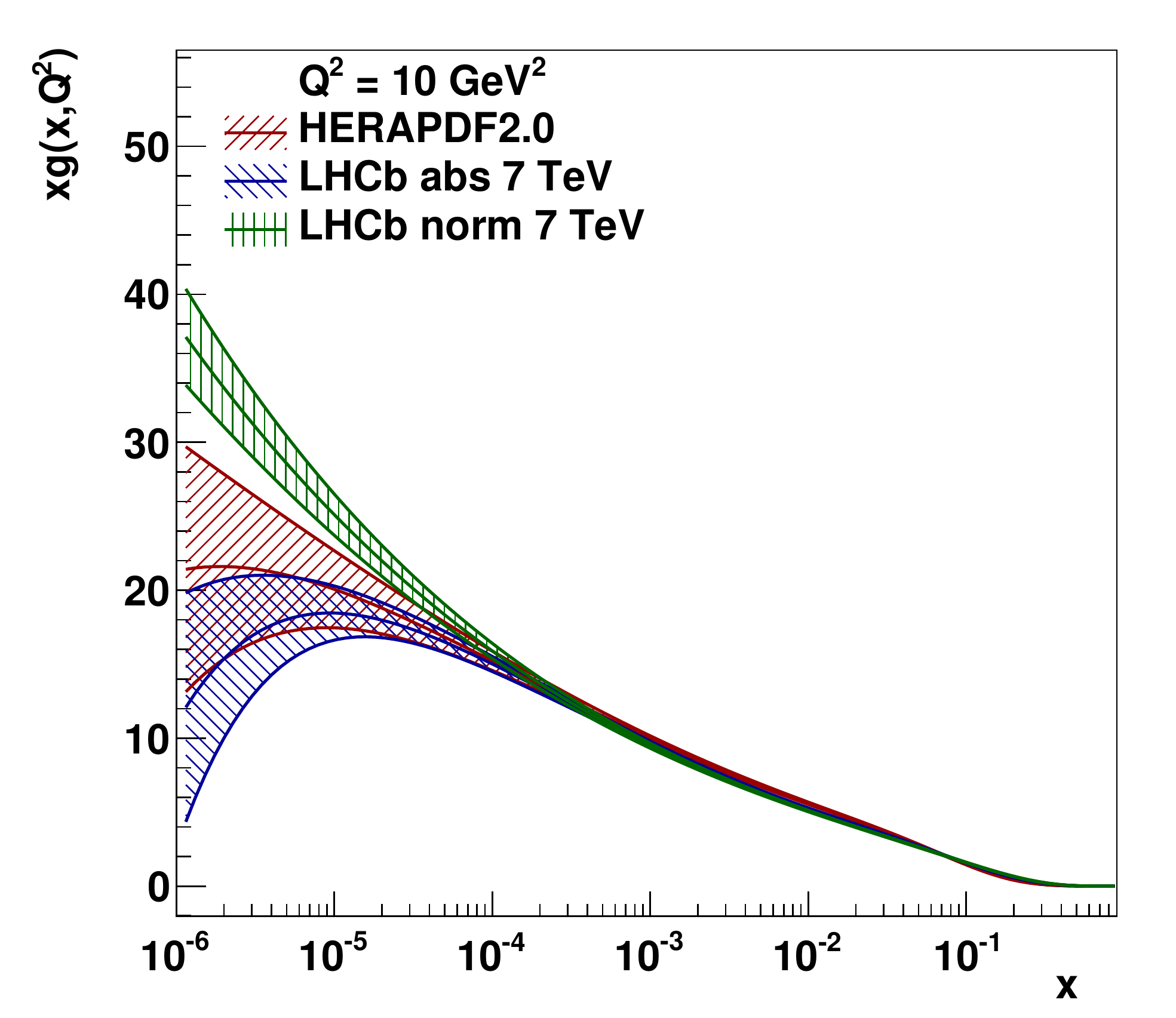}
\includegraphics[width=0.4\textwidth]{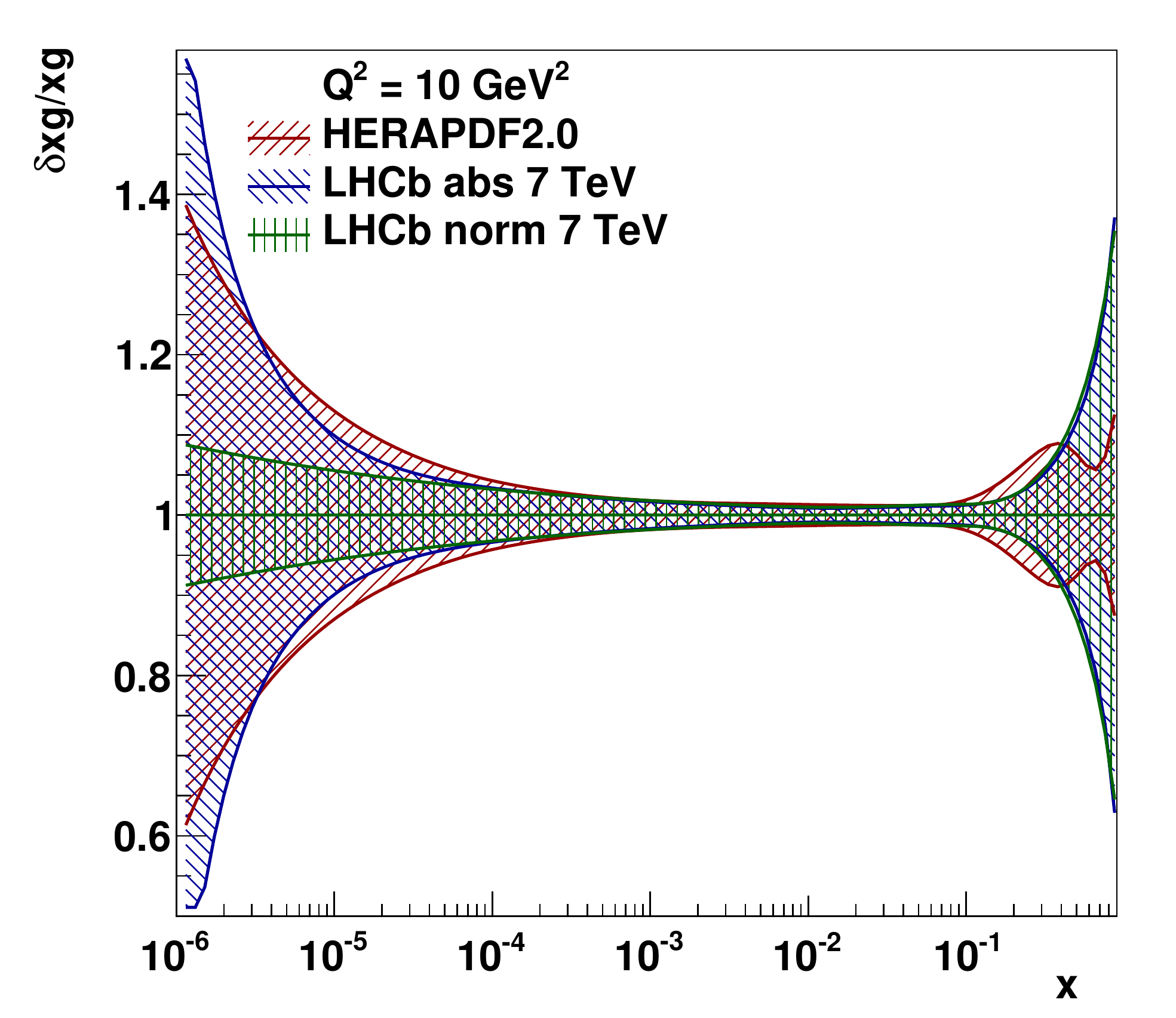}\\
\caption{  The extracted PDFs and its relative PDF uncertainties at the scale $Q^{2}$= 10 GeV$^{2}$, as a function of $x$ for $xu_{v}$,  $xd_{v}$, $xg$, and $xsea$, determined
with a fit to the HERA DIS data (red), adding with absolute LHCb 7 TeV data (blue)~\cite{Aaij:2017qml}, and finally adding with normalised 7 TeV (green) data sets~\cite{Aaij:2017qml}.  The widths of the bands represent the total uncertainties.} \label{fig:six}
\end{figure}

\begin{figure}
\center
\includegraphics[width=0.4\textwidth]{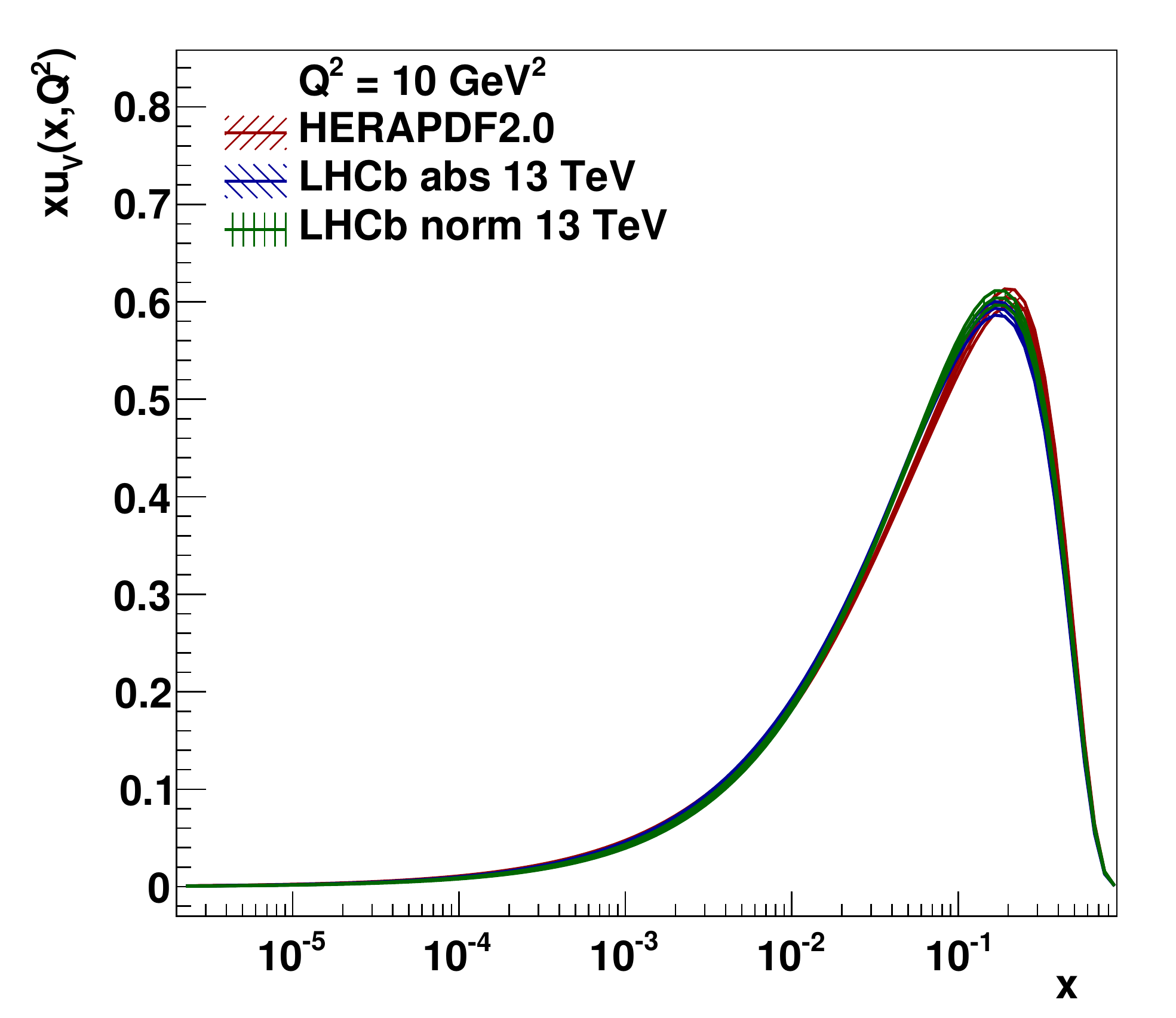}
\includegraphics[width=0.4\textwidth]{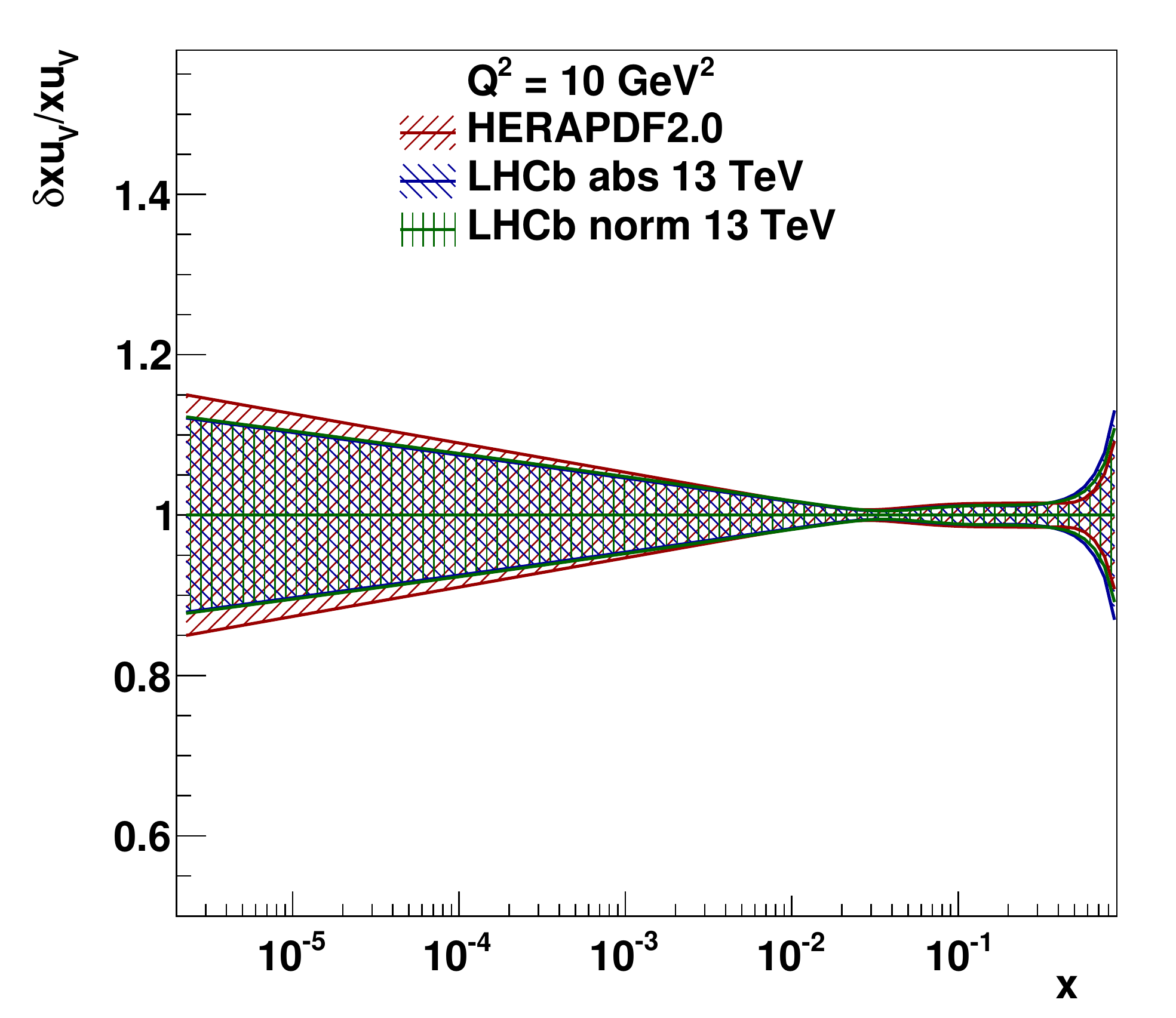}\\
\includegraphics[width=0.4\textwidth]{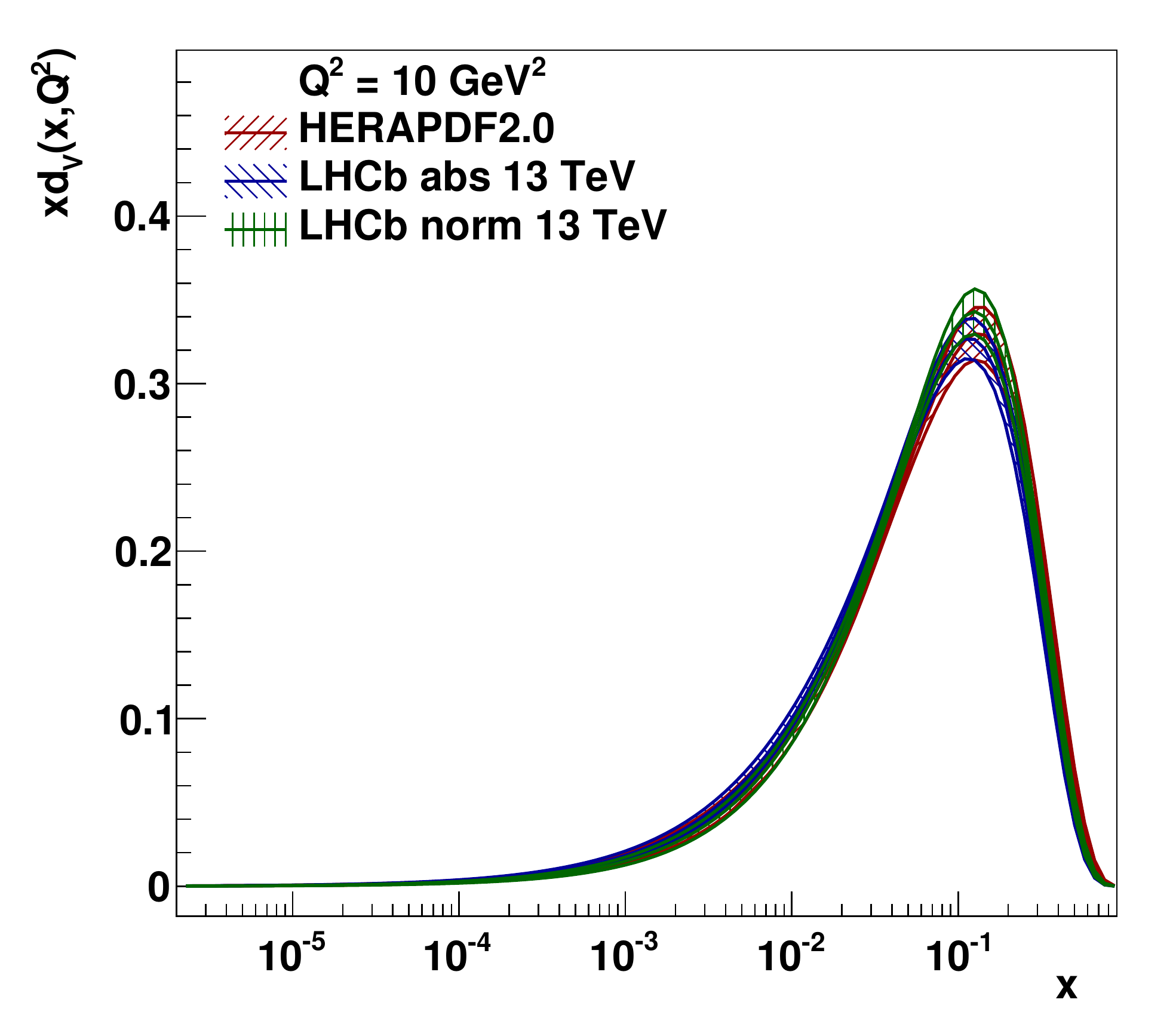}
\includegraphics[width=0.4\textwidth]{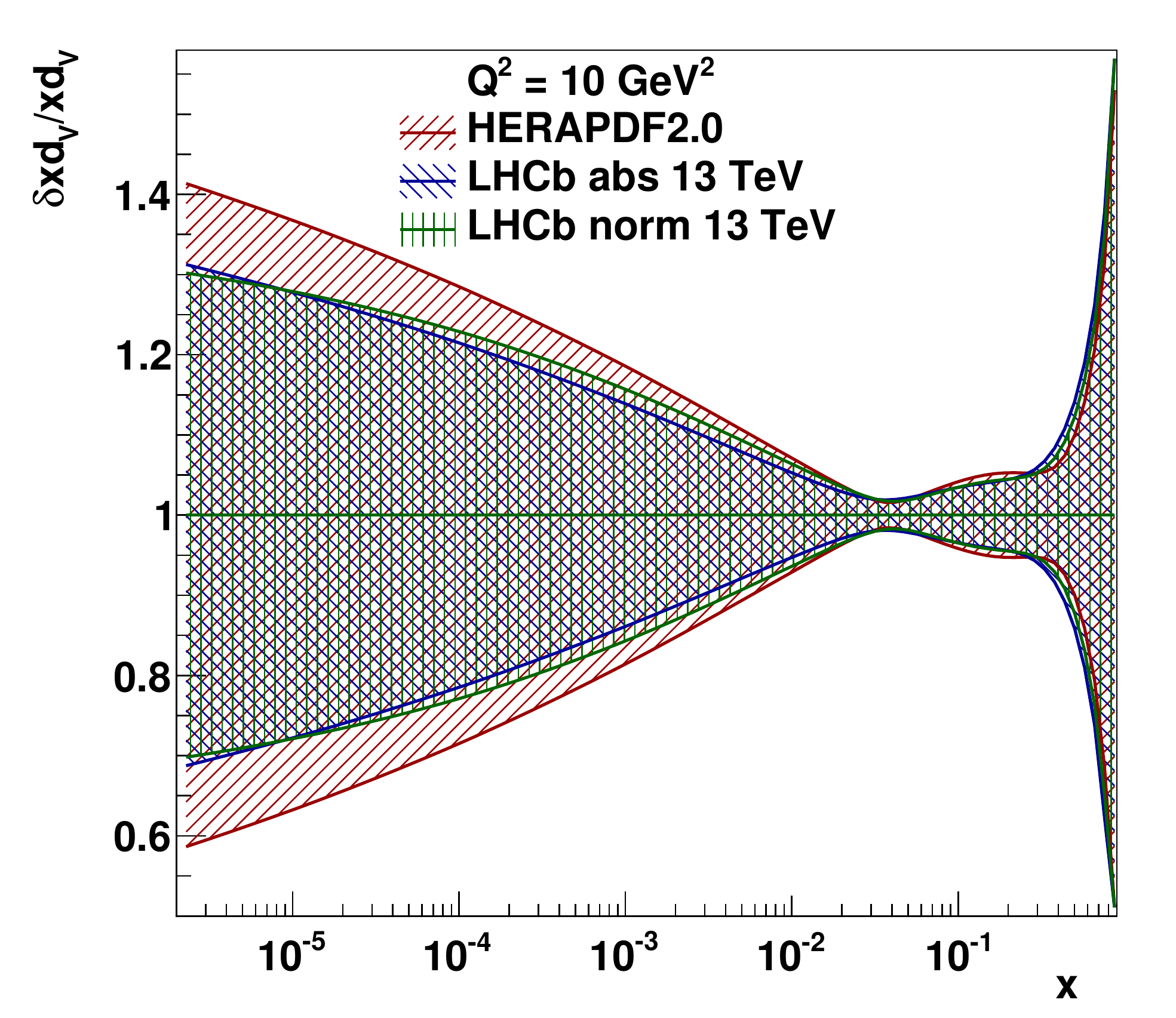}\\
\includegraphics[width=0.4\textwidth]{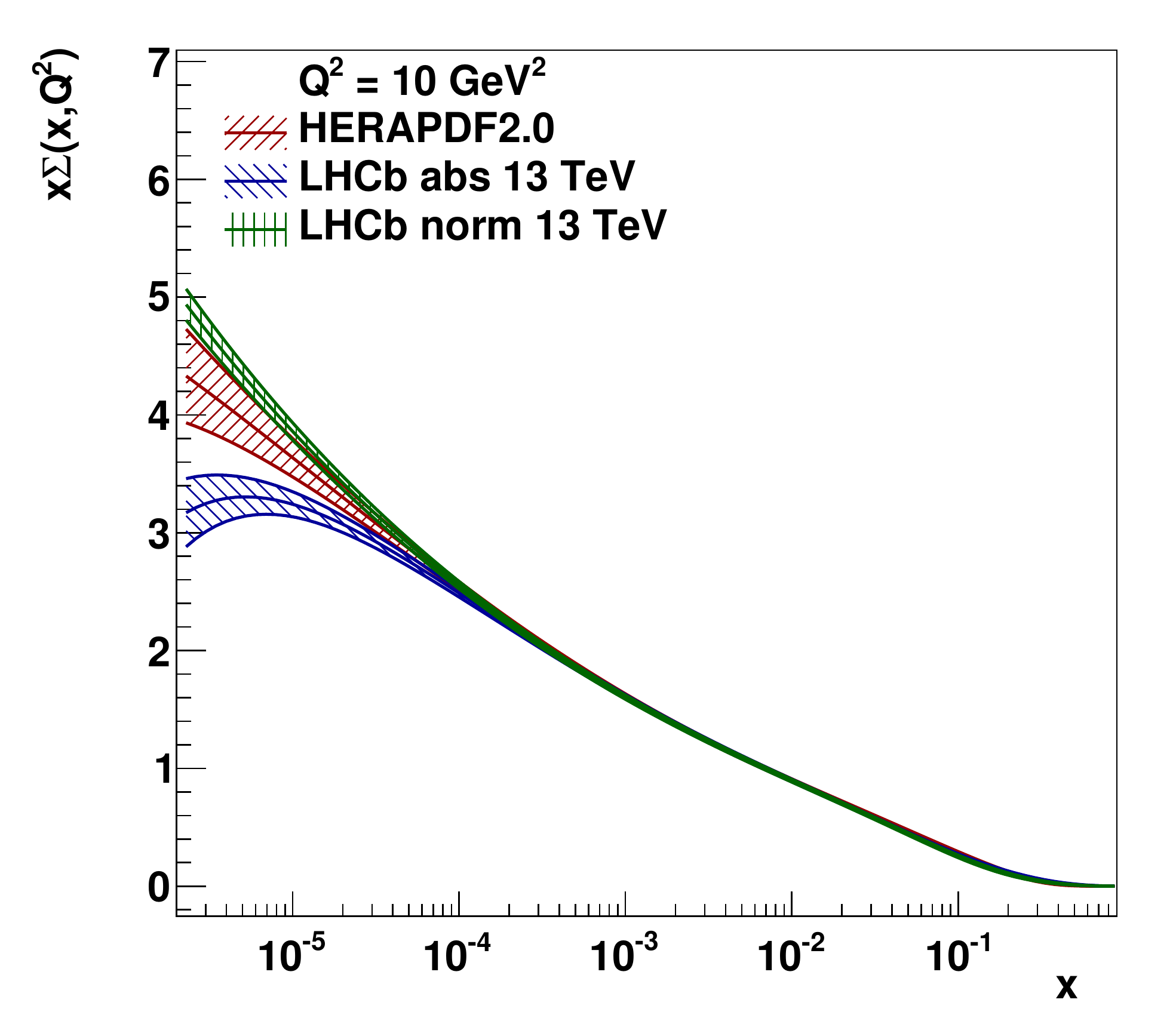}
\includegraphics[width=0.4\textwidth]{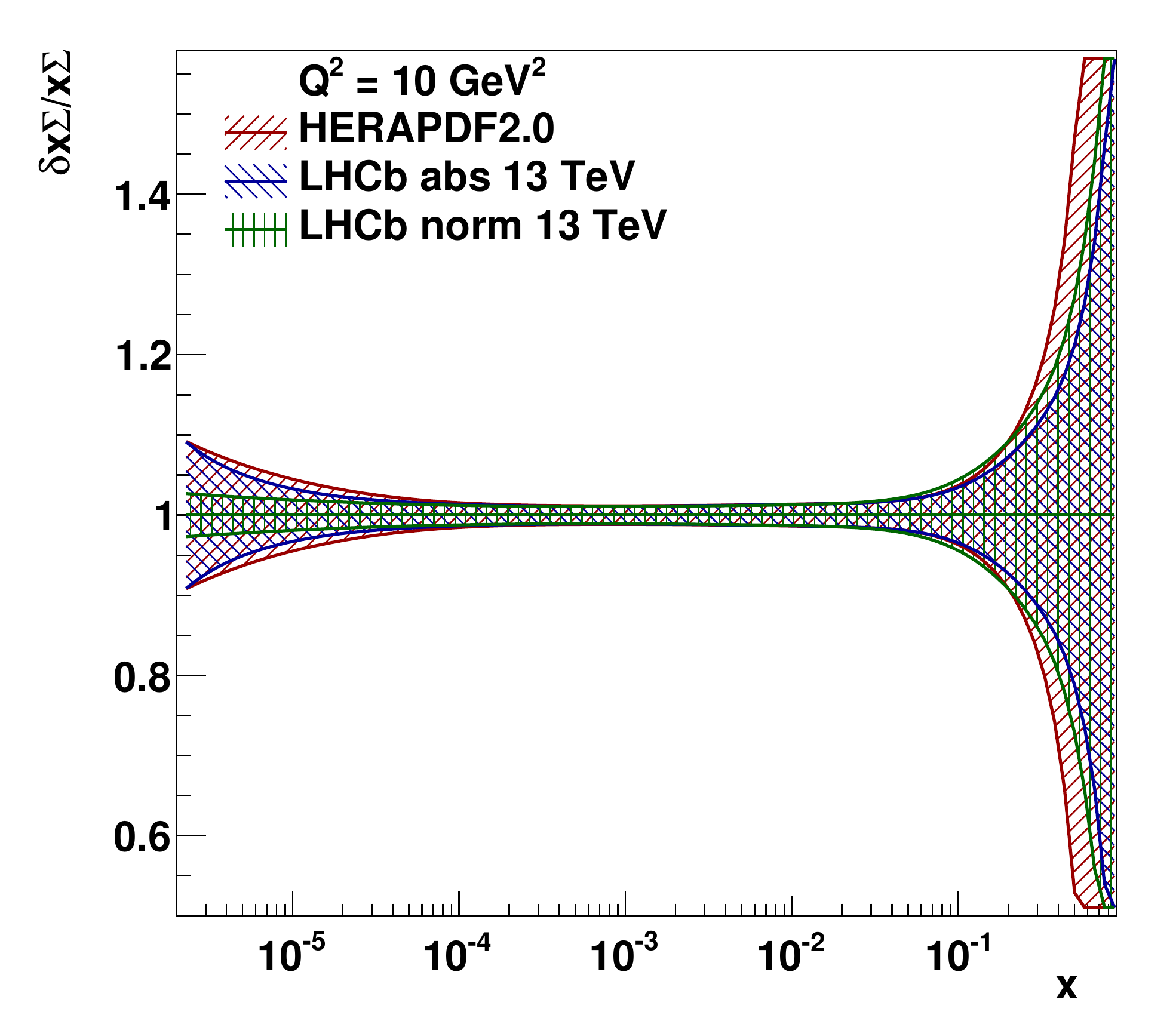}\\
\includegraphics[width=0.4\textwidth]{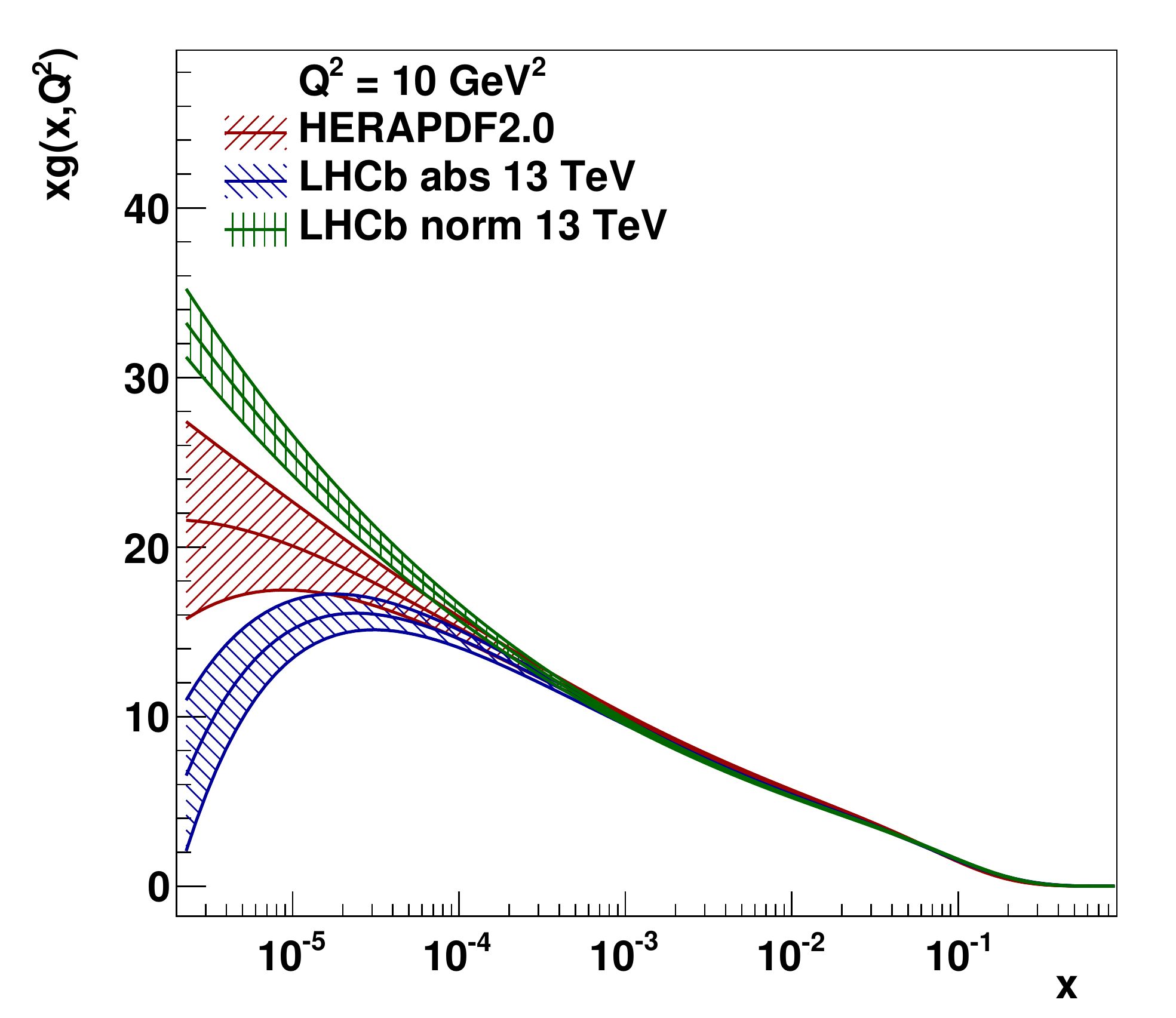}
\includegraphics[width=0.4\textwidth]{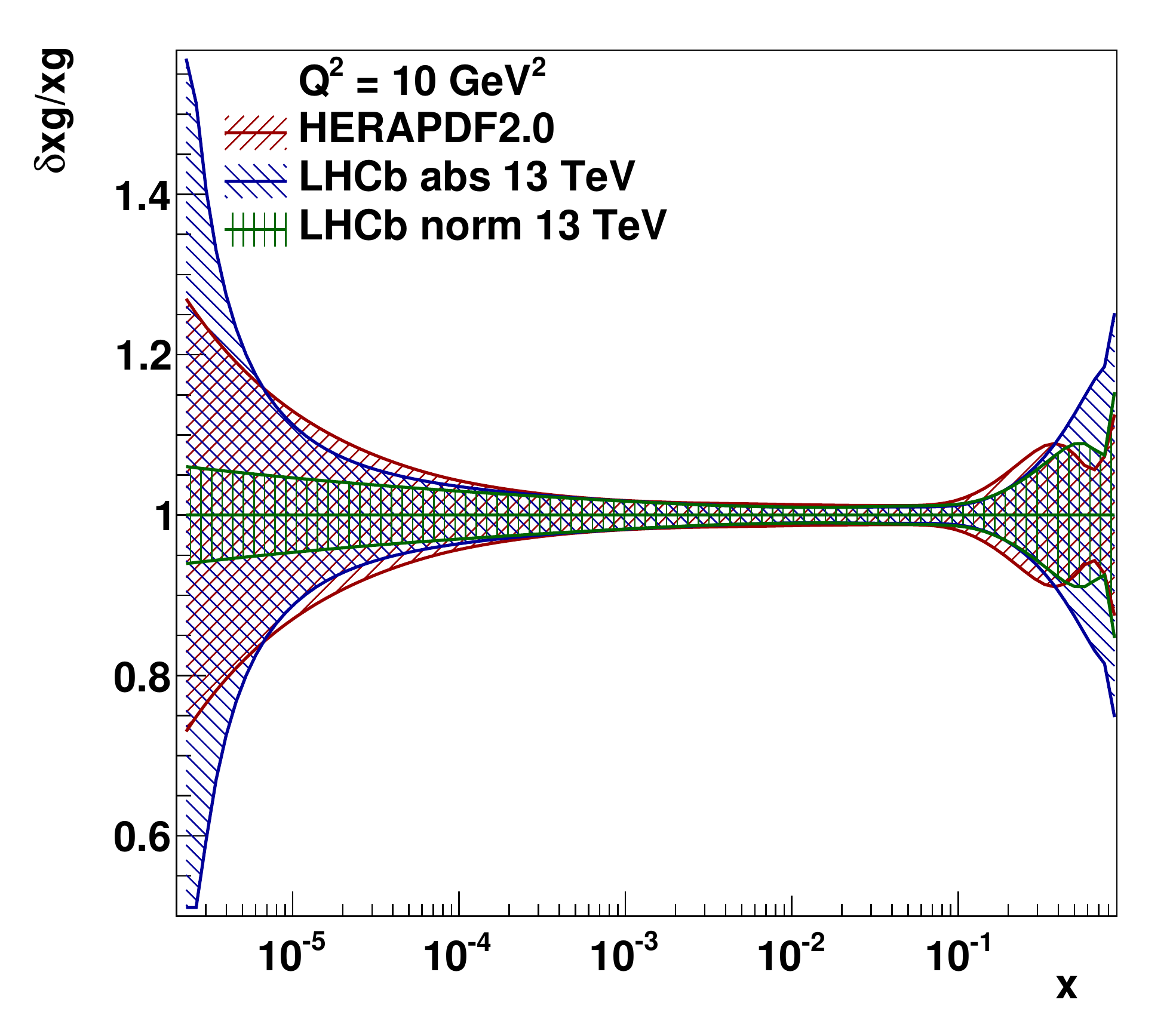}\\
\caption{  The extracted PDFs and its relative PDF uncertainties at the scale $Q^{2}$= 10 GeV$^{2}$, as a function of $x$ for $xu_{v}$,  $xd_{v}$, $xg$, and $xsea$, determined
with a fit to the HERA DIS data (red), adding with absolute LHCb 13 TeV data (blue)~\cite{Aaij:2017qml}, and finally adding with normalised LHCb 13 TeV (green) data sets~\cite{Aaij:2017qml}.  The widths of the bands represent the total uncertainties.} \label{fig:seven}
\end{figure}

As mentioned before, in order to reduce the scale dependence of the theoretical prediction, we used the normalised cross-section data in the analyses explained above. But, to examine the effect of theoretical uncertainty, such as the factorization and renormalisation scale uncertainties, and the uncertainty on the PDFs due to the fragmentation model for heavy quarks, a comparison of constraints on the PDFs obtained with the absolute and normalised LHCb cross sections would be worthwhile. To this aim, as a next step, we have also performed new analyses including the absolute LHCb cross sections. 
The values of the $\chi ^2$ and number of data points for each data sets, and the total $\chi ^2$ for each analyses have been illustrated in Table~\ref{tab:three} (for the best values of fit parameters see the columns included label ``abs" of Table~\ref{tab:two}).
Fig.~\ref{fig:six} shows a comparison between the extracted PDFs for $ u_v $ and $ d_v $ valence-quarks, gluon $ g $, and sum of sea-quarks $ \Sigma $ with their  uncertainties at the scale $Q^{2}$= 10 GeV$^{2}$, as a function of $x$, for analyses with including absolute and normalised LHCb data at 7 TeV. These comparisons clearly show the effect of theoretical uncertainty of the absolute and normalized LHCb cross sections on the PDFs, since the resulted uncertainties in the gluon and sea distributions have decreased remarkably, specially at smaller values of $ x $, using the normalized LHCb cross sections. 
The corresponding results for comparison between the absolute and normalized LHCb cross sections
at 13 TeV have been shown in Fig.\ref{fig:seven}. As can be seen, similar results are also gained  
at 13 TeV which confirm that using the normalized LHCb cross sections leads to less uncertainty
for PDFs.

\begin{figure}
\center
  \includegraphics[width=0.8\textwidth]{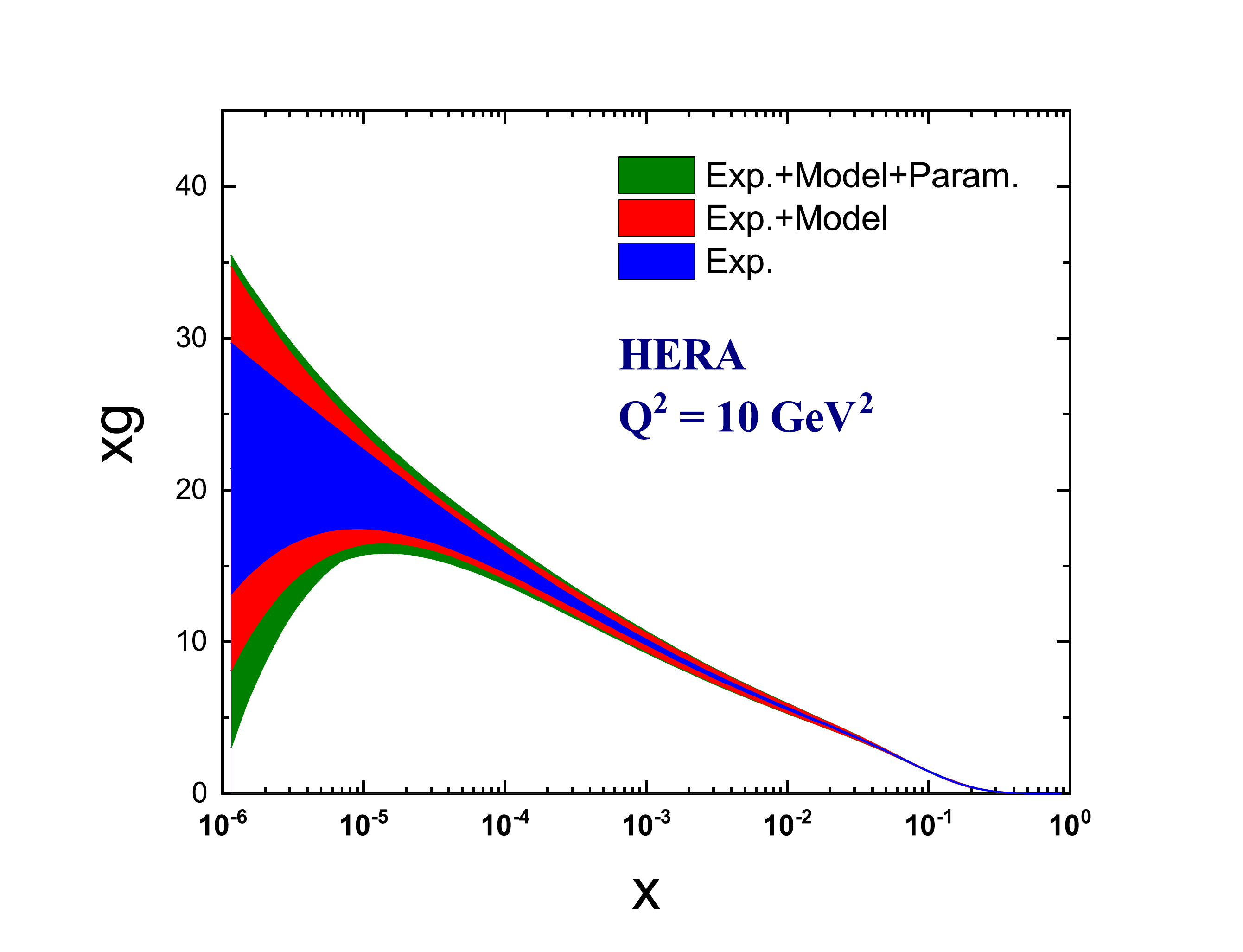}\\
  \includegraphics[width=0.8\textwidth]{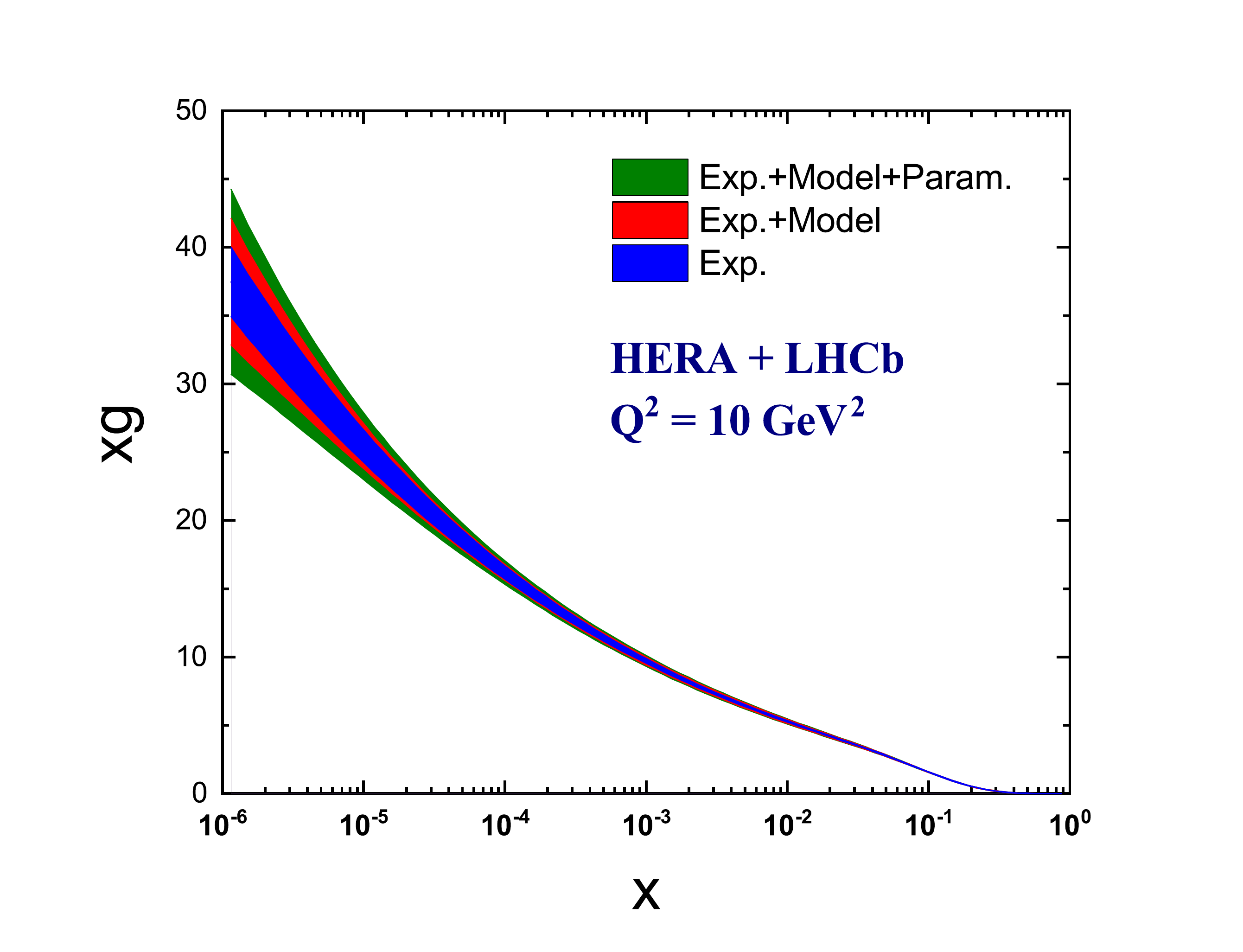}\\
\caption{A comparison between the experimental, model and parametrization uncertainties on the gluon distribution extracted from the analysis of HERA data solely (upper panel) and the analysis including the normalised LHCb data at $ \sqrt s= $13 TeV (lower panel). The results are related to $ Q^2=10 $ GeV$ ^2 $. } \label{fig:Errors}
\end{figure}

As a last step, for examining the stability of our results with different assumptions, we have also studied the model and parametrization uncertainties. The results obtained, both for analysis of HERA data solely and the analysis including the normalised LHCb data at $ \sqrt s= $13 TeV, have been shown in Fig.~\ref{fig:Errors} where the experimental, model and parametrization uncertainties have been compared with each other at $ Q^2=10 $ GeV$ ^2 $. It should be noted that the most effective sources in theoretical uncertainties of heavy-flavour production cross section are the renormalisation and factorisation scales dependence and the value of heavy-quark masses and coupling constant. However, the uncertainties due to considering the fraction of strange quarks $ f_s $ and also parameter $C_g^\prime$ in the gluon parametrization as fixed parameters should be also considered. In the present study, the contributions of the model uncertainties are calculated by considering the value of coupling constant $\alpha_S(m_Z)$, heavy quark masses $ m_c $ and $ m_b $, fraction of strange quarks $ f_s $, and parameter $C_g^\prime$ as free parameters of the fit and comparing the results obtained with the central fit. Moreover, the uncertainties due to the variation of the QCD scales are also included to model uncertainties as follows. The renormalisation and factorisation scales $\mu_r$ and $\mu_f$ are set to $\mu_f = \mu_r =C Q$ for inclusive cross section 
and $\mu_r = \mu_f=C \sqrt{Q^2+4m^2_Q}$ for heavy flavor production at HERA, where $ C $ is varying around unity by a factor 0.5 and 2. For the case of heavy flavour production at the LHCb, the uncertainties due to the variation of the QCD scales are also calculated by varying $\mu_r = \mu_f=\sqrt{p^2_T+m^2_Q}$ by a factor of 2 up and down. The optimal values of the free parameters
of the model introduced above and their variations have been summarized in Table~\ref{tab:model} both for analysis of HERA data solely and the analysis including the LHCb data at $ \sqrt s= $13 TeV. A very interesting point can be found from the results obtained is concerning the fraction of strange quarks $ f_s $, and parameter $C_g^\prime$. In fact, as can be seen, the values of these parameters become closer to their known values, namely 0.31 and 25, when the LHCb data are included in the analysis. To be more precise, $ f_s $ changes from 0.57 to 0.36 and $C_g^\prime$ changes from 16.0 to 21.4.

For calculating the parametrization uncertainty we follow a similar procedure described in Refs.~\cite{Aaron:2009aa,Abramowicz:1900rp,Abramowicz:2014zub}. According to this procedure, additional parameters should be added one by one in the functional form of the parametrisations which are used in the analysis. By comparing the maximal differences resulted in the distributions with results of the central fit, one can calculate the parametrisation uncertainties. In this regard, we tried to consider, as much as possible, more flexibility for PDF parametrisations, especially for sea quarks and gluon, to estimate the parametrisation uncertainty. We found that, for the analysis of HERA data solely, it does not show a significant impact on the valence quarks distributions but changes considerably the lower limit of the error bands of the gluon and sea quarks distributions. However, the results are somewhat different for the gluon and sea quark distributions when the normalised LHCb data at $ \sqrt s= $13 TeV are included in the analysis. To be more precise, considering more flexibility for gluon and sea quark distributions in this case has a moderate impact on the resulted distributions for them and then the parametrisation uncertainty. Note that the total uncertainty can be calculated as usual by the experimental, model and parametrisation uncertainties in quadrature.

%%%%%%%%%%%%%%%%%%%%%%%%%%%%%%%%%%%%%%%%%%%%%%%%%%%%%%%%%%%%%%%%%%%%%%%%%%%%%%%%%%%%%
\section{Summary and conclusions}
\label{sec:six}
It is well established now that the heavy-flavour production at forward rapidities in $pp$ collisions at the LHC is very sensitive to the gluon PDF at low $x$ and can put tighter constraints on it in this kinematic region.
In this work, we studied, for the first time, the impact of newest LHCb measurements for bottom production
in the forward region~\cite{Aaij:2017qml}, at center-of-mass energy of $ \sqrt s= $7 and 13 TeV on the PDFs, utilizing the xFitter framework~\cite{Alekhin:2014irh}.
In this respect, we separately included both absolute and normalised LHCb data into a NLO global analysis of the inclusive
and heavy-flavour production measurements at HERA. For making theoretical predictions, we used the massive NLO calculations in the FFNS~\cite{Nason:1987xz,Nason:1989zy,Beenakker:1990maa} with the number of flavours $ N_f=3 $ which is implemented 
in xFitter by using original routines from the MNR calculations~\cite{Mangano:1991jk,MNRcode}. 
As a result, we found that there is a good agreement 
between the LHCb data and theoretical predictions, so that the values of $ \chi^2 $ per data point obtained for them are about equal to unity.
We showed that both LHCb 7 and 13 TeV normalised data have almost a same impact on PDFs
and lead to a significant reduction in sea quarks and gluon uncertainties at low $x$. However, the LHCb measurements cannot lead to remarkable changes in valence-quarks uncertainties.
 %Moreover, particular attention was paid to study the effects of the LHCb data at large $x$, and it was indicated that in this region, PDFs are also affected with inclusion of LHCb data, especially at 7 TeV. 
We have also investigated the impact of the LHCb data on the mass of heavy quarks.
We shown that the inclusion of these data leads to a decrease in the mass of heavy quarks in both cases, but the LHCb data at 7 TeV will affect the mass of charm and beauty more than 13 TeV data.
In order to do further research, we studied, for instance, the correlations between the differential cross section of $B^{\pm}$ production measured by LHCb at $ \sqrt{s}=13 $ TeV, and $ xg(x) $ distribution for two different rapidity values $ y=2 $ and 4.5. We showed that the correlation results are consistent with the changes observed in gluon distribution after the inclusion of the LHCb data.
In order to examine the effect of theoretical uncertainties, such as the factorization and renormalisation scale uncertainties, and the uncertainty on the PDFs due to the fragmentation model for heavy quarks, we repeated the analyses by replacing the absolute LHCb data and compared the results with corresponding ones obtained using the normalised LHCb cross sections. We demonstrated that the resulted uncertainties in the gluon and sea distributions are decreased remarkably, specially at smaller values of $ x $, using the normalized data. Moreover, we estimated the amount of the model and parametrization uncertainties due to various sources both for analysis of HERA data solely and the analysis including the LHCb data at $ \sqrt s= $13 TeV. The results obtained illustrate that by inclusion of the LHCb data in a QCD analysis of PDFs, the gluon distribution at low $x$ is directly probed. This causes that the parametrization uncertainty of gluon distribution decreases significantly. 

Overall, the results obtained confirm that the inclusion of the LHCb measurements to a global analysis of PDFs  
imposes tighter constraints on the gluon and sea distributions at low $x$,
down to $x \sim 10^{-6}$, just like the results obtained in the analysis of PROSA Collaboration~\cite{Zenaiev:2015rfa}. However, our results show that the inclusion of the LHCb beauty production data at $ \sqrt s= $13 TeV can put somewhat more tighter constraints on the gluon and sea distributions rather than the data at $ \sqrt s= $7 TeV. On the other hand, since these data include a region that is currently not covered by other experimental data which have used in the global analysis of PDFs, the inclusion of them can provide more accurate PDFs for the LHC era.

\section{Acknowledgement}
We thank J. Rojo, L. Rottoli and R. Gauld for useful comments on the fragmentation model for heavy quarks. We are grateful to R. Placakyte for useful discussion related to the xFitter framework.

%%%%%%%%%%%%%%%%%%%%%%%%%%%%%%%%%%%%%%%%%%%%%%%%%%%%%%%%%%%%%%%%%%%%%%%%%%%%%%%%%%%%%

%%%%%%%%%%%%%%%%%%%%%%%%%%%%%%%%%%%%%%%%%%%%%%%%%%%%%%%%%%%%%%%%

\newpage

\begin{table}[h!]
\caption{The list of experimental data 
included in three global fits. The first analysis dose not include neither LHCb 7 TeV nor 13 TeV~\cite{Aaij:2017qml} data.
The second and third analyses include the normalised LHCb 7 and 13 TeV data of the $B^{\pm}$ production cross-section, respectively. For each data set, we presented the $ \chi^2 $/number of points.}
\label{tab:one}
\center
\footnotesize
\begin{tabular}{lccc}
\hline  \hline
  Dataset     & HERAPDF2.0   & LHCb 7 norm   & LHCb 13 norm \\ 
 \hline     
  Charm cross section H1-ZEUS~\cite{Abramowicz:1900rp}  & 52 / 52& 50 / 52& 51 / 52  \\ 
Beauty cross section ZEUS~\cite{Abramowicz:2014zub}  & 12 / 17& 12 / 17& 12 / 17  \\   
  HERA1+2 NCep 820~\cite{Abramowicz:2015mha} & 66 / 70& 70 / 70& 69 / 70  \\ 
  HERA1+2 NCep 920~\cite{Abramowicz:2015mha} & 433 / 377& 480 / 377& 459 / 377  \\ 
  HERA1+2 NCep 460~\cite{Abramowicz:2015mha} & 221 / 204& 223 / 204& 221 / 204  \\ 
  HERA1+2 NCep 575~\cite{Abramowicz:2015mha} & 220 / 254& 229 / 254& 225 / 254  \\ 
  HERA1+2 CCep\cite{Abramowicz:2015mha} & 55 / 39& 69 / 39& 66 / 39  \\ 
  HERA1+2 CCem\cite{Abramowicz:2015mha} & 50 / 42& 52 / 42& 54 / 42  \\ 
  HERA1+2 NCem\cite{Abramowicz:2015mha} & 224 / 159& 229 / 159& 225 / 159  \\ 
  LHCb 7 TeV~\cite{Aaij:2017qml}  & - & 57 / 108& -   \\ 
  LHCb 13 TeV~\cite{Aaij:2017qml}  & - & - & 56 / 108  \\ 
   \hline 
  Correlated $\chi^2$  & 93& 113& 102  \\ 
  Log penalty $\chi^2$  & -4.36& -35.13& -16.31  \\ 
  Total $\chi^2$ / dof  & 1422 / 1198& 1550 / 1307& 1522 / 1307  \\ 
\hline  \hline
\end{tabular}
\end{table}

\newpage

\begin{table}[h!]
\caption{The optimal values of the input PDF parameters at $Q^{2}$= 1.4 GeV$^{2}$ and the charm and beauty quark masses determined from the various global analyses. The fixed parameters have been shown with blue color.}
\label{tab:two}
\center
\tiny
\begin{tabular}{lccccc}
\hline\hline
 Parameter   & HERAPDF2.0 & LHCb 7 norm & LHCb 13 norm &  LHCb 7 abs  &  LHCb 13 abs  \\ 
  \hline
  $ B_g $ & $-0.10 \pm 0.16$& $0.17 \pm 0.14$& $0.03 \pm 0.15$ & $-0.2341 \pm 0.0073$ &$-0.307 \pm 0.076$ \\ 
  $ C_g $ & $8.7 \pm 1.0$& $5.68 \pm 0.64$& $7.41 \pm 0.80$ & $5.15 \pm 0.12$ &  $5.72 \pm 0.42$\\ 
  $ A'_g $ & $2.05 \pm 0.89$& $0.13 \pm 0.43$& $1.90 \pm 0.64$ & $1.146 \pm 0.047 $ & $1.28 \pm 0.19$\\ 
  $ B'_g $ & $-0.188 \pm 0.086$& $-0.18 \pm 0.23$& $-0.06 \pm 0.11$ &  $-0.2947 \pm 0.0049$ & $-0.346 \pm 0.047$\\ 
  $ C'_g $ & $\textcolor{blue}{ 25.00 }$& $\textcolor{blue}{ 25.00 }$& $\textcolor{blue}{ 25.00 }$ & $\textcolor{blue}{ 25.00 }$ & $\textcolor{blue}{ 25.00 }$ \\ 
  $ B_{u_v} $ & $0.696 \pm 0.044$& $0.696 \pm 0.022$& $0.745 \pm 0.022$ & $0.713 \pm 0.011$& $0.725 \pm 0.025$ \\ 
  $ C_{u_v} $ & $4.772 \pm 0.080$& $4.705 \pm 0.072$& $4.637 \pm 0.081$ & $4.683 \pm 0.058$ &   $4.655 \pm 0.096$\\ 
  $ E_{u_v} $ & $14.7 \pm 2.7$& $12.0 \pm 1.3$& $9.4 \pm 1.1$  &$10.60 \pm 0.69$ & $9.6 \pm 1.4$\\
  $ B_{d_v} $ & $0.872 \pm 0.094$& $0.935 \pm 0.066$& $0.944 \pm 0.073$ & $0.908 \pm 0.046 $& $0.872 \pm 0.064$ \\ 
  $ C_{d_v} $ & $4.34 \pm 0.38$& $5.03 \pm 0.34$& $5.05 \pm 0.35$ & $5.04 \pm 0.26 $&  $5.02 \pm 0.35$ \\ 
  $ C_{\bar U} $ & $8.45 \pm 0.71$& $3.20 \pm 0.46$& $3.07 \pm 0.42$ & $2.79 \pm 0.25$ & $2.58 \pm 0.32$ \\ 
  $ D_{\bar U} $ & $18.5 \pm 3.5$& 0 & 0  & 0 &  0\\ 
  $ A_{\bar D} $ & $0.130 \pm 0.011$& $0.171 \pm 0.011$& $0.1672 \pm 0.0098$ & $0.1623 \pm 0.0067$&   $0.1581 \pm 0.0088$ \\ 
  $ B_{\bar D} $ & $-0.176 \pm 0.010$& $-0.1408 \pm 0.0079$& $-0.1457 \pm 0.0074$ &$-0.1485 \pm 0.0052$ & $-0.1536 \pm 0.0071$ \\ 
  $ C_{\bar D} $ & $4.7 \pm 1.5$& $11.8 \pm 2.5$& $7.0 \pm 1.5$ &  $8.1 \pm 1.2$ & $5.5 \pm 1.1$ \\ 
  $ \alpha_s $ & $\textcolor{blue}{ 0.1059 }$& $\textcolor{blue}{ 0.1059 }$& $\textcolor{blue}{ 0.1059 }$ & $\textcolor{blue}{ 0.1059 }$ & $\textcolor{blue}{ 0.1059 }$ \\ 
  $ f_s $ & $\textcolor{blue}{ 0.3100 }$& $\textcolor{blue}{ 0.3100 }$& $\textcolor{blue}{ 0.3100 }$ & $\textcolor{blue}{ 0.3100 }$ & $\textcolor{blue}{ 0.3100 }$ \\ 
  $ m_c $ & $1.431 \pm 0.052$& $1.332 \pm 0.053$& $1.379 \pm 0.048$ & $1.376 \pm 0.034$ & $1.412 \pm 0.045$ \\ 
  $ m_b $ & $4.40 \pm 0.15$& $4.20 \pm 0.15$& $4.29 \pm 0.17$  & $4.765 \pm  0.070 $ & $4.687 \pm 0.079$\\ 
%  $A_f$ & - & 1 & 1 &$0.0794 \pm  0.0039 $& $0.114 \pm 0.010$   \\ 
%  $A_r$ & - & 1 & 1 &$0.086 \pm  0.010 $& $0.087 \pm 0.013$  \\ 
%
 \hline \hline
\end{tabular}
\end{table}

\newpage

\begin{table}[h!]
\caption{The list of experimental data 
included in three global fits. The first analysis dose not include neither LHCb 7 TeV nor 13 TeV~\cite{Aaij:2017qml} data.
The second and third analyses include the absolute LHCb 7 and 13 TeV data of the $B^{\pm}$ production cross-section, respectively. For each data set, we presented the $ \chi^2 $/number of points.}
\label{tab:three}
\center
\footnotesize
\begin{tabular}{lccc}
\hline  \hline
  Dataset     & HERAPDF2.0   & LHCb 7 abs   & LHCb 13 abs  \\ 
 \hline     
  Charm cross section H1-ZEUS~\cite{Abramowicz:1900rp}  & 52 / 52& 50 / 52& 52 / 52  \\ 
Beauty cross section ZEUS~\cite{Abramowicz:2014zub}  & 12 / 17& 19 / 17& 17 / 17  \\   
  HERA1+2 NCep 820~\cite{Abramowicz:2015mha} & 66 / 70& 69 / 70& 69 / 70  \\ 
  HERA1+2 NCep 920~\cite{Abramowicz:2015mha} & 433 / 377& 469 / 377& 467 / 377  \\ 
  HERA1+2 NCep 460~\cite{Abramowicz:2015mha} & 221 / 204& 223 / 204& 222 / 204  \\ 
  HERA1+2 NCep 575~\cite{Abramowicz:2015mha} & 220 / 254& 226 / 254& 224 / 254  \\ 
  HERA1+2 CCep\cite{Abramowicz:2015mha} & 55 / 39& 68 / 39& 68 / 39  \\ 
  HERA1+2 CCem\cite{Abramowicz:2015mha} & 50 / 42& 53 / 42& 55 / 42  \\ 
  HERA1+2 NCem\cite{Abramowicz:2015mha} & 224 / 159& 228 / 159& 229 / 159  \\ 
  LHCb 7 TeV~\cite{Aaij:2017qml}  & - & 100 / 108& -   \\ 
  LHCb 13 TeV~\cite{Aaij:2017qml}  & - & - & 114 / 108  \\ 
   \hline 
  Correlated $\chi^2$  & 93& 112& 103  \\ 
  Log penalty $\chi^2$  & -4.36&  -88.58& -71.70 \\ 
  Total $\chi^2$ / dof  & 1422 / 1198& 1530 / 1332& 1546 / 1332  \\ 
\hline  \hline
\end{tabular}
\end{table}

\begin{table}[h!]
\renewcommand*{\arraystretch}{1.2}
\caption{The optimal values of the free parameters of model extracted from the QCD analysis of HERA data solely and with including the normalised LHCb data at $ \sqrt s= $13 TeV and their variation used for estimating the model uncertainties.}
\label{tab:model}
\vspace{0.5cm}
\centerline{
\begin{tabular}{|l|c|c|c|c|}
\hline
 & \multicolumn{2}{c|}{HERAPDF2.0}  & \multicolumn{2}{c|}{LHCb norm 13}    \\ 
    \cline{2-3}   \cline{4-5} 
  Parameters & Fix  & Free & Fix & Free  \\
\hline
$\alpha_S(m_Z)$         & 0.105 & 0.102 & 0.105 & 0.109 \\
\hline
$m_c$ [GeV]     & $1.25$    & $1.45$        & $1.25$& $1.40$  \\
\hline
$m_b$ [GeV]     & $4.19$   & $4.53$ & $4.19$ & $4.60$  \\
\hline
$f_s$           & $0.31$      & $0.57$ & $0.31$ & $0.36$   \\
\hline
$C_g^\prime$   & $25$      & $16.0$ & $25$ & $21.4$  \\
\hline
\end{tabular}}
\end{table}

\newpage

\newpage

\newpage

\newpage

\newpage

\newpage

%\begin{figure}
%\center
%\includegraphics[width=0.48\textwidth]{pic/pdf/q210pdfuvrelratio.pdf}
%\includegraphics[width=0.48\textwidth]{pic/pdf/q210pdfdvrelratio.pdf}\\
%\includegraphics[width=0.48\textwidth]{pic/pdf/q210pdfSearelratio.pdf}
%\includegraphics[width=0.48\textwidth]{pic/pdf/q210pdfgrelratio.pdf}\\
%\caption{ Relative PDF uncertainties at the scale $Q^{2}$= 10 GeV$^{2}$, as a function of $x$ for $u_{v}$,  $d_{v}$, $g$, and $sea$, determined
%with a fit to the HERA DIS data (red), adding with LHCb 7 TeV data (blue), and finally adding with 13 TeV (green) data sets. The sea-quark distribution is defined as $\Sigma=2 \cdot (\bar{u}+\bar{d}+\bar{s})$.} \label{fig:five}
%\end{figure}

\newpage

\newpage

\newpage

\newpage

\end{document}